\documentclass[fleqn,usenatbib]{mnras}

\usepackage{newtxtext,newtxmath}

\usepackage[T1]{fontenc}
\usepackage{xfrac}

\DeclareRobustCommand{\VAN}[3]{#2}
\let\VANthebibliography\thebibliography
\def\thebibliography{\DeclareRobustCommand{\VAN}[3]{##3}\VANthebibliography}

\usepackage{graphicx}	
\usepackage{amsmath}	
\usepackage{amssymb}	
\usepackage{makecell}
\usepackage{multirow}
\usepackage{ulem}
\usepackage{xcolor}
\usepackage{float}

\newcommand{\kms}{km s$^{-1}$}
\newcommand{\dego}{$^\circ$}

\title[Evidence for episodic and patchy mass ejection in the circumstellar envelope of R Leo]{Evidence for episodic and patchy mass ejection in the circumstellar envelope of AGB star R Leonis}

\author[D.T. Hoai et al.]{D.T. Hoai \thanks{E-mail: dthoai@vnsc.org.vn},
  P.T. Nhung \thanks{E-mail: pttnhung@vnsc.org.vn},
  M.N. Tan,
  P. Darriulat,
  P.N. Diep,
  N.B. Ngoc,
  T.T. Thai
  and P. Tuan-Anh
\\
Department of Astrophysics, Vietnam National Space Center, Vietnam Academy of Science and Technology, \\
18, Hoang Quoc Viet, Nghia Do, Cau Giay, Ha Noi, Vietnam
}

\date{Accepted XXX. Received YYY; in original form ZZZ}

\pubyear{2022}

\begin{document}
\label{firstpage}
\pagerange{\pageref{firstpage}--\pageref{lastpage}}
\maketitle

\begin{abstract}
The morpho-kinematics of the circumstellar envelope of oxygen-rich AGB star R Leonis is probed using ALMA (Atacama Large Millimeter/submillimeter Array) observations of the emission of molecular lines, including in particular CO(2-1) and $^{29}$SiO(5-4). Evidence is found for an episode of enhanced mass loss, a few centuries ago, that produced a broad expanding shell of mean radius $\sim$6 arcsec and mean radial expansion velocity $\sim$5.5 \kms. Detailed scrutiny of its structure, as displayed by the emission of the CO(2-1) line, reveals strong inhomogeneity and patchy morphology. Evidence is also found, in particular from the morpho-kinematics of the emission of SiO, SO and SO$_2$ lines probing the close neighbourhood of the star, for distinct gas outflows covering broad solid angles in the south-eastern, south-western and north-western quadrants, suggesting significant contribution of the convective cell granulation in defining the pattern of mass ejection. A study of relative molecular abundances in these outflows suggests that a Local Thermal Equilibrium (LTE) description applies approximately beyond $\sim$10 stellar radii from the centre of the star but not at the smaller angular separations where the SO and SO$_2$ molecules are found to be confined. Near the stellar disc, masers of the vibrationally excited SiO lines are found to probe north-western parts of a layer of hot gas, consistent with the earlier observation of an asymmetric expanding shell within 1-2 stellar radii from the centre of the star. Globally, a picture dominated by episodic and patchy mass ejections is found to prevail.
\end{abstract}




\section{Introduction}
High resolution observations of the circumstellar envelopes of oxygen rich AGB stars, in particular using the Very Large Telescope (VLT) or the Atacama Large Millimeter/submillimeter Array (ALMA), have revealed very complex structures. Recent state-of-the art hydro-dynamical simulations \citep{Hofner2019,Freytag2017,Liljegren2018} suggest that such complexity is the result of shocks induced by a combination of convection and stellar pulsations. They induce an effective line broadening for ALMA observations of the emission of molecular lines in the close neighbourhood of the star. Typical examples are R Dor \citep{Hoai2020, Nhung2021}, $o$ Cet \citep{Nhung2022a} or R Hya \citep{Nhung2022b}. In the present article we study the case of R Leo, for which ALMA observations of the emissions of the $^{12}$CO($\nu$=0,$J$=2-1) and $^{29}$SiO($\nu$=0,$J$=5-4) (hereafter abbreviated as CO(2-1) and $^{29}$SiO(5-4) respectively) lines have been the subject of preliminary studies by \citet{Fonfria2019a,Fonfria2019b}. We contribute additional and complementary analyses of the same data as well as new analyses of other ALMA observations, including the emissions of molecular lines $^{28}$SiO($\nu$=1,$J$=5-4), $^{28}$SiO($\nu$=1,$J$=8-7), $^{28}$SiO($\nu$=2,$J$=5-4), SO($^3\Sigma$,$\nu$=0,$N_J$=5$_5$-4$_4$), SO$_2$($\nu$=0,$J_{\text{Ka,Kc}}$=22$_{2,20}$-22$_{1,21}$) and SO$_2$($\nu$=0,$J_{\text{Ka,Kc}}$=13$_{2,12}$-12$_{1,11}$), hereafter abbreviated as SiO($\nu$=1,5-4), SiO($\nu$=1,8-7), SiO($\nu$=2,5-4), SO(5$_5$-4$_4$), SO$_2$(22$_{2,20}$-22$_{1,21}$) and SO$_2$(13$_{2,12}$-12$_{1,11}$) respectively.

R Leo is an oxygen-rich AGB star. Its distance to Earth is not well measured, varying from 71 pc \citep{Gaia2018} to 114 pc \citep{vanLeeuwen2007}. In our analysis we use 114 pc, a value consistent with the distance obtained from the period-luminosity relation \citep{Haniff1995}. We note that after submission of the present paper for publication, an article of utmost relevance to this issue has been published \citep{Andriantsaralaza2022}, arguing in favour of a distance of 100$\pm$5 pc. The pulsation period is 310 days \citep{Feast1996, Watson2006} and the mass-loss rate, evaluated from single dish observations, is $\sim$1.0 10$^{-7}$ solar masses per year \citep{Danilovich2015}. The star effective temperature is $\sim$2500-3000 K and its angular diameter is 25-30 mas \citep{Perrin1999, Fedele2005, Wittkowski2016, Matthews2018}. \citet{Vlemmings2019} using ALMA observations to probe the extended atmosphere within 1-2 stellar radii, measured an asymmetric, radial expansion of the radio photosphere at a mean velocity of 10.6$\pm$1.4 \kms\ at stellar phases between 0.41 and 0.46.

The close neighbourhood of the star has been observed in the near-infrared by \citet{Fedele2005} using VLTI/VINCI and in the mid-infrared by \citet{Paladini2017} using VLTI/MIDI. SiO masers have been the subject of numerous VLBA observations of outstanding quality by \citet{SoriaRuiz2007}, \citet{Cotton2004, Cotton2008, Cotton2009} and \citet{Desmurs2014}. \citet{Herpin2006}, using the IRAM 30 m telescope, have analysed single-dish SiO maser Zeeman observations to estimate the value of the magnetic field in the near CSE at the level of 4.2-4.6 G; quasi-periodic polarization fluctuations have been interpreted as resulting from a precessing planetary magnetosphere, suggesting the presence of an evaporating Jovian planet \citep{Wiesemeyer2009}. Using the Stratospheric Observatory for Infrared Astronomy (SOFIA), \citet{Fonfria2020} have detected the fluorescence of hot carbon dioxide in the extended atmosphere of the star.

Millimetre observations of the expanding gas have measured a terminal velocity of 6-9 \kms\ \citep{DeBeck2010, Ramstedt2014} and have detected significant abundances of CO, H$_2$O, OH, SiO, SO, SO$_2$ and HCN molecules \citep{Hinkle1979, Bujarrabal1994, Etoka1997, Bieging2000, GonzalezDelgado2003, Ohnaka2004, Schoier2013, Danilovich2020}.

Of particular relevance to the present work are the ALMA observations of $^{29}$SiO(5-4) and CO(2-1) molecular emissions reported by \citet{Fonfria2019a,Fonfria2019b}. They interpret the observed morpho-kinematics in terms of an expanding, thin and dense partial shell behind the star that is being shocked by a faster, collimated matter ejection coming from the star. This shell would have been ejected in an apparently single strong event during the last 700 yr. Its thickness ($\sim$1 arcsec) suggests a mass-loss rate enhancement lasting for about 50 yr. They remark that the $^{29}$SiO(5-4) emission around the systemic velocity resembles a deformed spiral structure that suggests the existence of a companion of R Leo. Within $\sim$0.17 arcsec from the centre of the star, they find evidence for the presence of hot gas, elongated along the NE/SW direction, contributing to continuum emission. They remark that the molecular emission displays prominent red-shifted absorption in front of the star, typical of material in-fall, and lateral gas motions compatible with the presence of a possibly rotating torus-like structure.  

\section{Observations and data reduction}

The present work uses archival observations of R Leo from ALMA projects ADS/JAO.ALMA\#2016.1.01202.S (PI. J. Fonfr\'{i}a) in Band 6 and ADS/JAO.ALMA\#2017.1.00862.S (PI Kami\'{n}ski) in Band 7.

The Band 6 (211-275 GHz) observations were made on 1 October 2016, 22 March and 3 May 2017 with low angular resolution ($\sim$0.2 arcsec) and on 21 September and 27 October 2017 with high angular resolution ($\sim$0.03 arcsec). The baselines were between 15 m and 13.9 km long. The total time on source of the observations was $\sim$80 min.

The Band 7 (275-373 GHz) observations were made on 16 November 2017 with configuration C43-8 which has baselines between 92 m and 8.5 km. The total time on source of the observations was $\sim$22 min. Both Band 6 and Band 7 observations were made shortly after the minimum of the light curve, the stellar phase of the former being $\sim$0.50 \citep{Fonfria2019b}.

The data were calibrated using CASA\footnote{https://casa.nrao.edu/} with the standard script provided by the ALMA staff. The images of CO(2-1) in Section 4 were produced from the merged data of Band 6 observations using GILDAS\footnote{https://www.iram.fr/IRAMFR/GILDAS/}/IMAGER\footnote{https://imager.oasu.u-bordeaux.fr/} with robust weighting (threshold value of 1.5). The images of the continuum and the lines at high resolution in Band 6 were produced from the data observed on 27 October 2017 with baselines between 191 m and 13.9 km and total time on source of $\sim$15 min. The phase centre was shifted to the centre of the continuum emission and the visibilities were changed accordingly. Phase self-calibration was done for the continuum data and the results were applied to the line data. The cleaned images were produced using CASA with uniform weighting for the continuum and natural weighting for the lines. We also used CASA for imaging the data of Band 7 with robust parameter of 0.5. In all cases of line emission imaging, continuum emission has not been subtracted at visibility level in order to preserve the possibility to account properly for its contribution in the image plane.

Relevant parameters are listed in Tables \ref{tab1} and \ref{tab2}.

We use coordinates centred on the peak of continuum emission, $x$ pointing east, $y$ pointing north and $z$ pointing away from Earth. The projected angular separation to the star is calculated as $R$=$\sqrt{x^2+y^2}$. Position angles, $\omega$, are measured counter-clockwise from north. The Doppler velocity $V_\text{z}$ spectrum is centred on a star systemic velocity of $-$0.5 \kms\ \citep{Teyssier2006}, which, however, is not known to better than $\sim$$\pm$0.5 \kms.

\begin{table*}
  \caption{Observations: relevant parameters}
  \label{tab1}
  \begin{tabular}{cccccc}
    \hline
&Line&
Beam&
Pixel size&
Channel spacing (\kms)&
Noise (mJy beam$^{-1}$)\\
\hline
\multirow{8}{*}{Band 6}&
$^{29}$SiO(5-4)&
47$\times$44 mas$^2$, PA=40\dego&5$\times$5 mas$^2$&1.36&0.7\\
&SiO($\nu$=1,5-4)&46$\times$38 mas$^2$, PA=34\dego&{5$\times$5 mas$^2$}&1.36&0.6\\

&SiO($\nu$=2,5-4)&47$\times$44 mas$^2$, PA=39\dego&{5$\times$5 mas$^2$}&1.36&0.8\\

&SO(5$_5$-4$_4$)&46$\times$38 mas$^2$, PA=34\dego&{5$\times$5 mas$^2$}&1.36&0.5\\

&SO$_2$(22$_{2,20}$-22$_{1,21}$)&46$\times$37 mas$^2$, PA=34\dego&{5$\times$5 mas$^2$}&1.36&0.8\\

&CO(2-1) extended&43$\times$36 mas$^2$, PA=33\dego&{5$\times$5 mas$^2$}&0.635&0.8\\

&Continuum&28$\times$26 mas$^2$, PA=34\dego&{5$\times$5 mas$^2$}&$-$&0.06\\

&CO(2-1) merged&106$\times$86 mas$^2$, PA=71\dego&30$\times$30 mas$^2$&0.635&0.5\\
\hline

\multirow{3}{*}{Band 7}&SiO($\nu$=1,8-7)&49$\times$44 mas$^2$, PA=$-$12\dego&7.4$\times$7.4 mas$^2$&0.85&2.1\\

&SO$_2$(13$_{2,12}$-12$_{1,11}$)&49$\times$44 mas$^2$, PA=$-$11\dego&7.4$\times$7.4 mas$^2$&0.85&2.1\\ 

&Continuum&45$\times$40 mas$^2$, PA=$-$38\dego&7.4$\times$7.4 mas$^2$&$-$&0.33\\
\hline
  \end{tabular}
  \end{table*}

\begin{table*}
  \caption{Line parameters \citep{Muller2005}: frequency $f$, Einstein coefficient $A_\text{ji}$, energy $E_\text{u}$ and angular momentum $J$ of the upper level. The factors $k$ are the ratio of the temperature to the partition function $Q$. In general they depend on temperature and we give their values at 100 and 1000 K. }
  \label{tab2}
  \begin{tabular}{ccccccc}
    \hline
Line&
$f$ [GHz]&
$A_\text{ji}$ [10$^{-3}$ Hz]&
$E_\text{u}$ [K]&
2$J$+1&
$k$ (100 K)&
$k$ (1000 K)\\
\hline
CO(2-1)&
230.5380&
0.69 10$^{-3}$&
16.6 &
5&
2.75&
2.63\\
$^{29}$SiO(5-4)&
214.3858&
0.50 &
30.9 &
11&
1.03&
0.85\\
SiO($\nu$=1,5-4)&
215.5959&
0.50 &
1789 &
11&
1.01&
0.85\\
SiO($\nu$=1,8-7)&
344.9165&
2.2 &
1844 &
17&
1.01&
0.85\\
SiO($\nu$=2,5-4)&
214.0885&
0.51 &
3552 &
11&
1.04&
0.86\\
SO(5$_5$-4$_4$)&
215.2206&
0.12 &
44.1 &
11&
0.37&
0.35\\
SO$_2$(13$_{2,12}$-12$_{1,11}$)&
345.3385&
0.24 &
93 &
27&
0.087&
0.021\\
SO$_2$(22$_{2,20}$-22$_{1,21}$)&
216.6433&
0.093&
248.4&
45&
0.087&
0.021\\
H$^{13}$CN(4-3)&
344.2001&
1.9&
41.4&
9&
2.04&
0.87\\
\hline
  \end{tabular}
  \end{table*}

\section{Continuum emission} 
\begin{figure*}
  \includegraphics[height=4.3cm,trim=.0cm 1cm 2cm .5cm,clip]{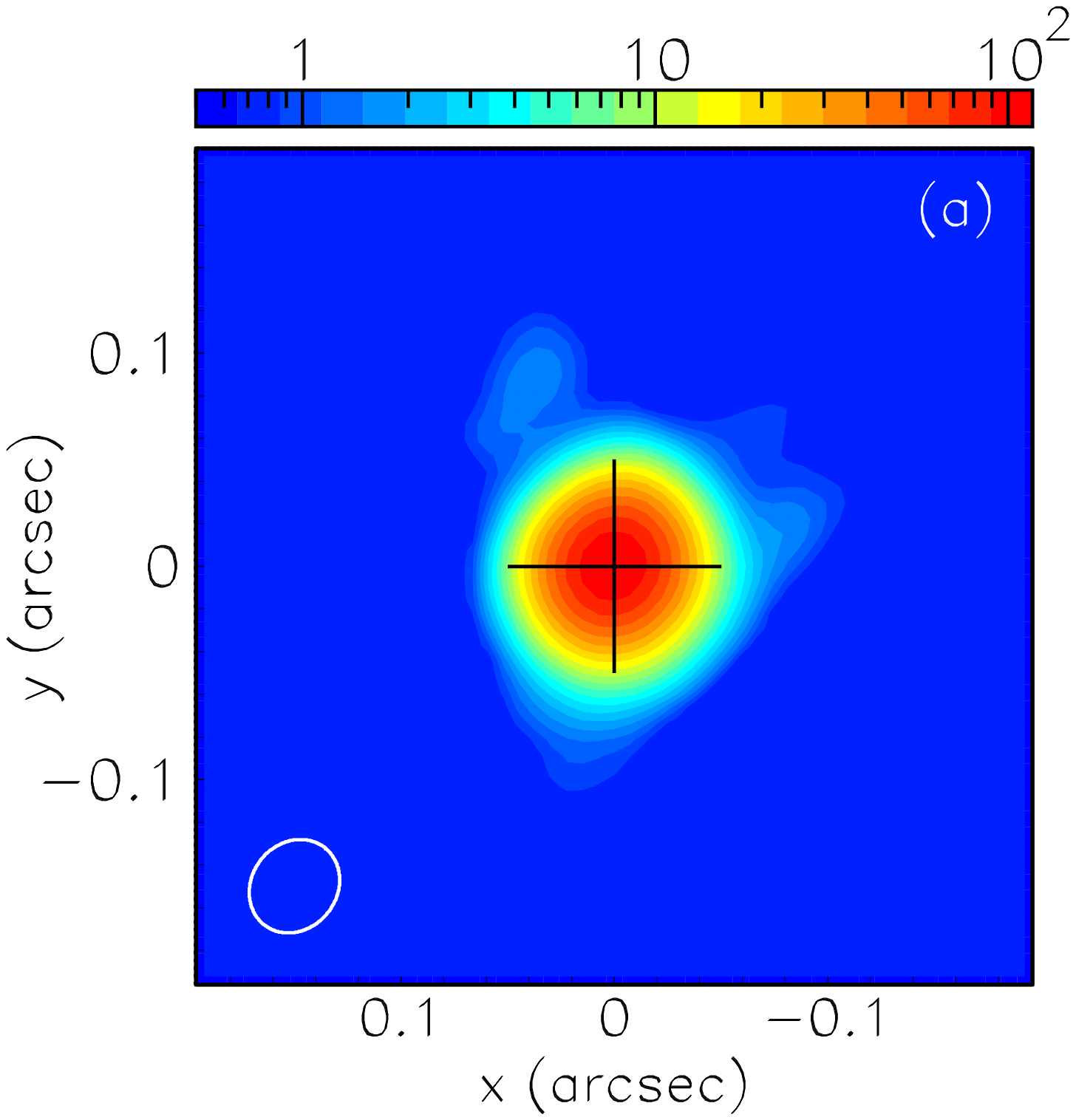}
  \includegraphics[height=4.3cm,trim=.0cm 1cm 2cm .5cm,clip]{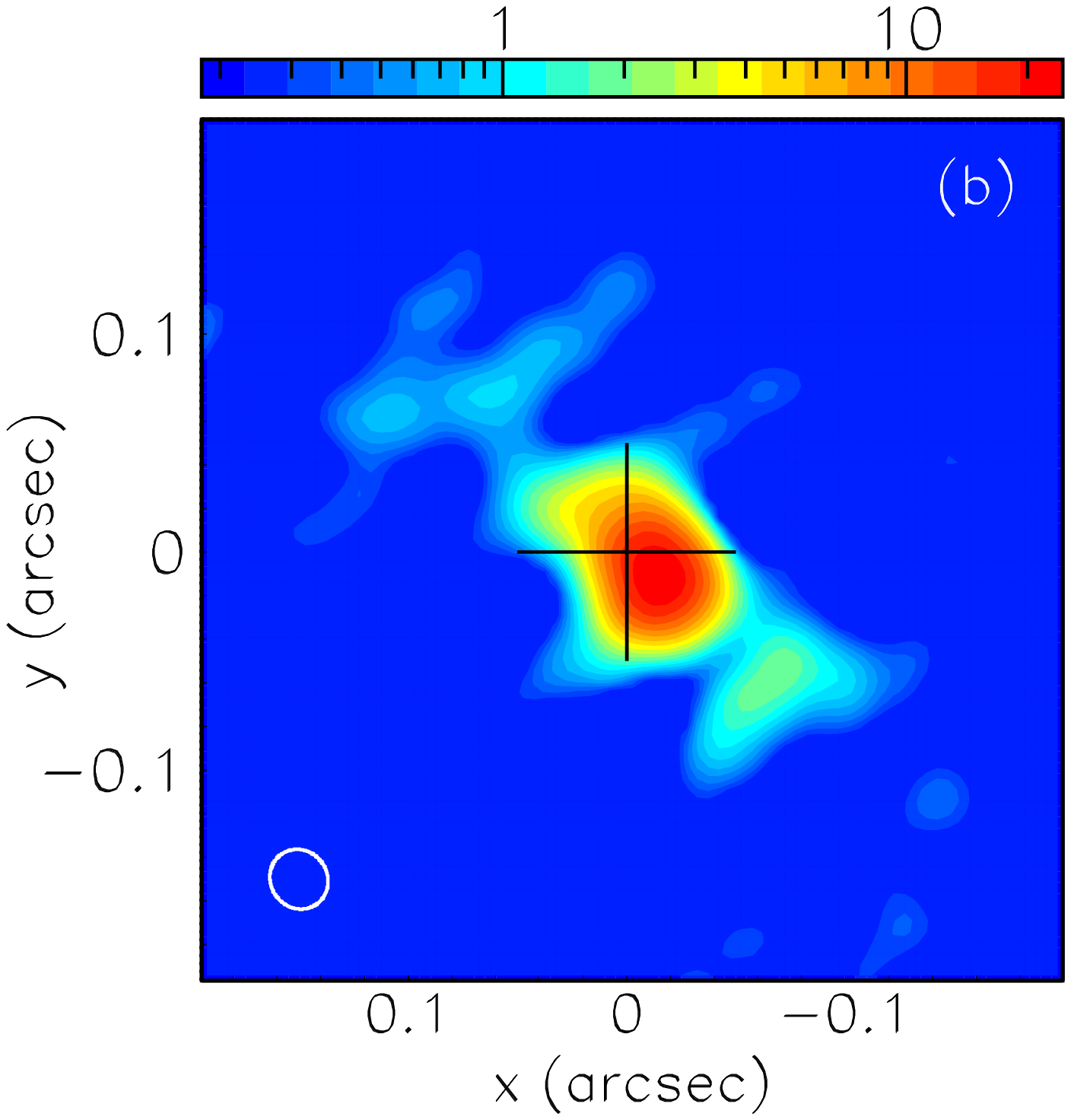}
  \includegraphics[height=4.3cm,trim=.0cm 1cm 2cm .5cm,clip]{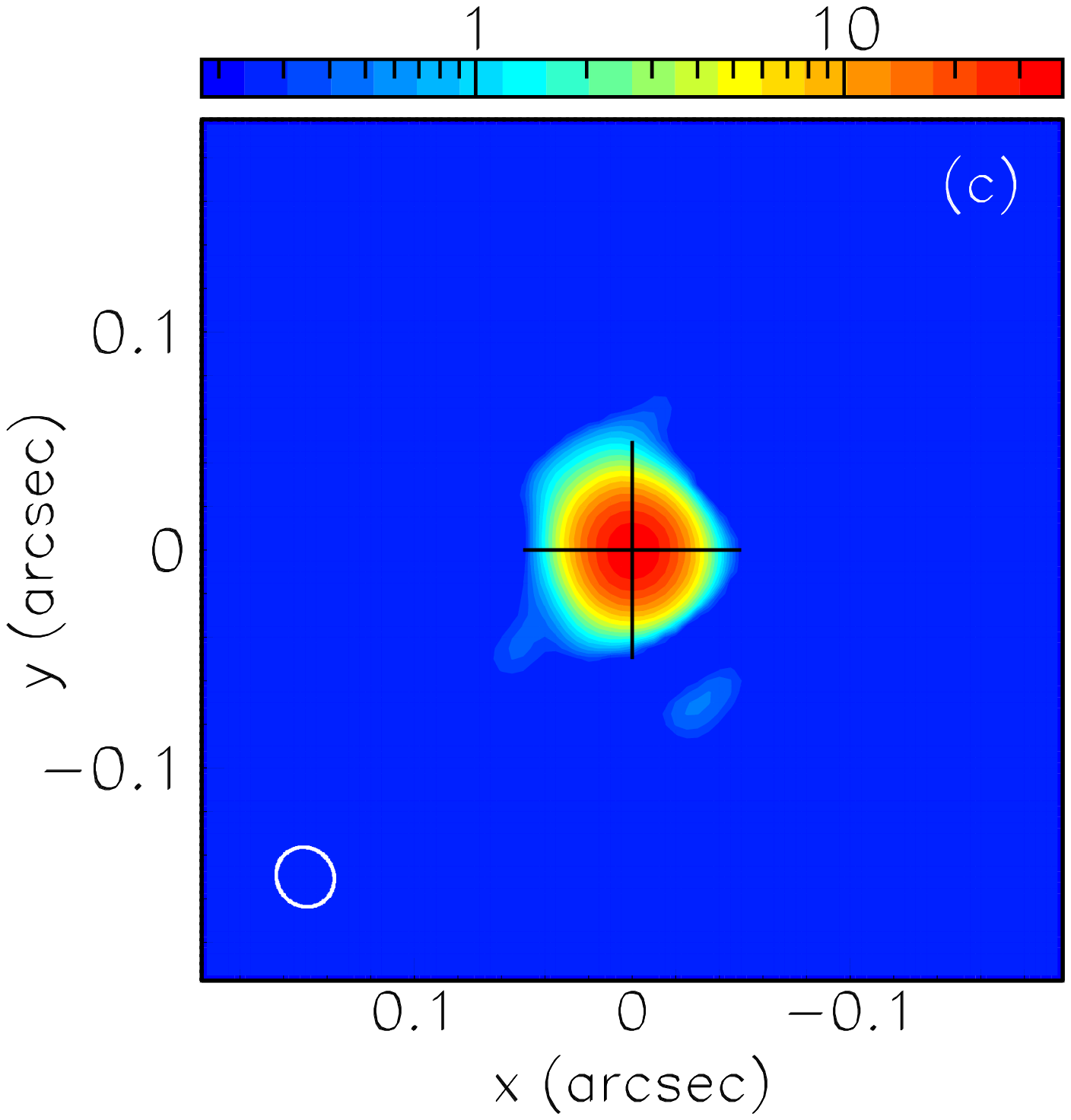}
  \includegraphics[height=4.3cm,trim=.0cm 1cm 0cm .5cm,clip]{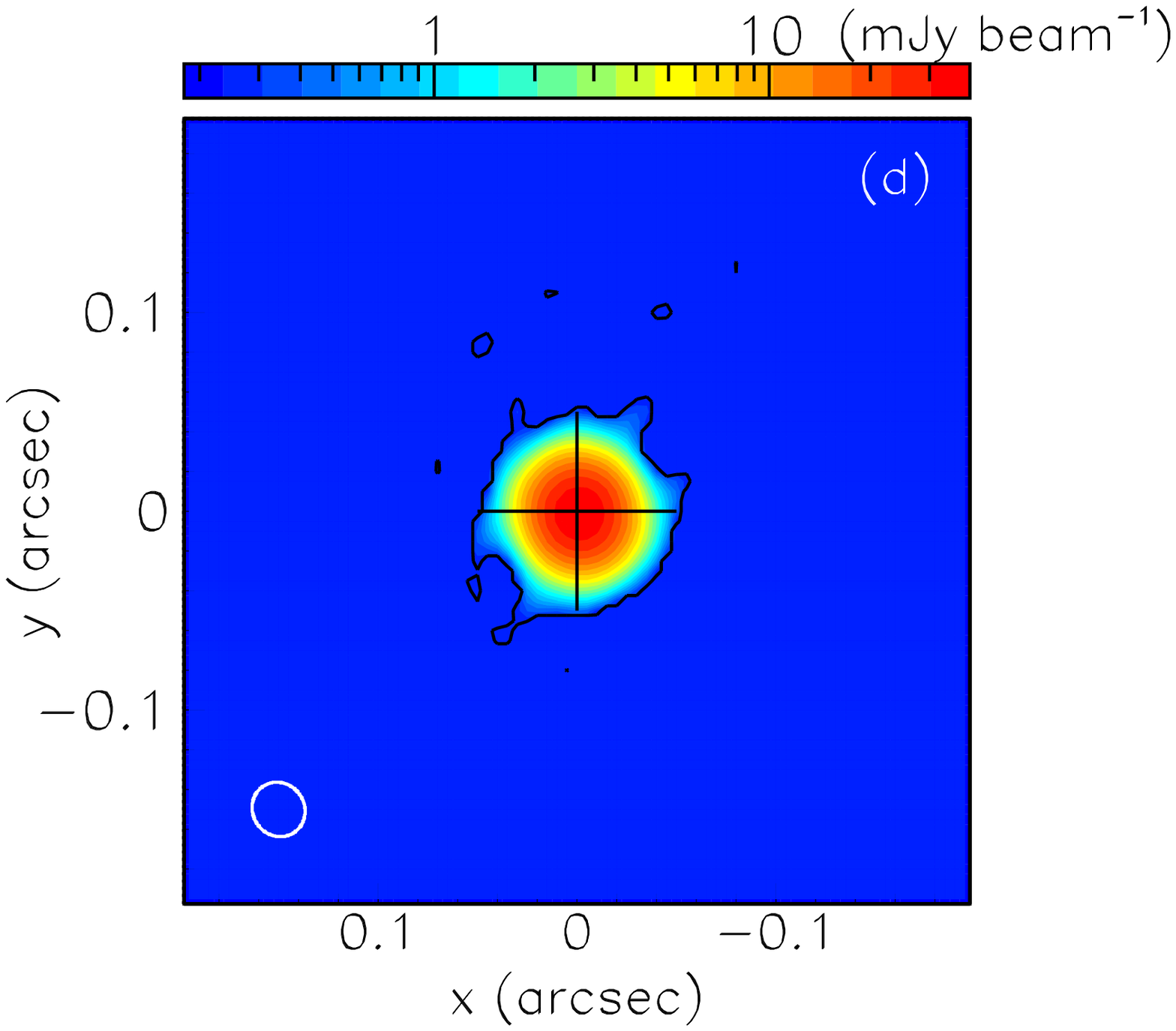}

  \caption{Continuum intensity maps for Band 7 (panel (a), beam of 45$\times$40 mas$^2$) and Band 6 (uniform weighting with beam of 28$\times$26 mas$^2$; panel (b) before phase-centre correction, panel (c) after phase-centre correction and panel (d) after self-calibration). The contour shown in panel (d) is at 3-$\sigma$ noise level.}
    \label{fig2}
\end{figure*}

While Band 7 data display a nearly circular continuum emission (Figure \ref{fig2}), as do lower frequency data \citep[e.g.][at VLA frequencies]{Matthews2018}, Band 6 data were shown by \citet{Fonfria2019b} to display clear elongation in the NE/SW direction, which they interpreted as evidence for hot gas and dust around the star, probably in a nearly edge-on torus configuration. However, this elongation is caused by a significant offset of the phase-centre and disappears after having corrected for it (Figure \ref{fig2}). The coordinates of the origin are: RA=09h47m33.492315s and Dec=11\dego25'42.9081536''. This confirms results obtained by \citet{Vlemmings2019}. These authors performed a very careful and detailed analysis of ALMA Band 6 observations and found evidence for asymmetric expansion with significant residuals from a uniform stellar disc towards the north-east. They describe it in terms of an increased opacity layer that might in reality be a directional shock wave propagating through the extended atmosphere. They argue that the observed motion is likely due to density or ionisation changes and remark that it means an expansion of material with a size of up to a quarter of the stellar disc, not inconsistent with the size of a large convective cell.

\section{CO(2-1) line emission: a radially expanding shell}

\begin{figure*}
  \includegraphics[height=7cm,trim=.5cm 0cm 0cm 0cm,clip]{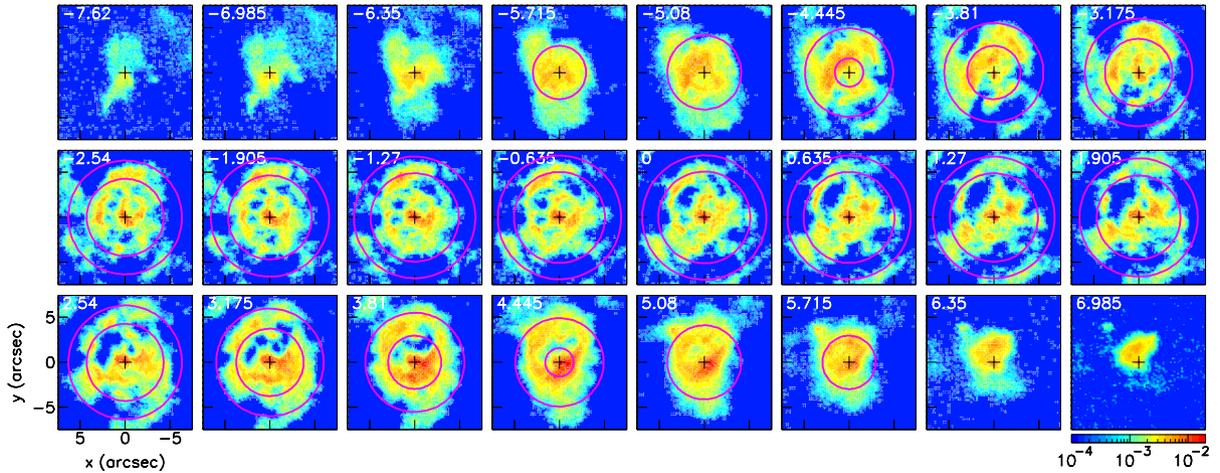}
  \caption{Channel maps of the CO(2-1) line emission. The circles are drawn for $\rho$=0.85 and 1.15, respectively. The colour scale is in units of Jy beam$^{-1}$.}
  \label{fig3}
\end{figure*}

\begin{figure*}

  \includegraphics[height=4cm,trim=.5cm 1cm 1.7cm 1.9cm,clip]{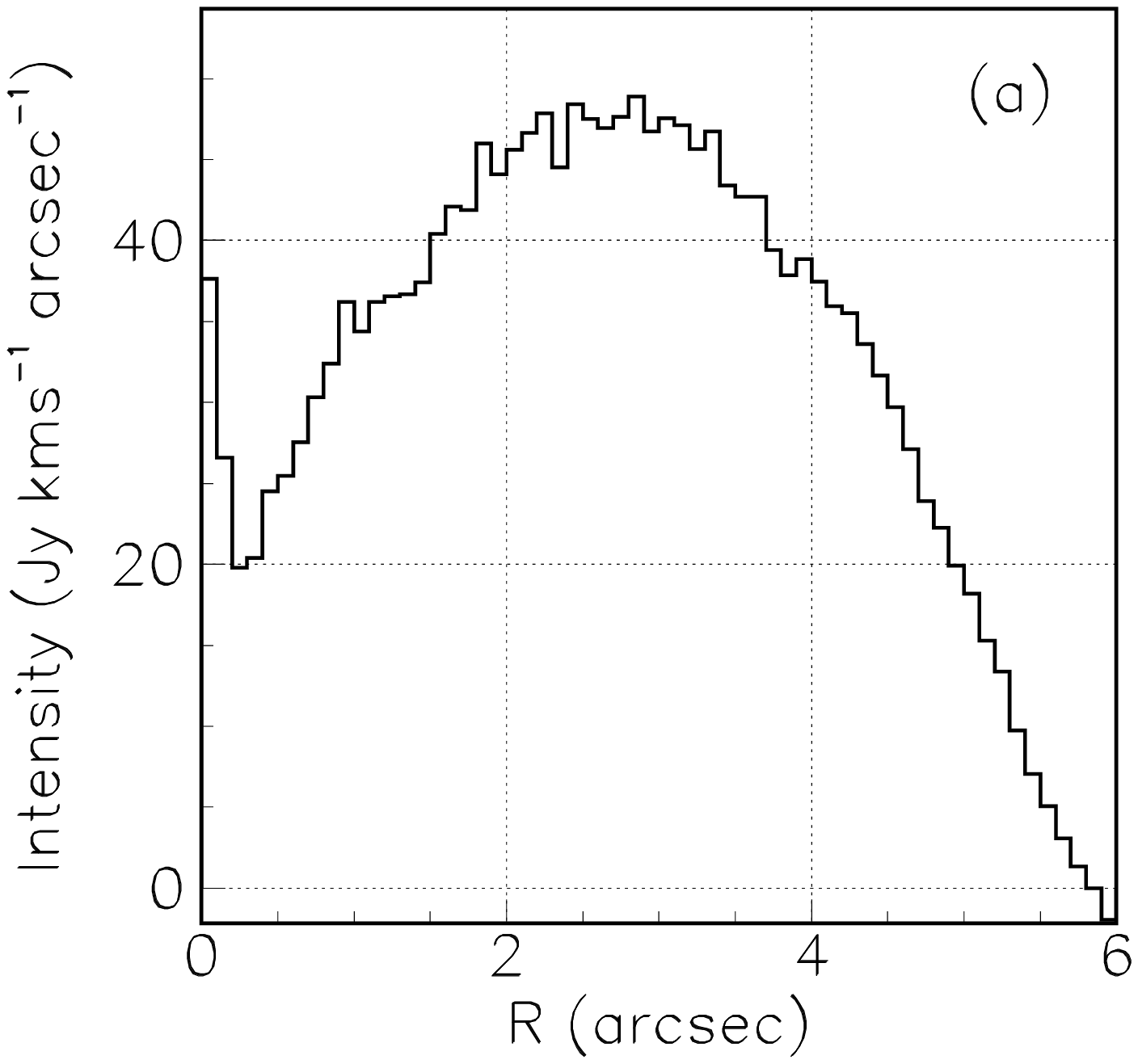}
  \includegraphics[height=4cm,trim=.5cm 1cm .0cm 1.9cm,clip]{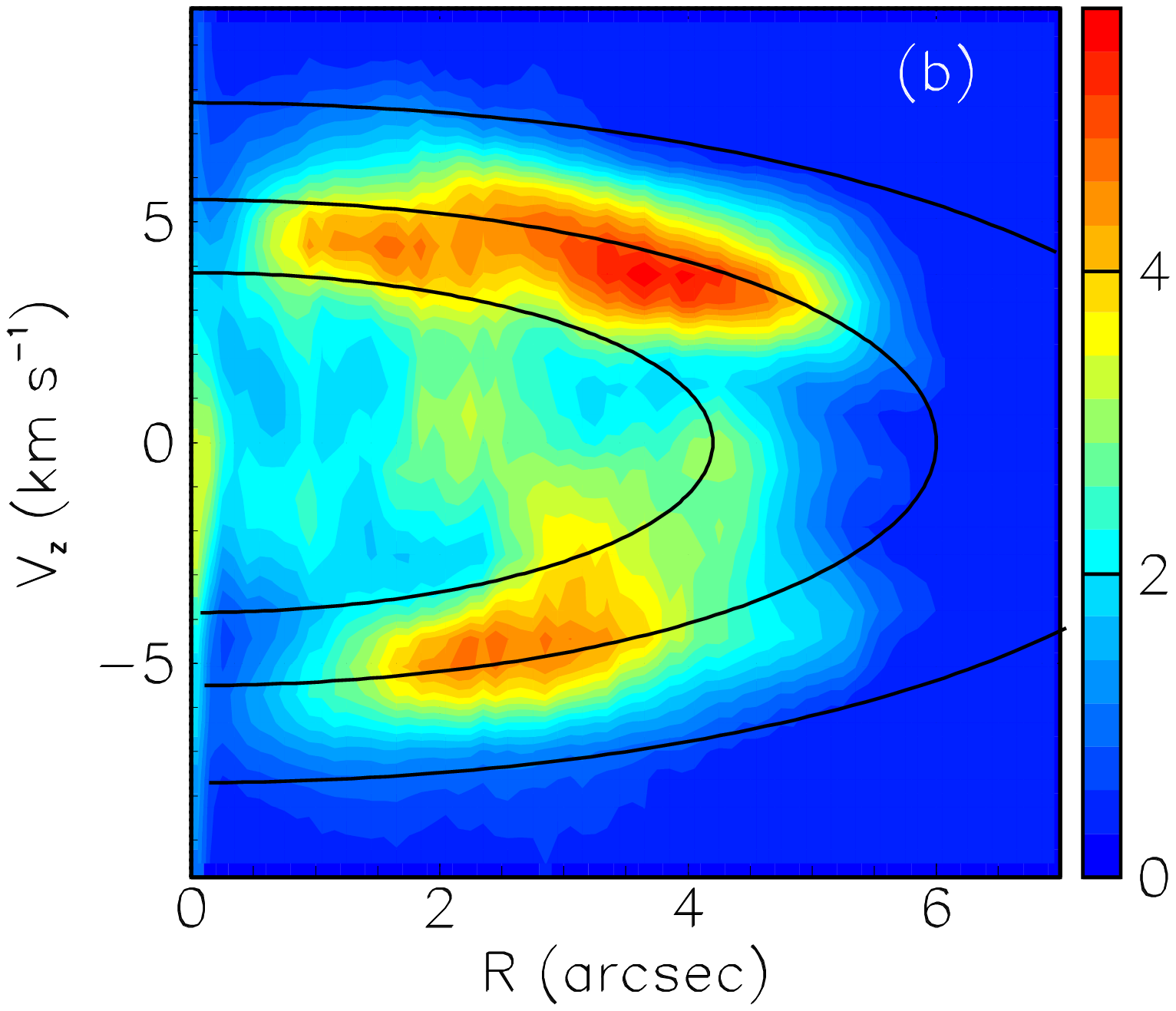}
  \includegraphics[height=4cm,trim=.5cm 1cm 1.7cm 1.9cm,clip]{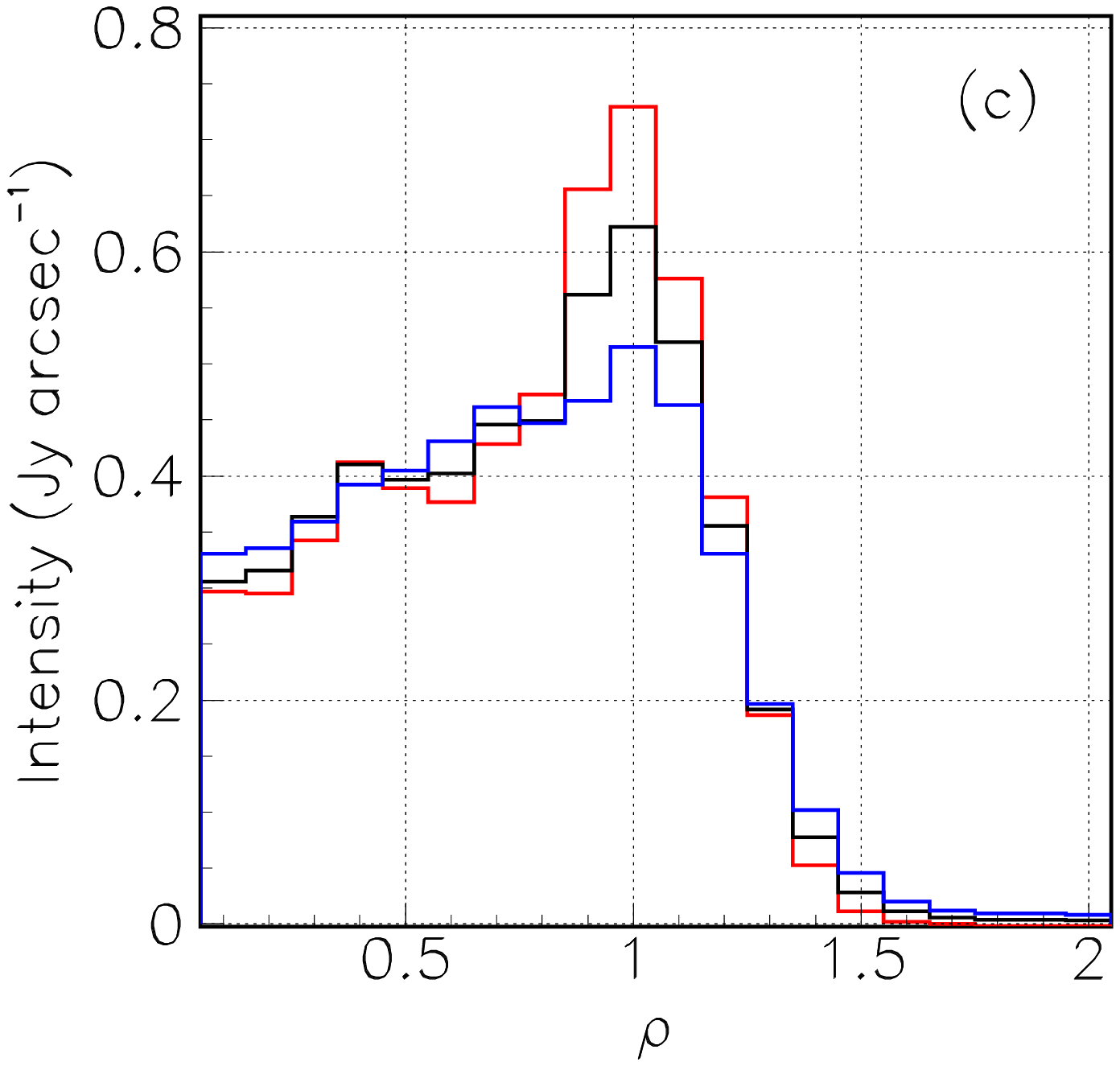}
  \includegraphics[height=4cm,trim=.5cm 1cm 1.7cm 1.9cm,clip]{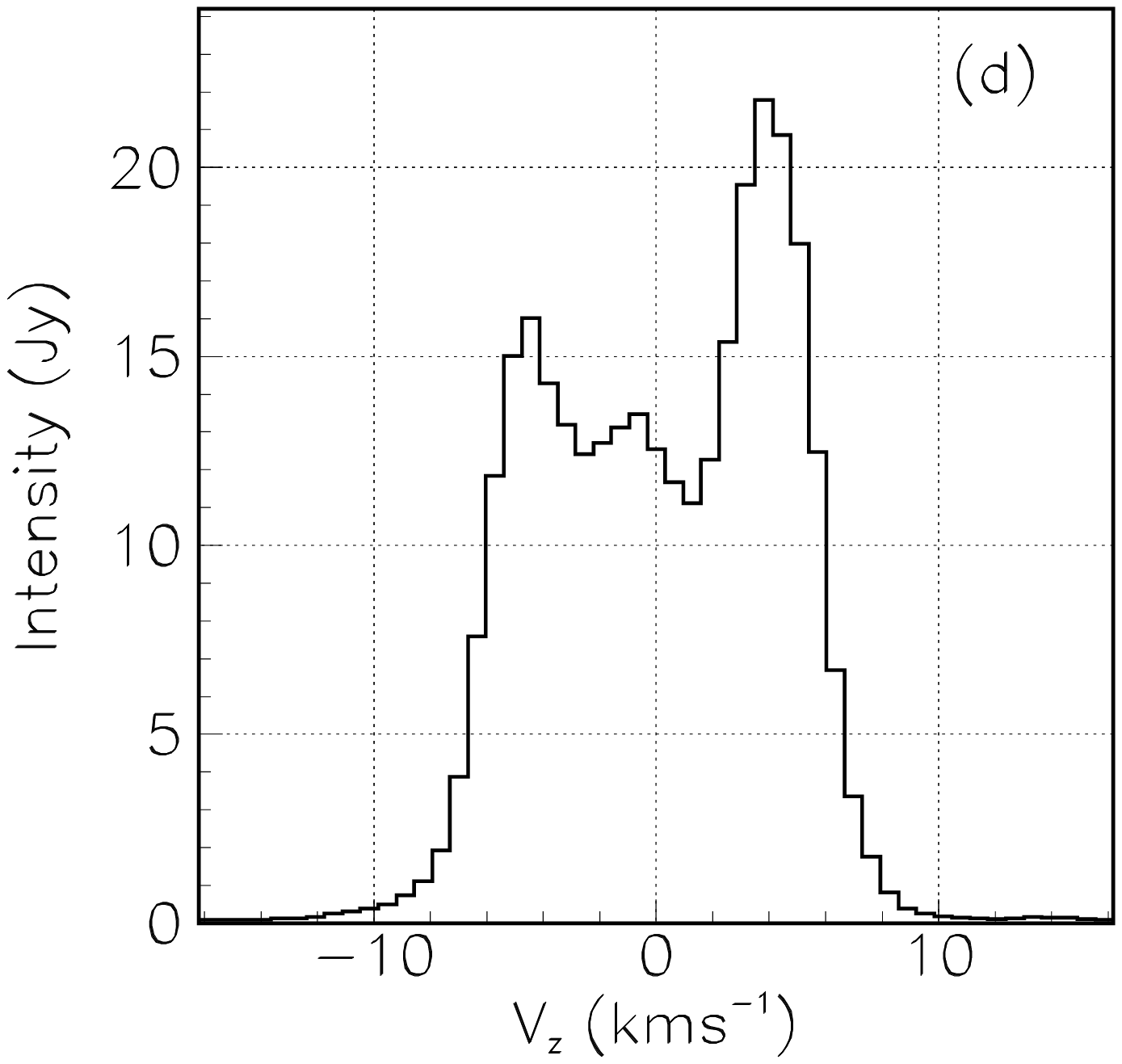}
  \caption{CO(2-1) emission. (a): $R$ distribution of the intensity integrated over |$V_\text{z}$|<10 \kms. (b): $V_\text{z}-R$ map of intensity (Jy arcsec$^{-1}$) integrated over position angles. Three ellipses are defined by $\rho$=1/$\sqrt{2}$, 1 and $\sqrt{2}$. (c): dependence on $\rho$ of the intensity averaged and multiplied by $R$ over the surface area of the red-shifted (red) hemisphere, blue-shifted (blue) hemisphere and the whole velocity range (black). (d): Doppler velocity spectrum integrated over the circle $R$<6 arcsec. Note that on the pair of leftmost panel the contribution of continuum emission is clearly visible.}
  \label{fig4}
\end{figure*}

Channel maps of the CO(2-1) emission are displayed in Figure \ref{fig3}. The bulk of the emission is well contained within $R$<6 arcsec and |$V_\text{z}$|<10 \kms.   Figure \ref{fig4} gives evidence for a significant enhancement of emission in a shell having a mean radius of some 6 arcsec and expanding at a mean velocity of some 5.5 \kms, namely having been ejected 6-7 centuries ago. Panel (a) displays the $R$ distribution of the intensity integrated over |$V_\text{z}$|<10 \kms\ and panel (b) an intensity map on the $V_\text{z}$ vs $R$ plane, integrated over position angles. The latter illustrates clearly the shell structure: the data are compared with three ellipses defined as $\rho$=$\sqrt{(V_\text{z}/5.5)^2+(R/6)^2}$=1/$\sqrt{2}$, 1 and $\sqrt{2}$, respectively; they correspond to spherical shells of radius $6 \times \rho$ arcsec and expansion velocities $5.5\times\rho$ \kms. Panel (c) displays the $\rho$ distribution of the emission for each of the blue- and red-shifted hemispheres separately as well as for the whole range of Doppler velocities. The shape is the same in both hemispheres, the blue-shifted emission being some 30\% weaker than the red-shifted one. It gives evidence for a clear enhancement at $\rho$$\sim$0.8-1.2. The Doppler velocity spectrum, shown in panel (d), displays the double hump profile expected from the shell structure. Globally, Figure \ref{fig4} illustrates and quantifies the qualitative interpretation of the channel maps that \citet{Fonfria2019a} had suggested.

\begin{figure*}
  \includegraphics[height=3.8cm,trim=.5cm 1cm 1.7cm 1cm,clip]{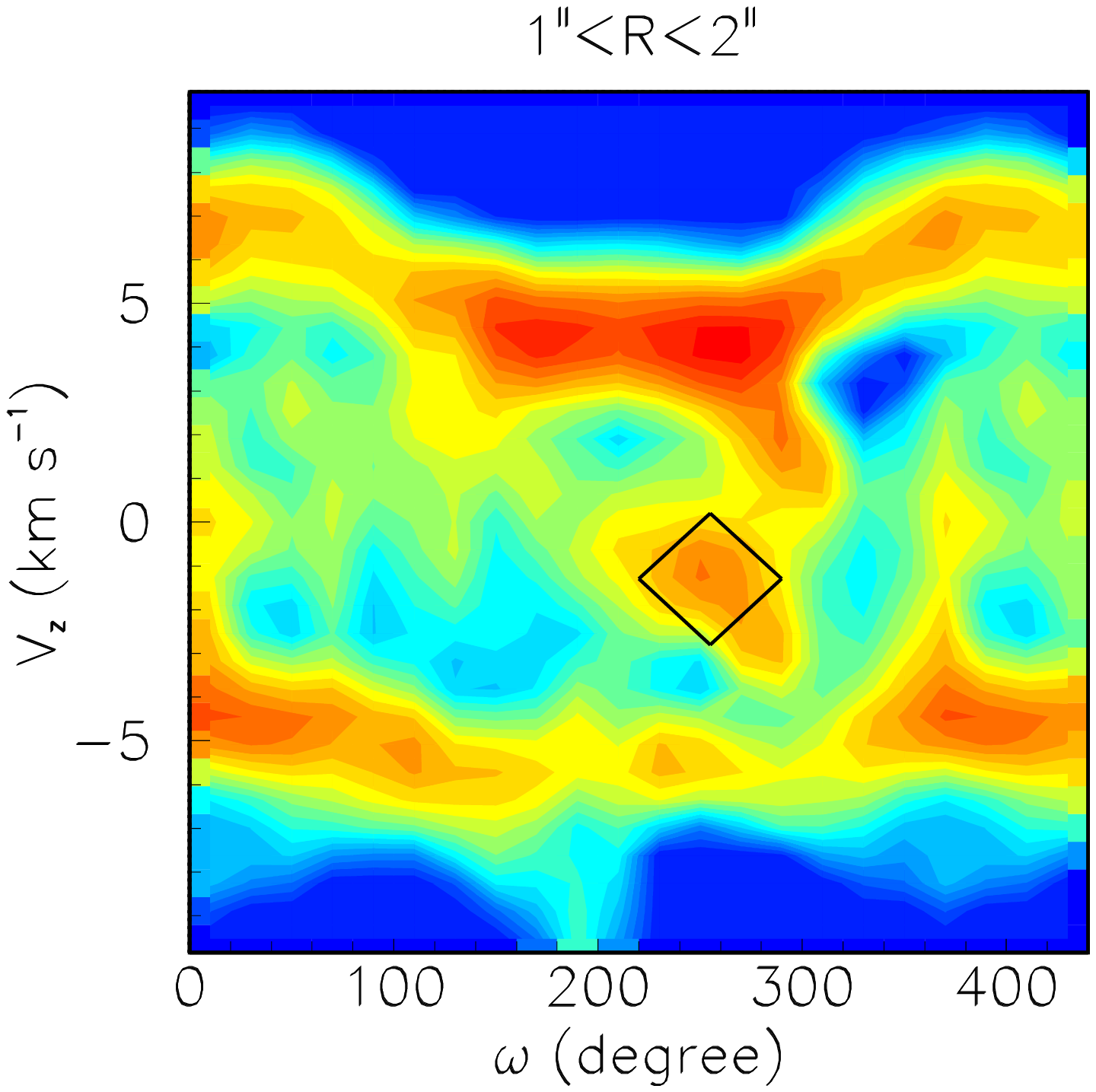}
  \includegraphics[height=3.8cm,trim=.5cm 1cm 1.7cm 1cm,clip]{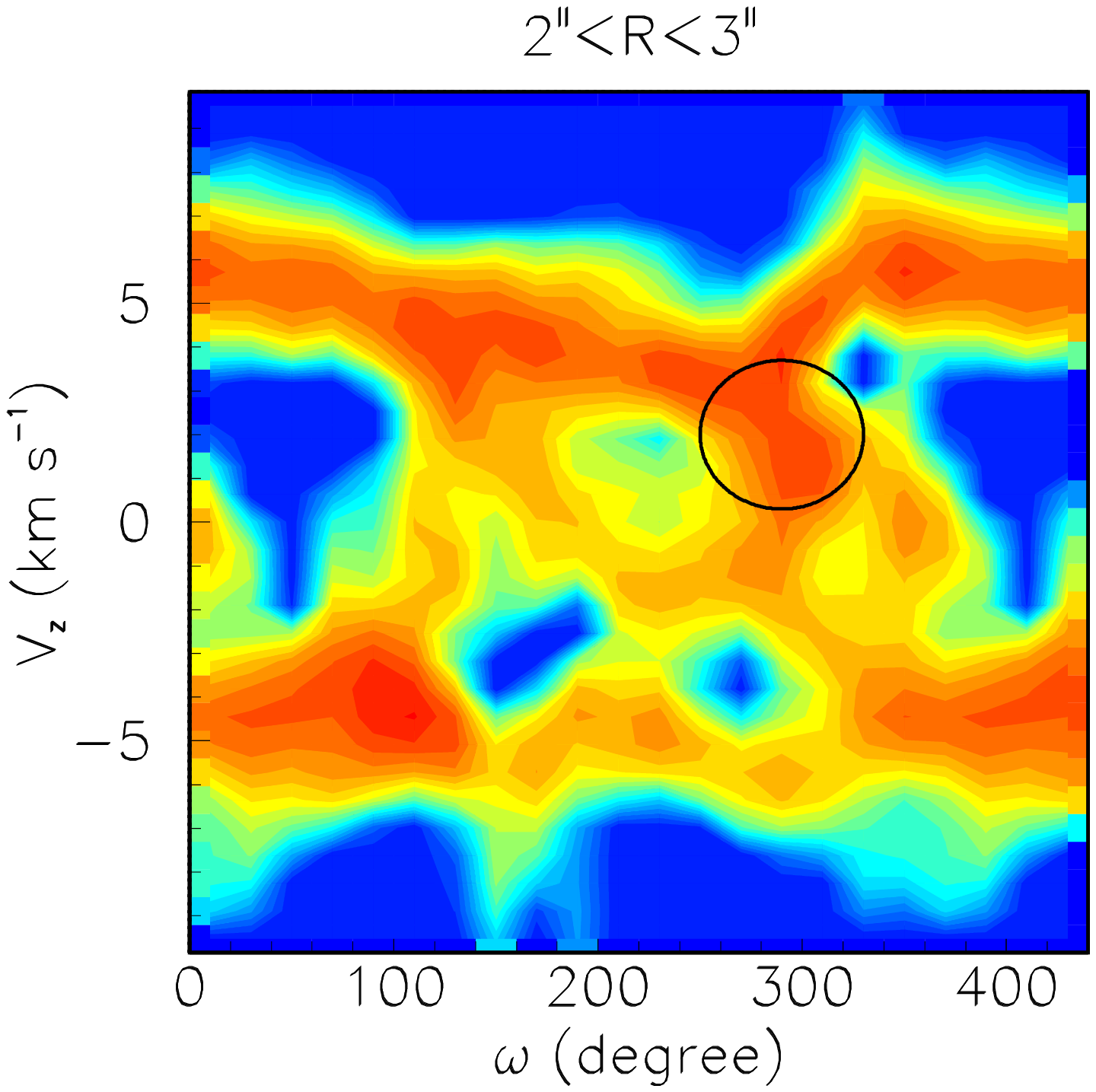}
  \includegraphics[height=3.8cm,trim=.5cm 1cm 0.cm 1cm,clip]{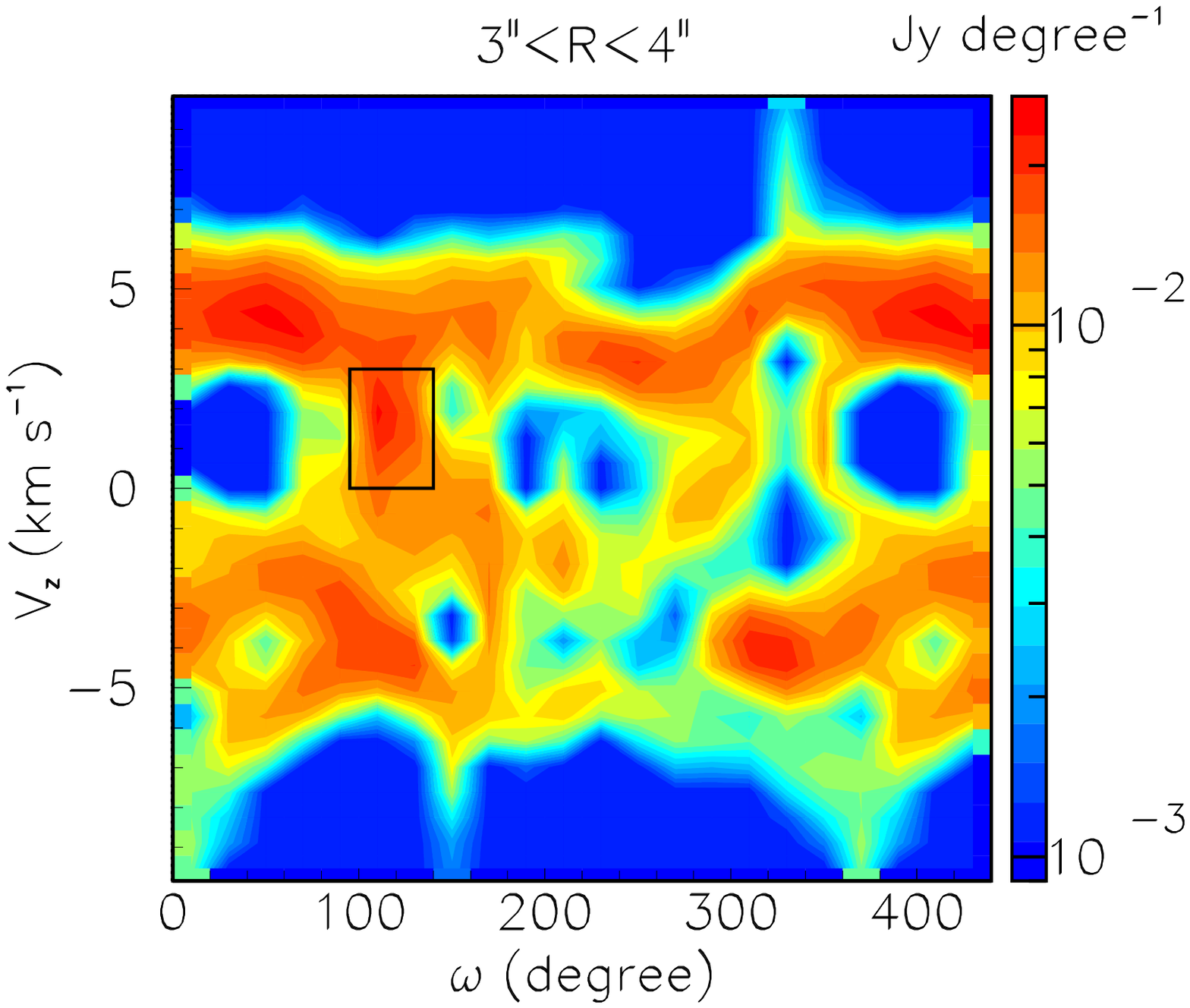}\\
  \includegraphics[height=3.5cm,trim=.5cm 1cm .0cm 1.cm,clip]{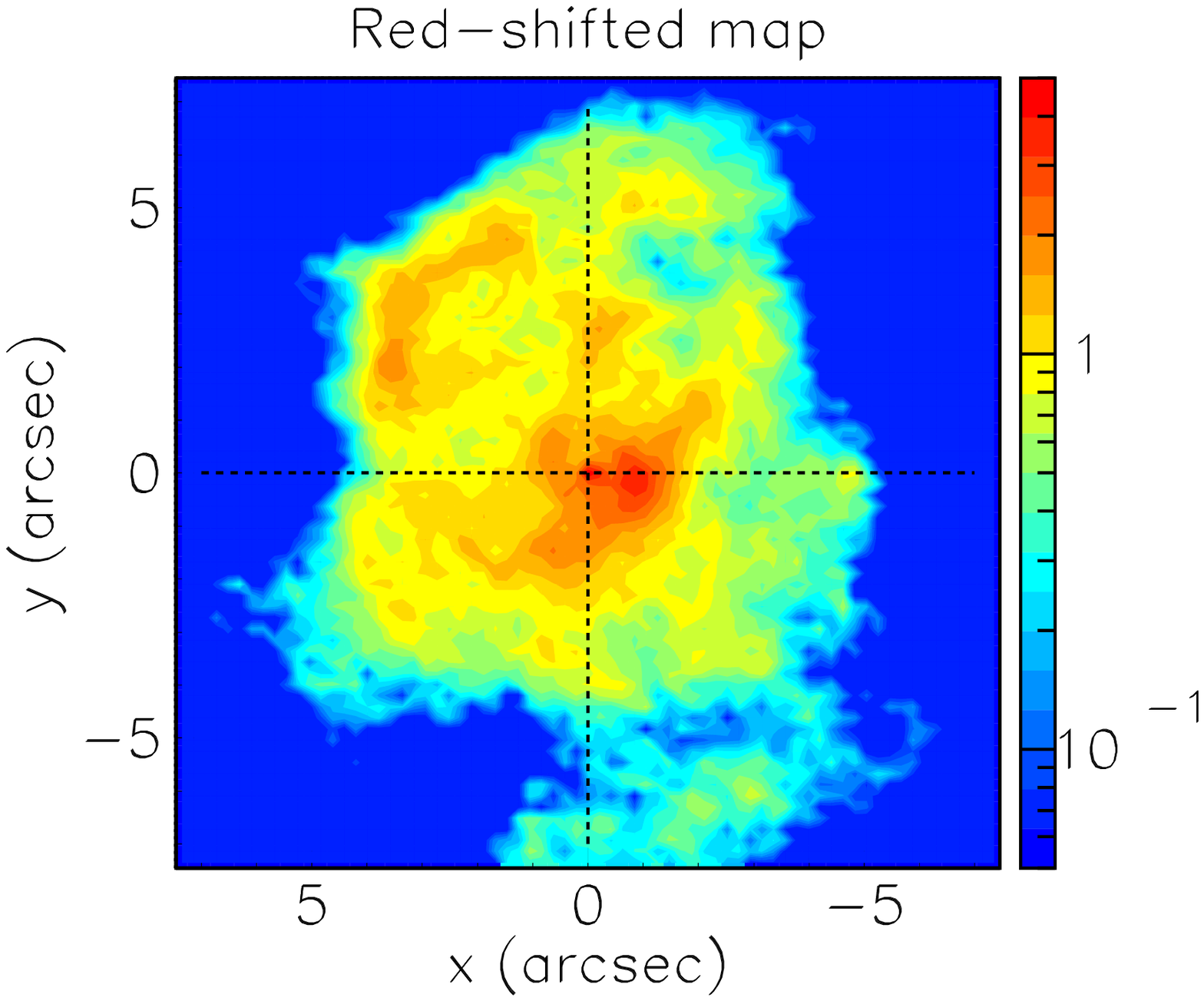}
  \includegraphics[height=3.5cm,trim=.0cm 1cm .0cm 1.cm,clip]{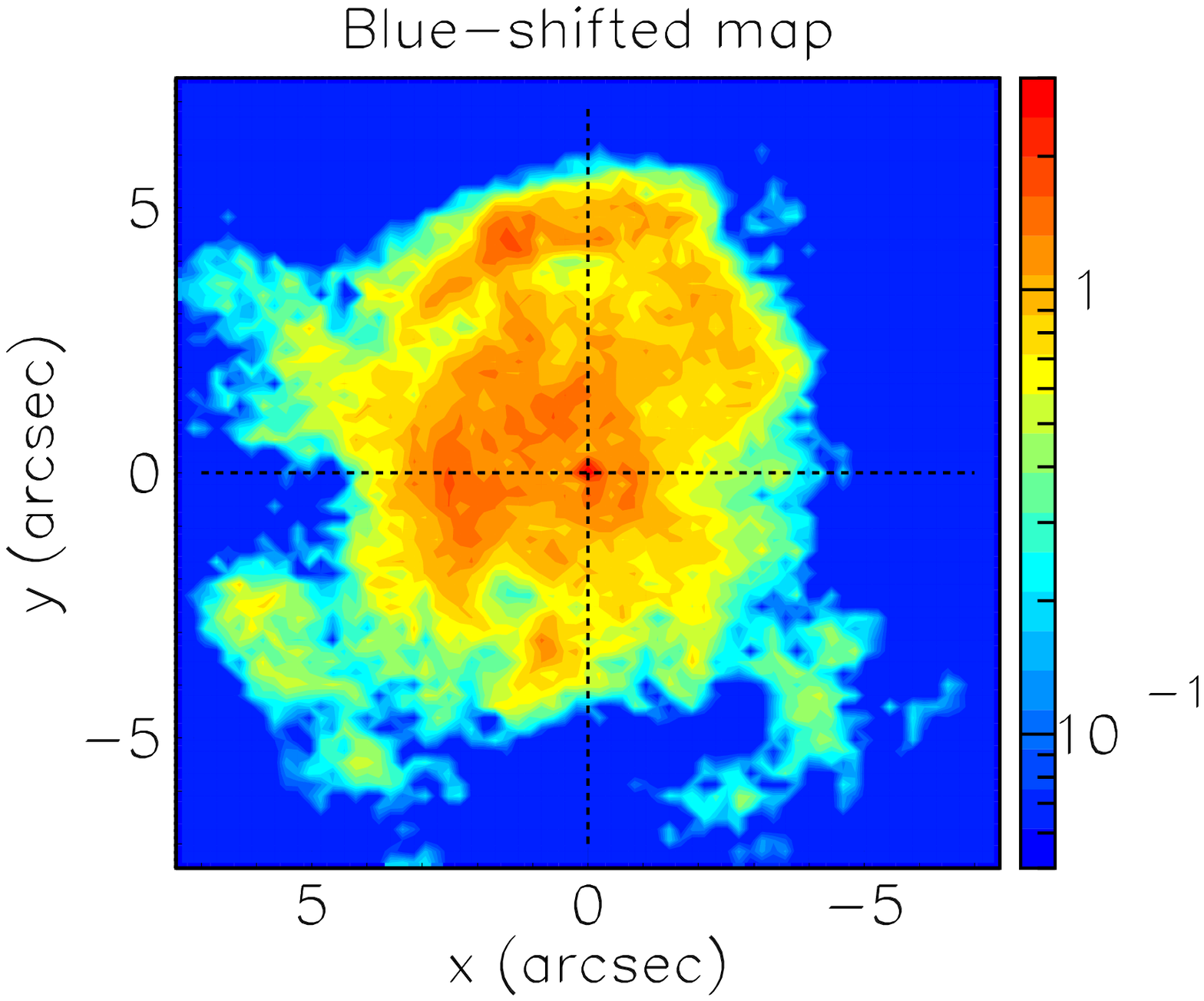}
  \includegraphics[height=3.5cm,trim=.0cm 1cm .0cm 1.cm,clip]{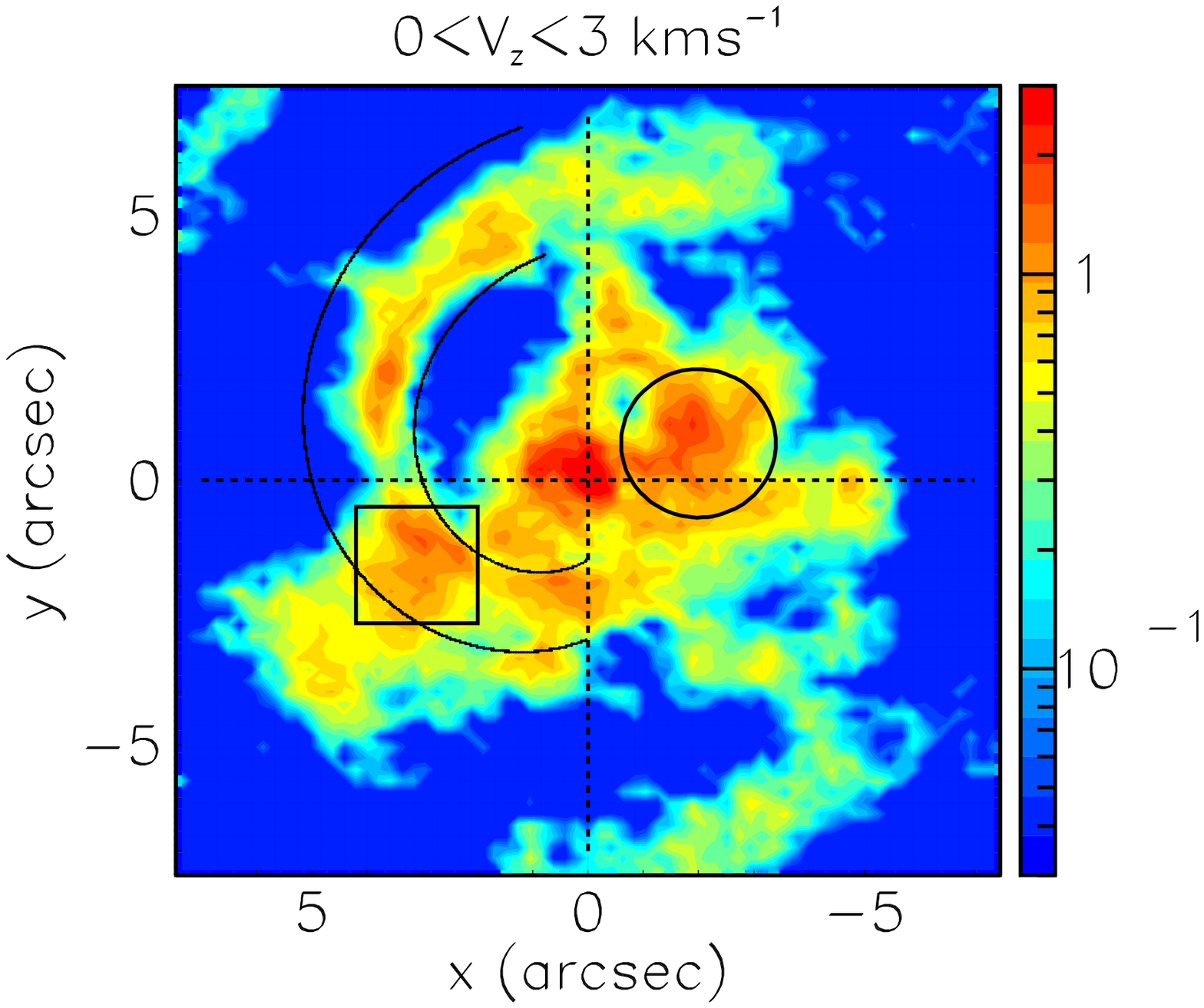}
  \includegraphics[height=3.5cm,trim=.0cm 1cm .0cm 1.cm,clip]{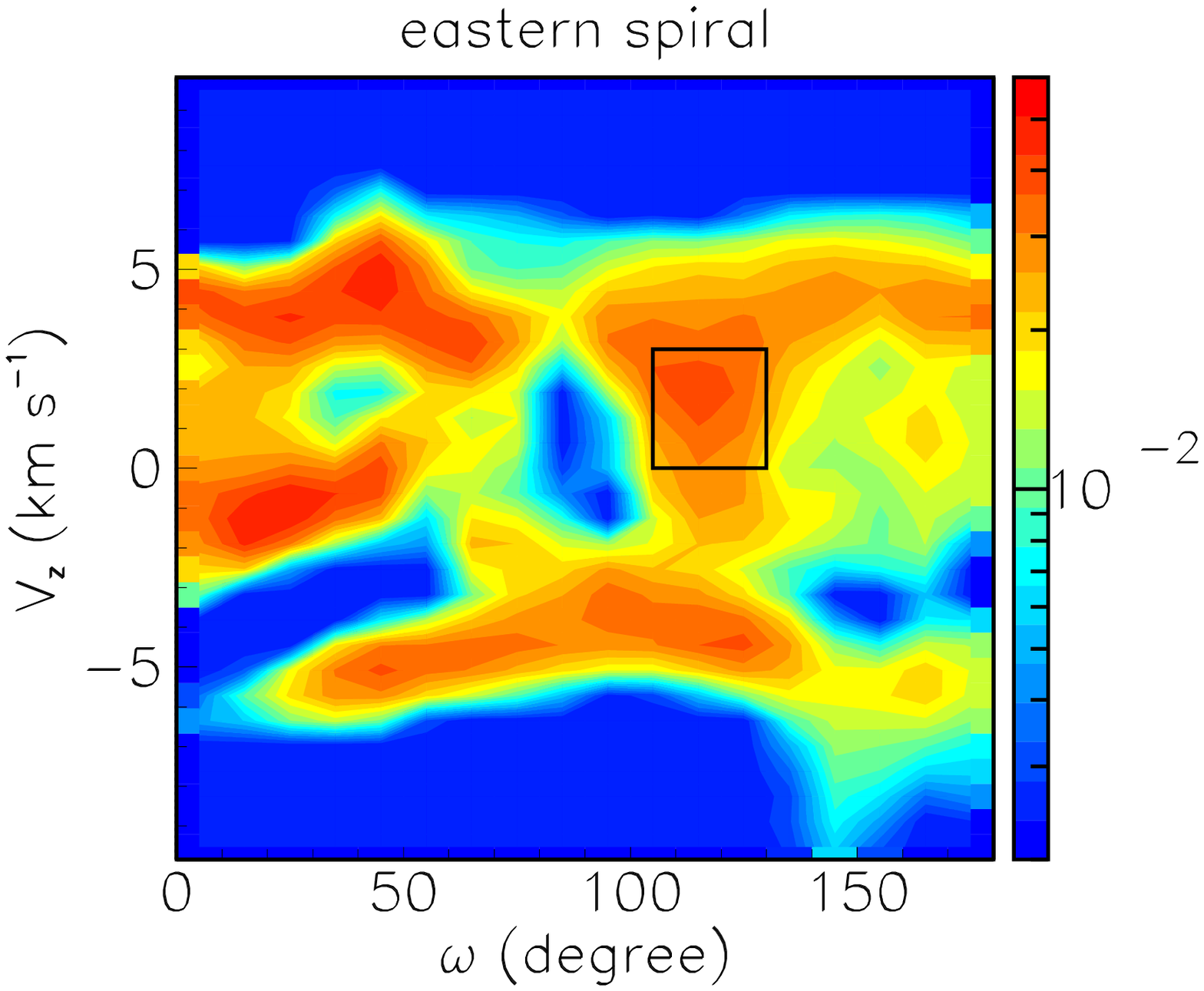}
  
  \caption{CO(2-1) emission. Upper row: PV maps, $V_\text{z}$ vs $\omega$ for 1<$R$<2 arcsec (left), 2<$R$<3 arcsec (centre) and 3<$R$<4 arcsec (right). Lower row: intensity maps (Jy arcsec$^{-2}$ \kms): from left to right, red-shifted, blue-shifted, 0<$V_\text{z}$<3 \kms\ and $V_\text{z}$ vs $\omega$ map (Jy degree$^{-1}$) on the eastern ``spiral'' (10\dego<$\omega$<180\dego\ \& 4.5''$-\omega$/60\dego<$R$<7.0''$-\omega$/45\dego, as shown on the centre-right lower panel). }
  
  \label{fig5}
\end{figure*}

\begin{figure*}
  \includegraphics[height=4.2cm,trim=.5cm 1cm 1.9cm 1.9cm,clip]{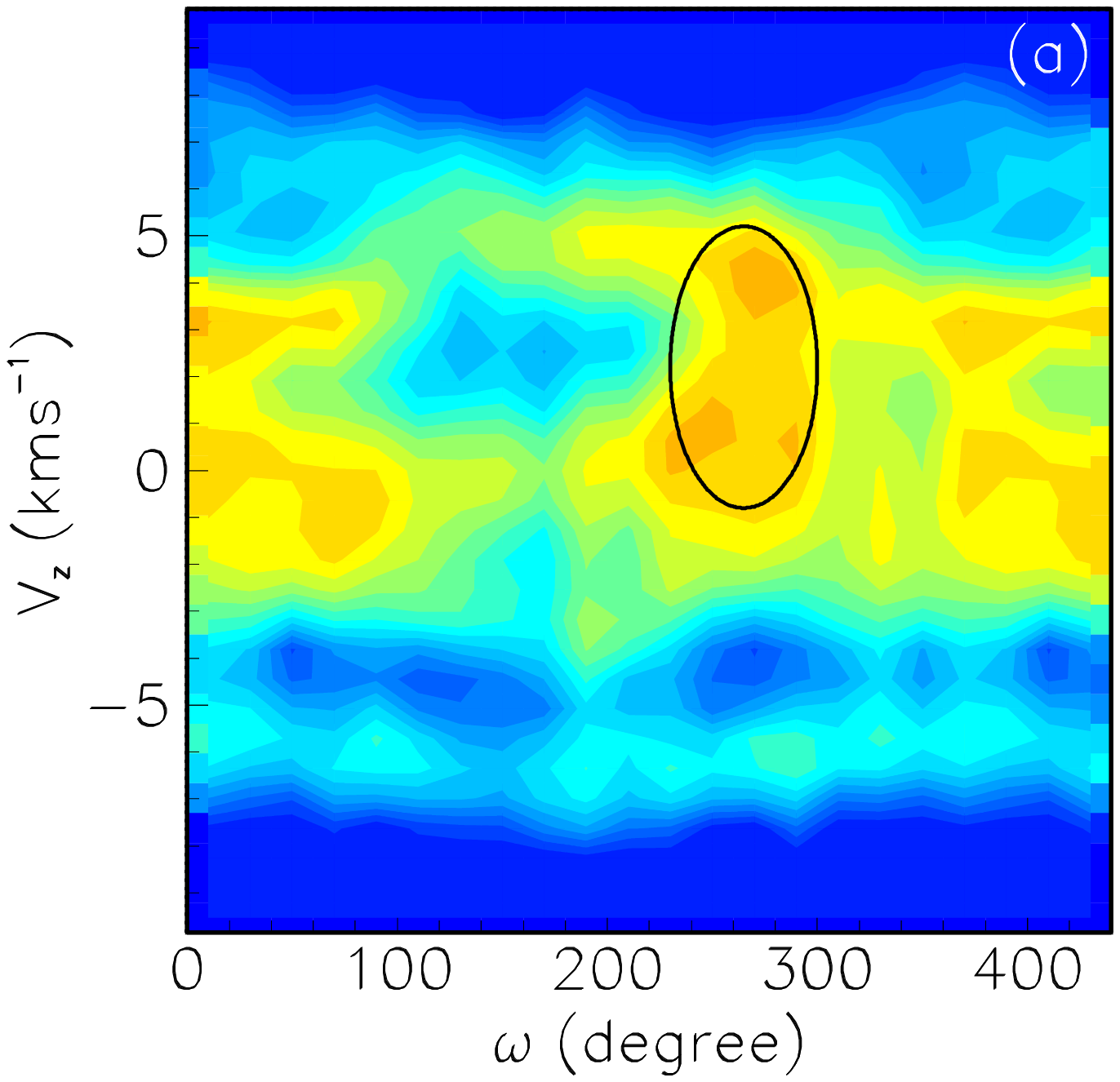}
  \includegraphics[height=4.2cm,trim=1.4cm 1cm 1.9cm 1.9cm,clip]{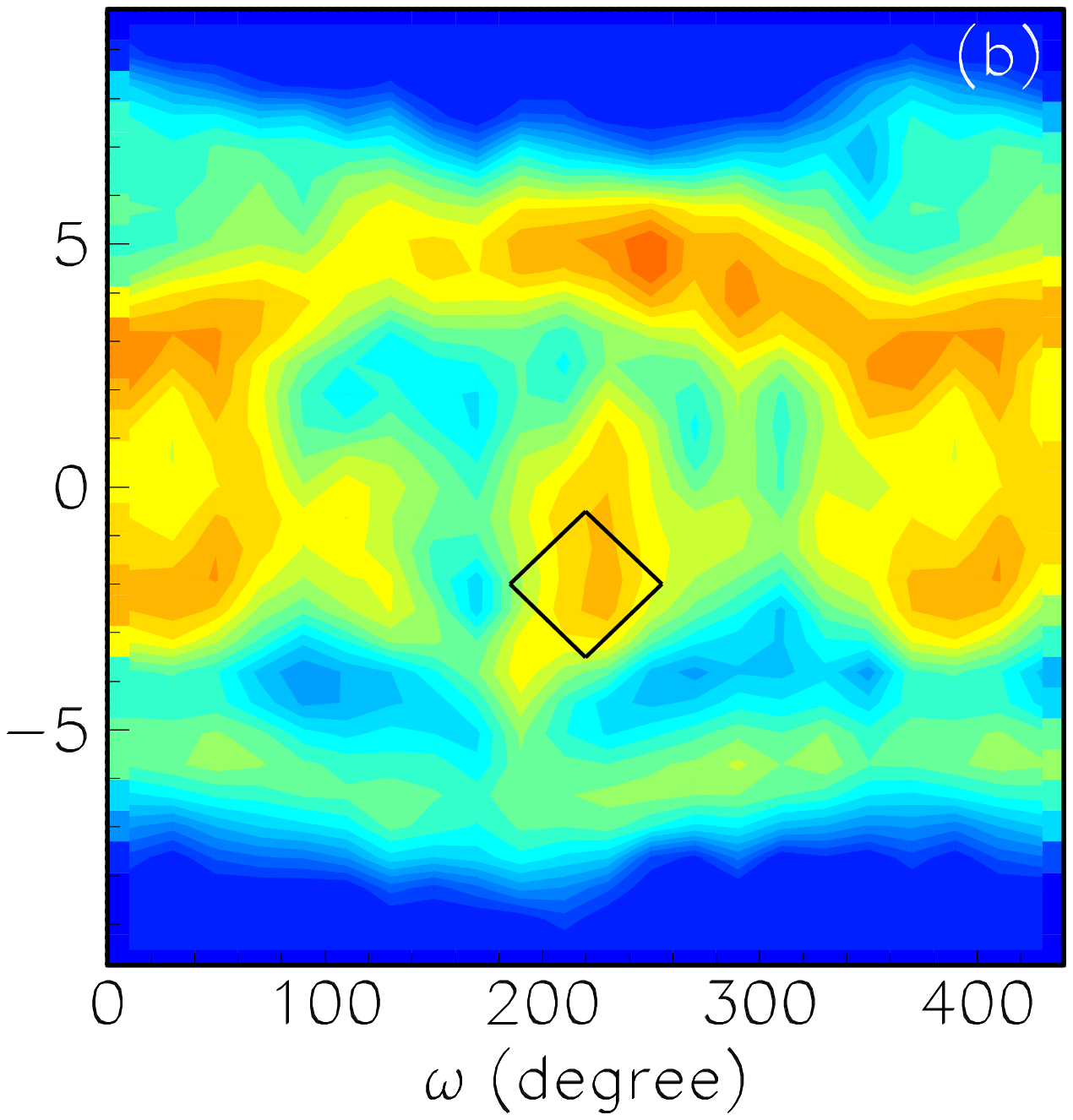}
  \includegraphics[height=4.2cm,trim=1.4cm 1cm 1.9cm 1.9cm,clip]{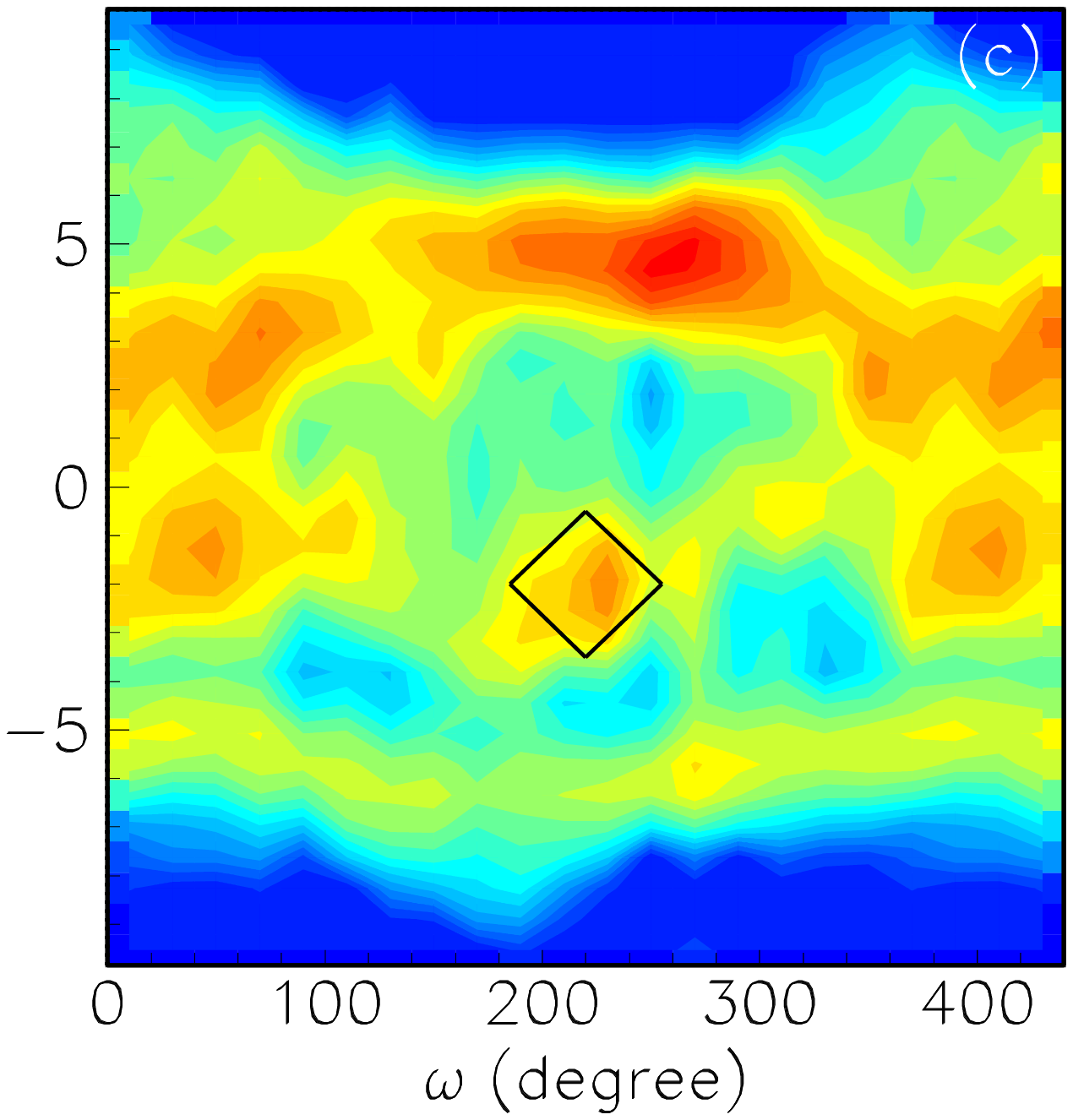}
  \includegraphics[height=4.2cm,trim=1.4cm 1cm 0cm 1.9cm,clip]{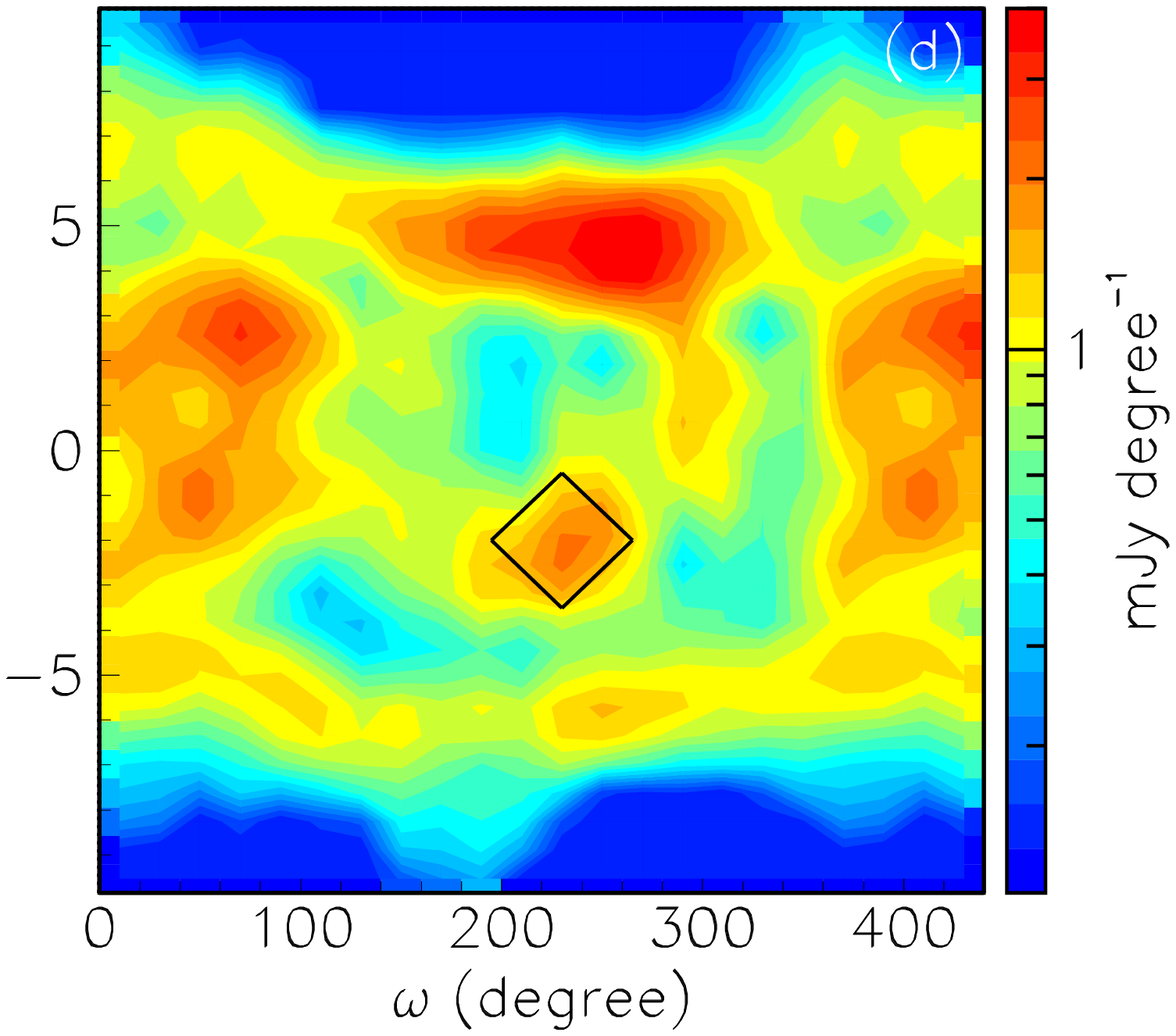}
  \caption{CO(2-1) emission. Maps of intensity on the $V_\text{z}$ vs $\omega$ plane for $R$ between 0.2 and 0.4 (a), 0.4 and 0.6 (b), 0.6 and 0.8 (c) and 0.8 and 1.0 (d) arcsec.}
   \label{fig6}
\end{figure*}

However, when inspecting in more detail the morpho-kinematics governing the structure of the shell, one finds evidence for very important inhomogeneity, the shell being in fact made of a collection of individual patches. Figures \ref{fig5} and \ref{fig6} illustrate the main features. Figure \ref{fig5} displays PV maps, $V_\text{z}$ vs $\omega$, of the CO(2-1) emission in three different intervals of $R$: 1 to 2, 2 to 3 and 3 to 4 arcsec, respectively. It gives evidence for the very irregular distribution of Doppler velocities in the shell, and for the patchy nature of its emission, in particular in the blue-shifted hemisphere. Moreover, patches of emission are seen inside the shell. In particular, at $V_\text{z}$$\sim$$-2$ \kms, $\omega$$\sim$260\dego\ (diamond) for 1<$R$<2 arcsec; at $V_\text{z}$$\sim$2-3 \kms, $\omega$$\sim$300\dego\ (circle) for 2<$R$<3 arcsec; and at $V_\text{z}$$\sim$2-3 \kms, $\omega$$\sim$120\dego\ (rectangle) for 3<$R$<4 arcsec. The first of these (diamond) corresponds to a blue-shifted gas flow that is further illustrated in Figure \ref{fig6} and starts being clearly identified at $R$$\sim$0.4 arcsec. The second (circle) is a blob of emission that is visible on the 0-3 \kms\ intensity map (centre-right panel in the lower row of Figure \ref{fig5}). The third (rectangle) is located near an arc of emission that can be seen on both the red-shifted and the 0-3 \kms\ intensity maps. A similar arc is visible on the blue-shifted intensity map. It is tempting to associate these arcs with possible spirals revealing the presence of a companion \citep{Fonfria2019a}. However, when mapping the $V_\text{z}$ vs $\omega$ map of the intensity along such a presumed spiral (right panel of the lower row of Figure \ref{fig5}), one sees that one cannot make simply sense of such an interpretation.

Figure \ref{fig6} displays PV maps on the $V_\text{z}$ vs $\omega$ plane at shorter distances from the star, between 0.2 and 1 arcsec. In addition to the previously identified gas flow (diamond), it shows, within $R$$\sim$0.4 arcsec, a blob of emission (ellipse) covering between 0 and 5 \kms\ at mean position angle $\omega$$\sim$270\dego, which is associated with a western, red-shifted mass ejection that we study in detail in the next section. Absorption in the blue-shifted part of the shell is seen to possibly extend up to $R$$\sim$0.5 arcsec, but cannot be invoked to explain its weaker emission when compared with the red-shifted part over a much broader range of $R$.

\begin{figure*}
   \includegraphics[height=5.5cm,trim=0cm 0.5cm 0cm .0cm,clip]{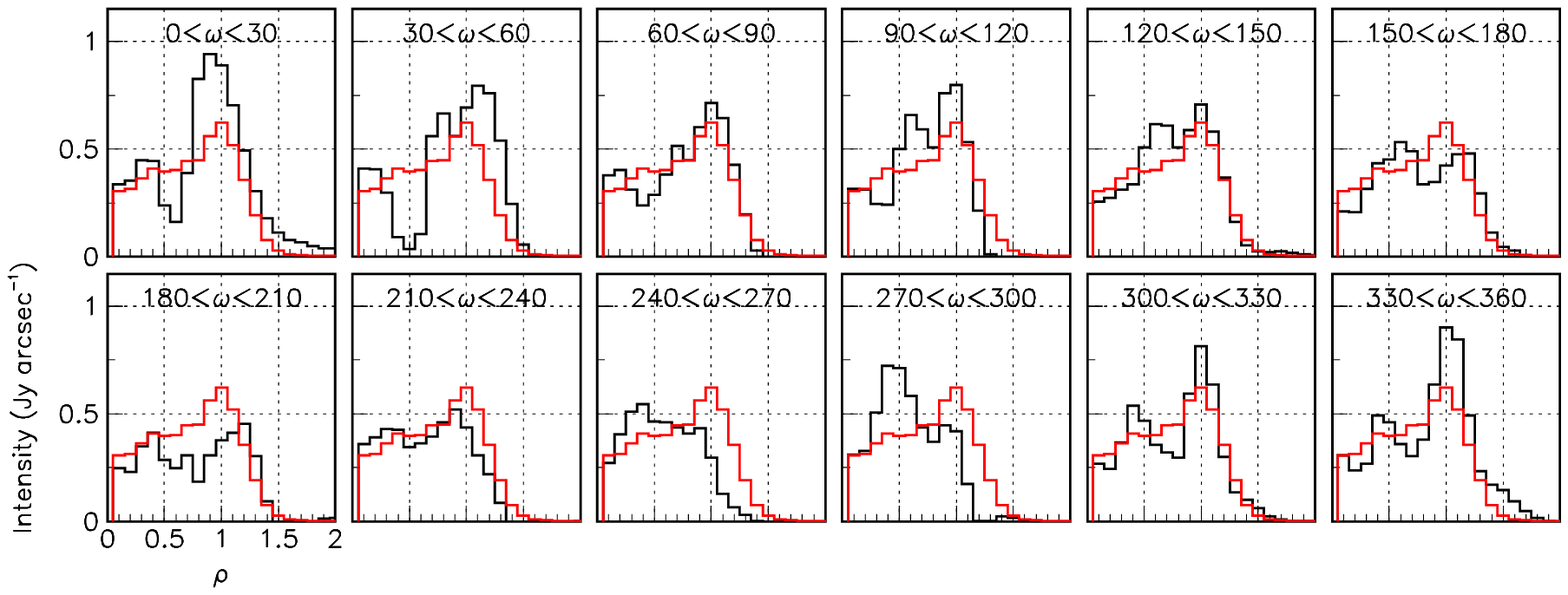}
  \caption{CO(2-1) distributions of $\rho$ in twelve 30\dego\ wide sectors of position angle, $\omega$. In each panel, the range of $\omega$ is inserted and the mean distribution is shown in red.}
   \label{fig7}
\end{figure*}

In order to quantify the inhomogeneity of the shell structure, we have compared the $\rho$ distributions of the CO(2-1) intensity measured in twelve sectors of position angle, each 30\dego\ wide, with the mean distribution in Figure \ref{fig7}. In each sector we measure the resulting rms deviation from unity of their ratio, evaluated in two different intervals of $\rho$. The results are listed in Table \ref{tab3}. On average, they reach 37\% and 73\% for |$\rho-1$|<0.25 and |$\rho-1$|<0.5, respectively, illustrating the importance of inhomogeneity.

\begin{table*}
  \caption{$\text{Rms}$ deviations from a uniform expanding spherical shell. In each 30\dego\ sector (labelled 1 to 12 counter-clockwise, starting from 0<$\omega$<30\dego) we calculate $\Delta$=$\sqrt{\frac{\sum_\text{i}(f_\text{i}/<f_\text{i}>-1)^2}{n}}$, where the sum runs over the $n$ bins of the relevant $\rho$ interval (first column) and where $f_\text{i}$ is the intensity measured in bin $i$.}
  \label{tab3}
  \begin{tabular}{ccccccccccccc}
    \hline
Sector&
1&
2&
3&
4&
5&
6&
7&
8&
9&
10&
11&
12\\
\hline
$\Delta$(0.5<$\rho$<1.5) [\%]&
109&
104&
54&
109&
25&
36&
47&
92&
67&
61&
43&
132\\
$\Delta$(0.75<$\rho$<1.25) [\%]&
57&
60&
16&
28&
10&
30&
41&
29&
56&
57&
24&
39\\
\hline
    \end{tabular}
\end{table*}

\section{The close neighbourhood of the star}

In the present section we study the emission of lines that probe the close neighbourhood of the star, within $\sim$ 0.3 arcsec angular separation from its centre. In addition to the high resolution CO(2-1) data, this includes the vibrational ground state emission of the $^{29}$SiO(5-4) line and the emissions of the sulphur oxide lines, SO(5$_5$-4$_4$), SO$_2$(22$_{2,20}$-22$_{1,21}$) and SO$_2$(13$_{2,12}$-12$_{1,11}$). The latter is blended with a contribution of H$^{13}$CN($\nu$=0,$J$=4-3) (hereafter abbreviated as H$^{13}$CN(4-3)) emission, which we discuss below. In the remaining of the article, we refer to it as the SO$_2$-HCN line. Figures \ref{fig8} to \ref{fig12} illustrate the main features using a common format: channel maps and PV maps in both $V_\text{z}$ vs $\omega$ and $V_\text{z}$ vs $R$, the latter being displayed in four separate quadrants.  

\subsection{Observational results}
\begin{figure*}
   \includegraphics[height=4.8cm,trim=.5cm 0cm 0cm .5cm,clip]{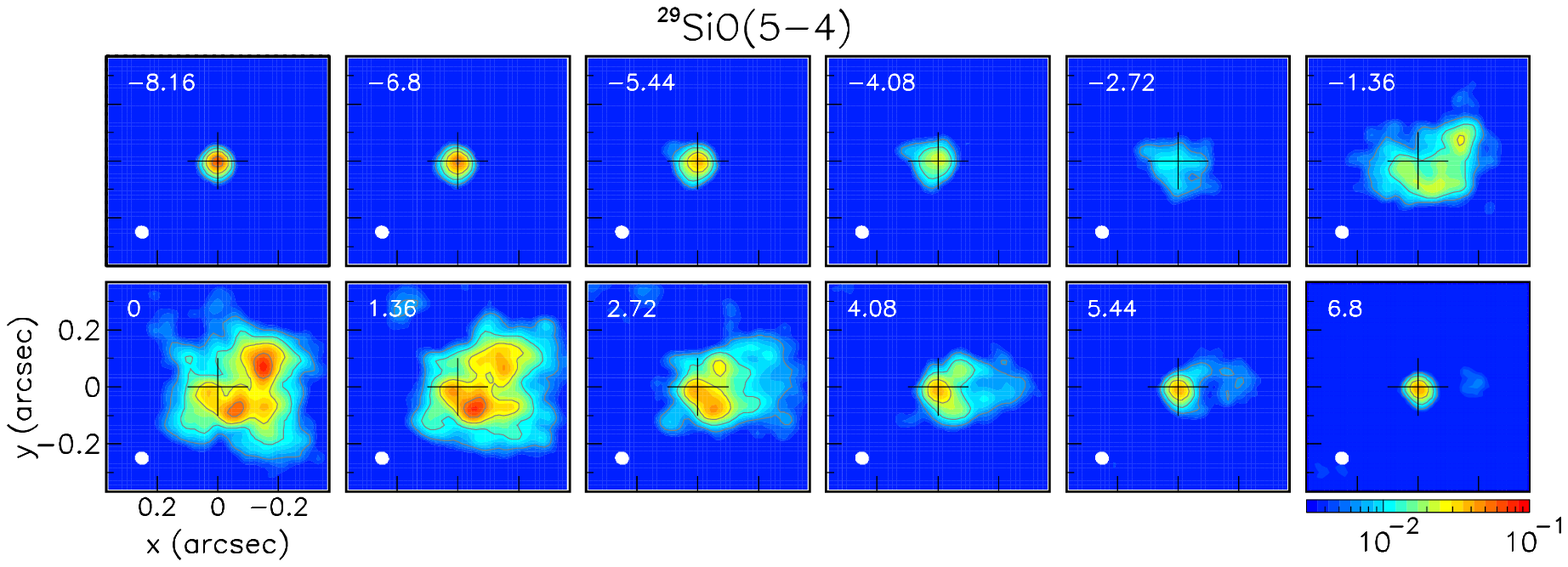}
  \includegraphics[height=4.55cm,trim=.0cm 0.cm 0cm 1.2cm,clip]{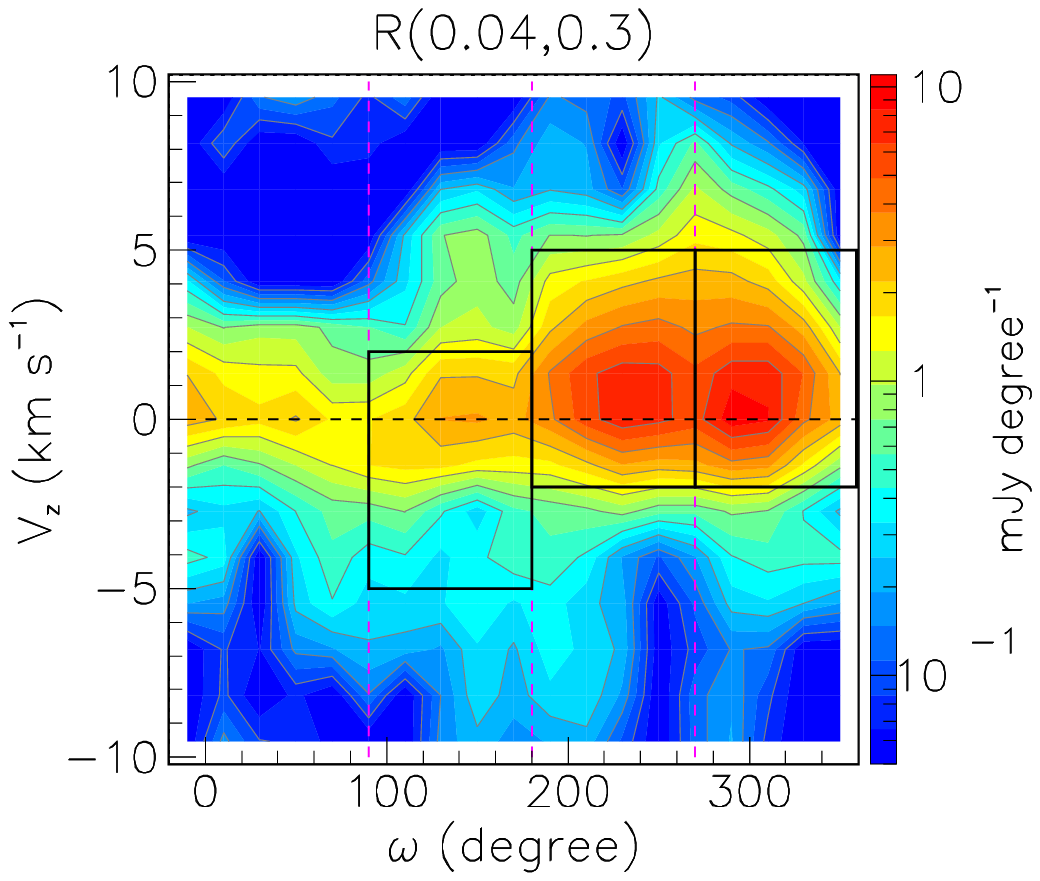}
  \includegraphics[height=3.7cm,trim=.0cm 1cm 0cm 1.3cm,clip]{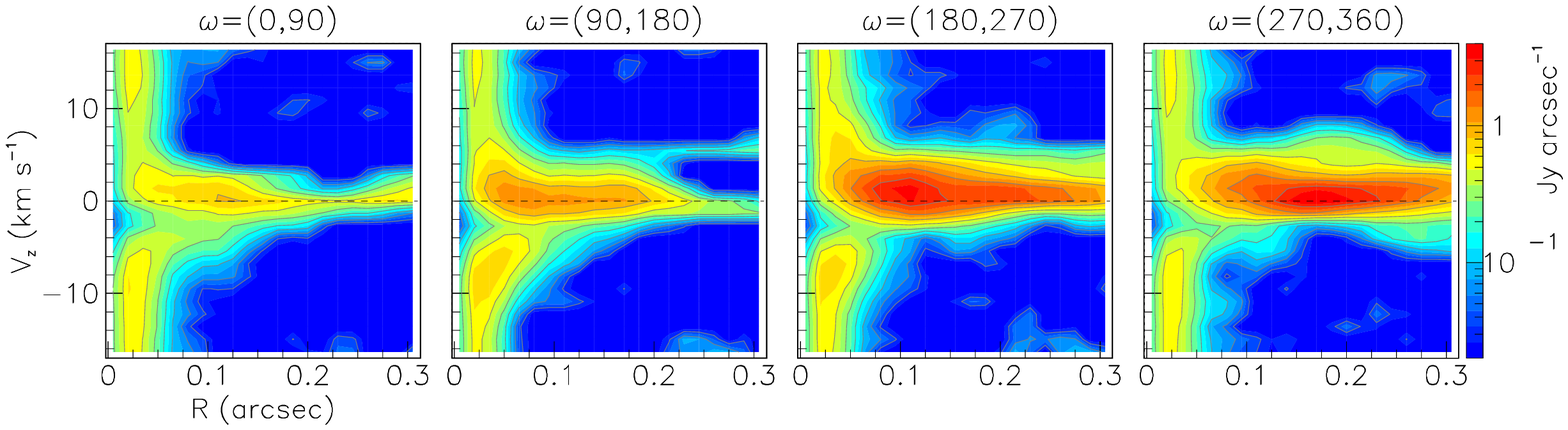}
  \caption{$^{29}$SiO(5-4) emission without continuum subtraction. Upper row: Left, channel maps. The crosses show the position of the star. The colour scale is in units of Jy beam$^{-1}$.  Right, PV map, $V_\text{z}$ vs $\omega$, for 0.04<$R$<0.3 arcsec. The rectangles delineate the regions defined in Table \ref{tab4}, from left to right, SE, NW and SW. Lower row: PV maps, $V_\text{z}$ vs $R$, for 0\dego<$\omega$<90\dego, 90\dego<$\omega$<180\dego, 180\dego<$\omega$<270\dego\ and 270\dego<$\omega$<360\dego, from left to right.}
\label{fig8}
\end{figure*}

\begin{figure*}
  \includegraphics[height=5.05cm,trim=.2cm 0cm 0.3cm .5cm,clip]{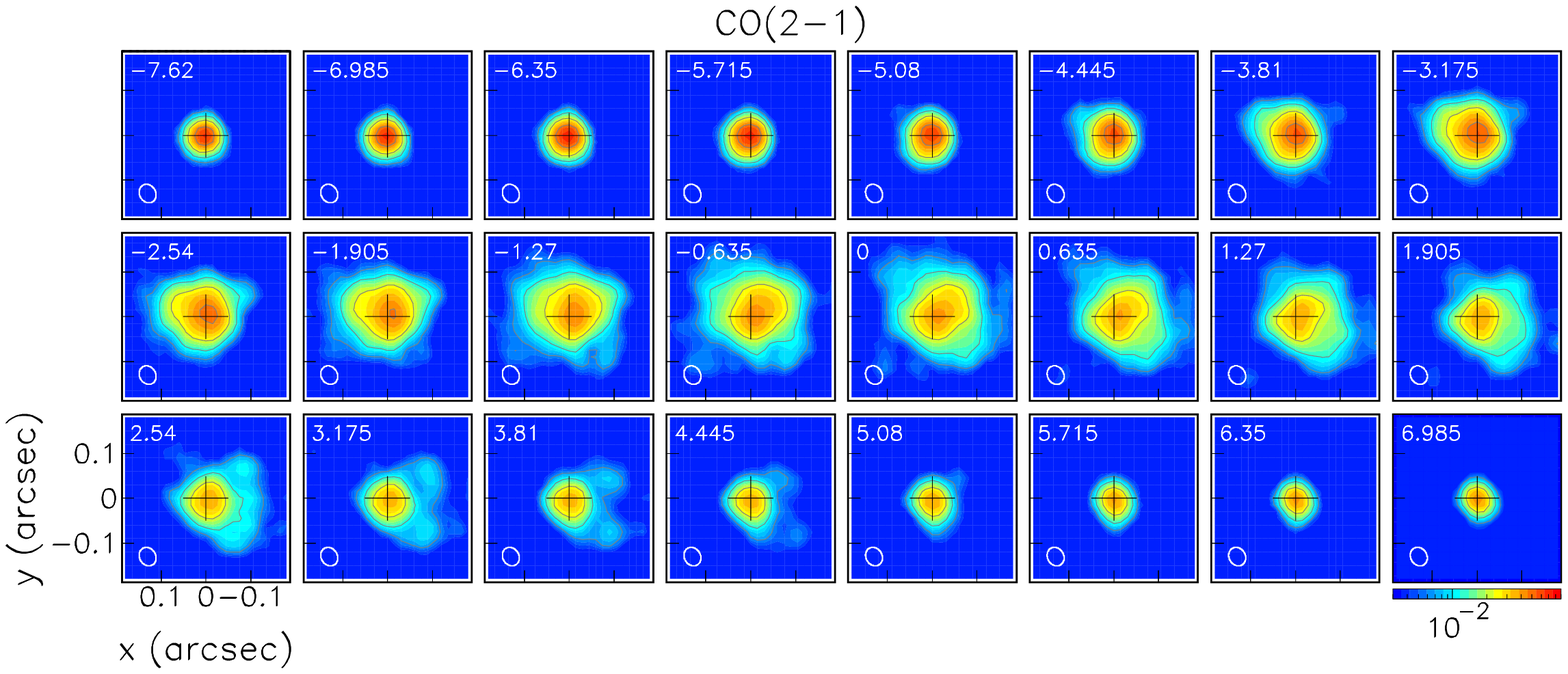}
  \includegraphics[height=5.cm,trim=.0cm 0.7cm 0cm 1.2cm,clip]{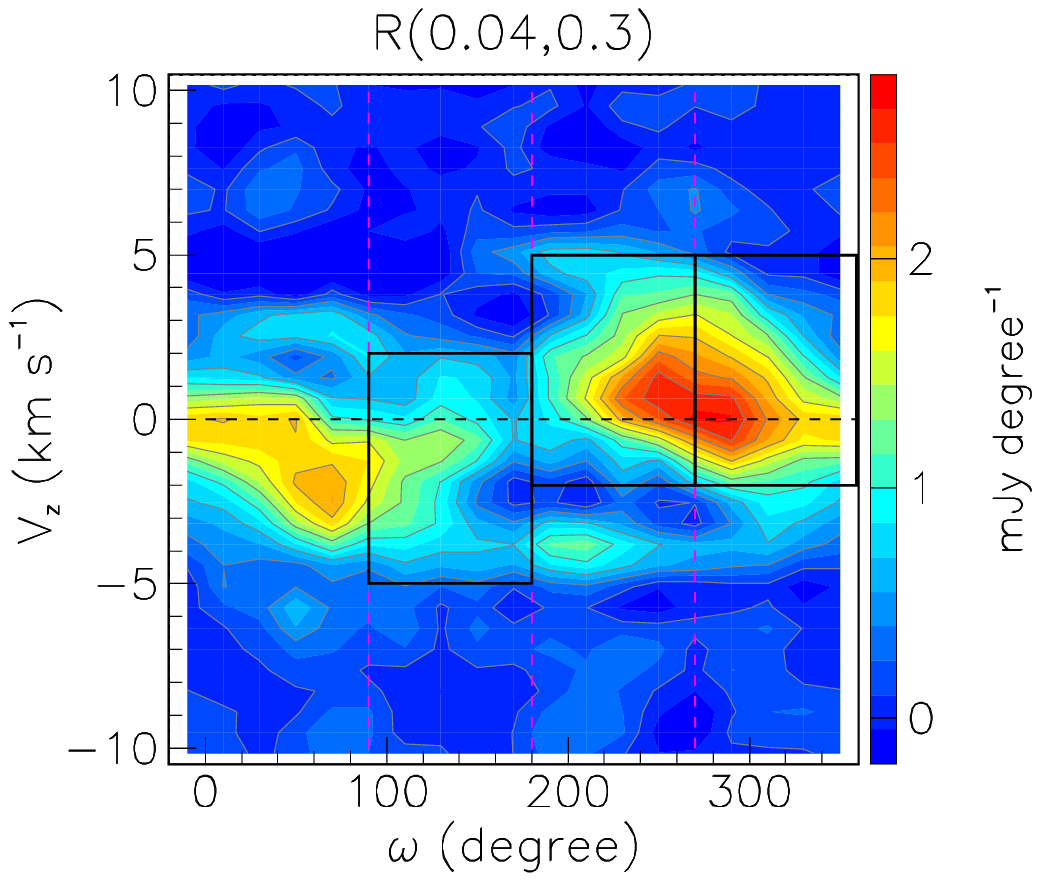}
  \includegraphics[height=4cm,trim=0cm 1cm 0cm 1cm,clip]{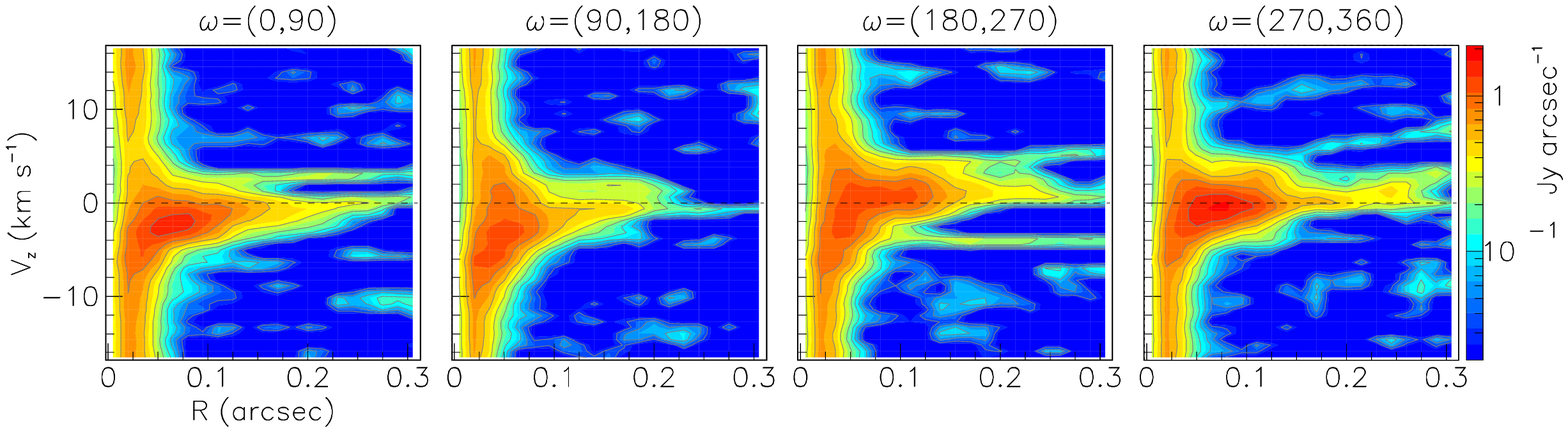}
\caption{CO(2-1) emission. Same as Figure \ref{fig8}.}
\label{fig9}
\end{figure*}

\begin{figure*}
   \includegraphics[height=4.8cm,trim=.5cm 0cm 0cm .5cm,clip]{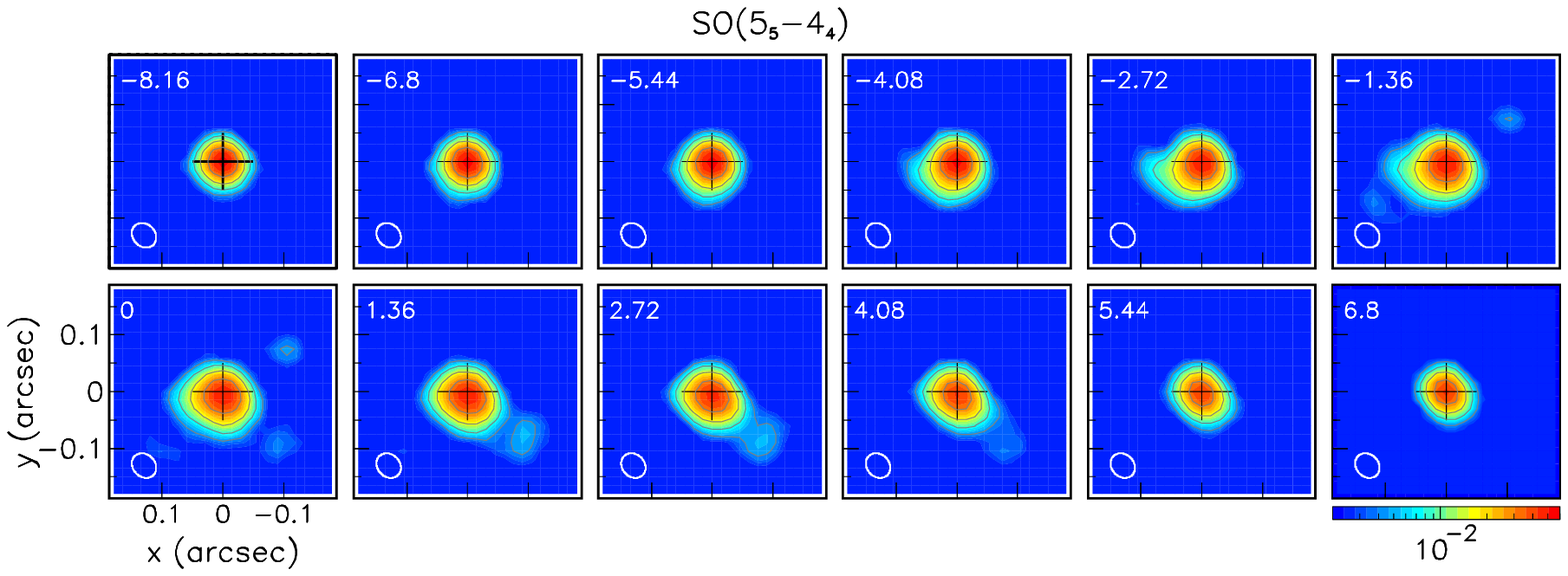}
  \includegraphics[height=4.55cm,trim=.0cm 0.cm 0cm 1.2cm,clip]{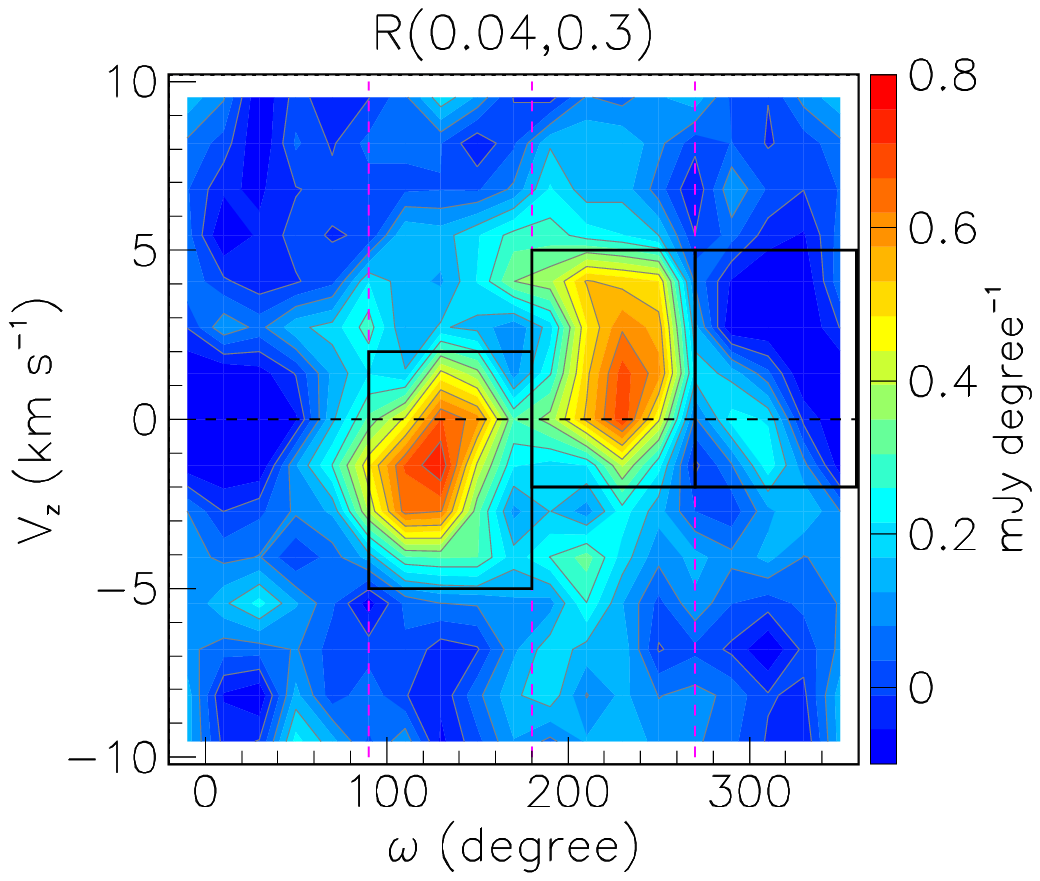}
  \includegraphics[height=3.5cm,trim=.0cm 1cm 0cm 1.3cm,clip]{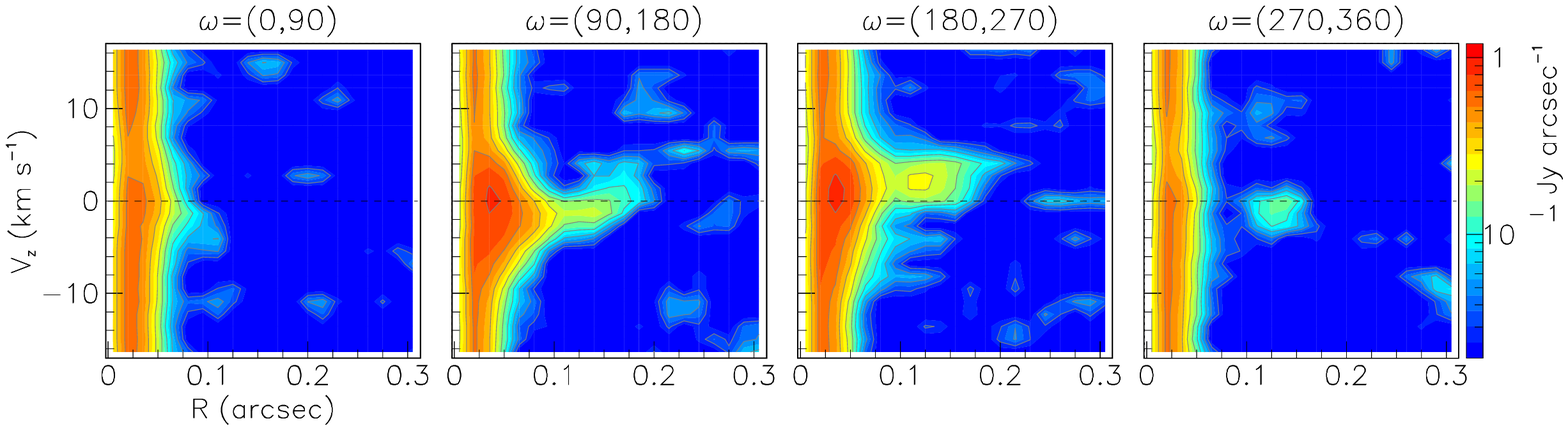}
 \caption{SO(5$_5$-4$_4$) emission. Same as Figure \ref{fig8}.}
\label{fig10}
\end{figure*}

\begin{figure*}
   \includegraphics[height=4.8cm,trim=.5cm 0cm 0cm .5cm,clip]{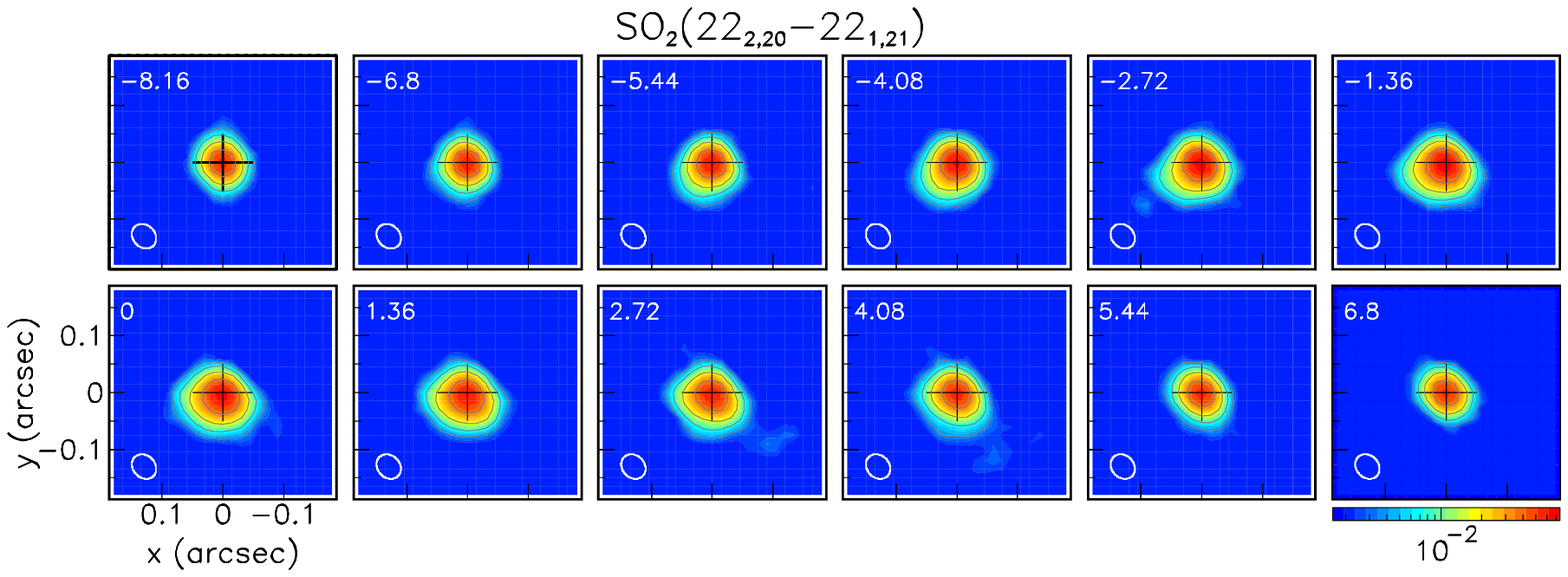}
  \includegraphics[height=4.55cm,trim=.0cm 0.cm 0cm 1.2cm,clip]{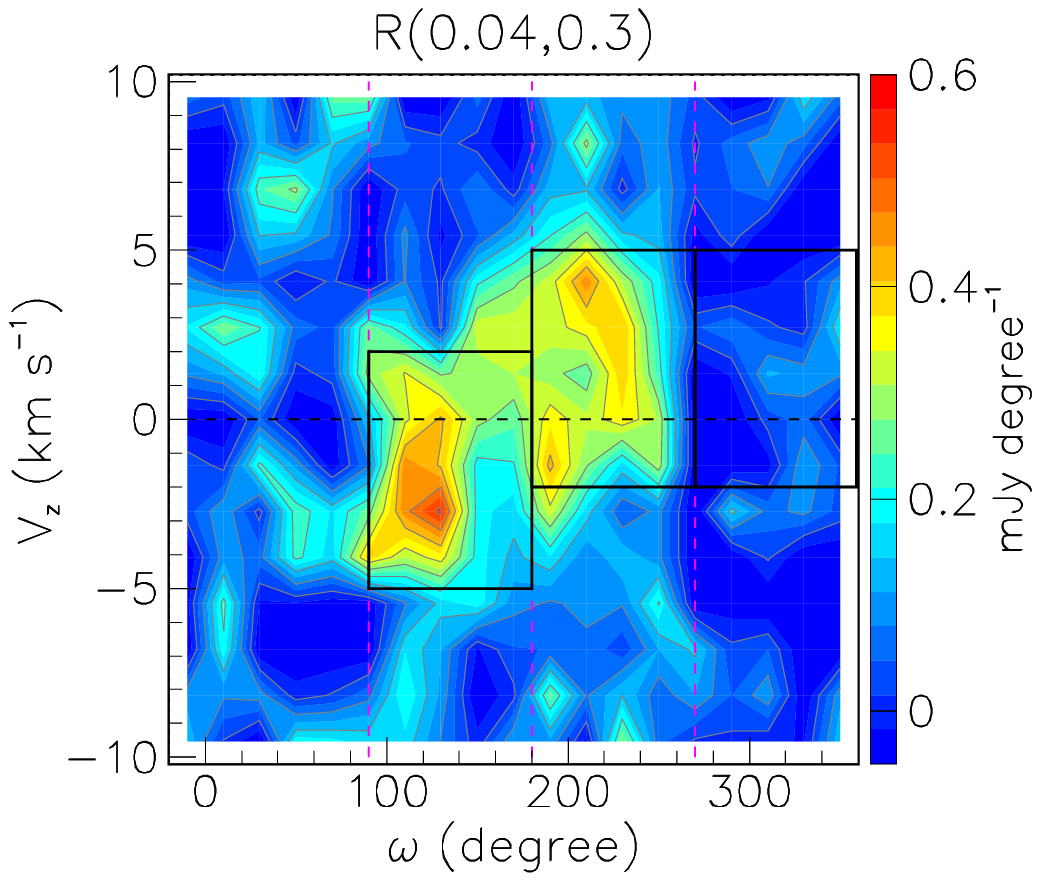}
  \includegraphics[height=3.5cm,trim=.0cm 1cm 0cm 1.3cm,clip]{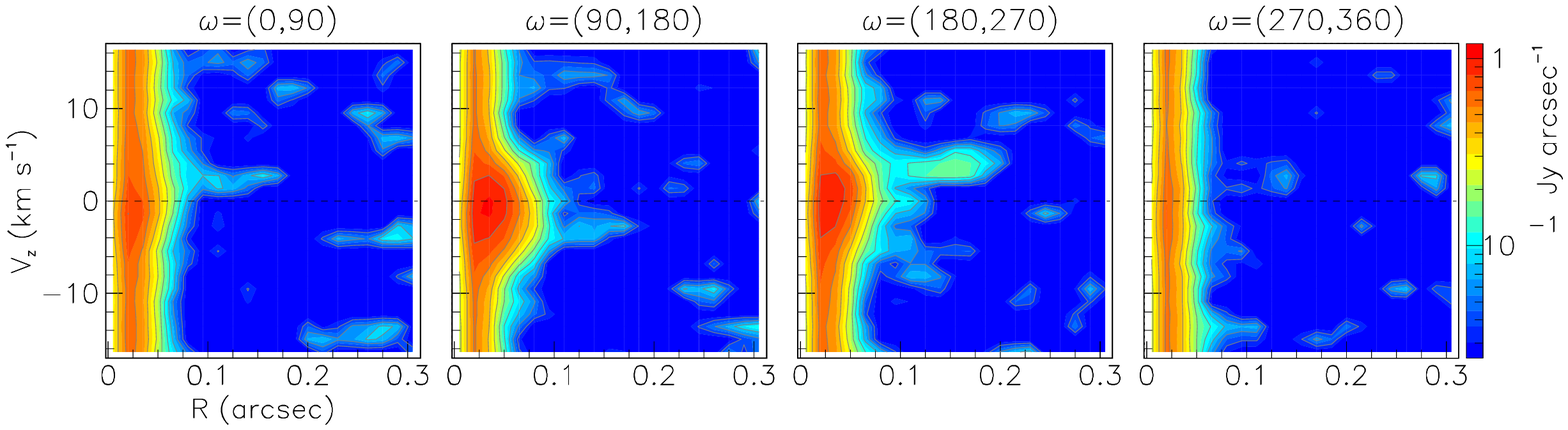}
  \caption{SO$_2$(22$_{2,20}$-22$_{1,21}$) emission. Same as Figure \ref{fig8}.}
\label{fig11}
\end{figure*}

\begin{figure*}
  \includegraphics[height=4.2cm,trim=.0cm 0cm 0.3cm .0cm,clip]{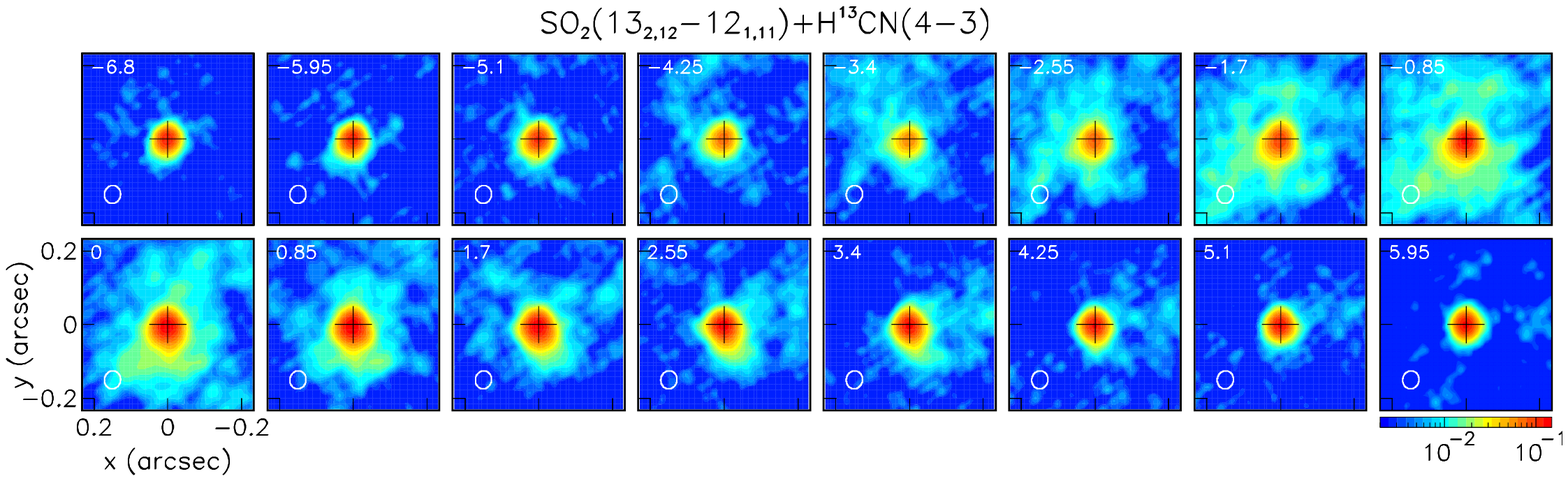}
  \includegraphics[height=4cm,trim=.0cm 0.2cm 0cm 1.1cm,clip]{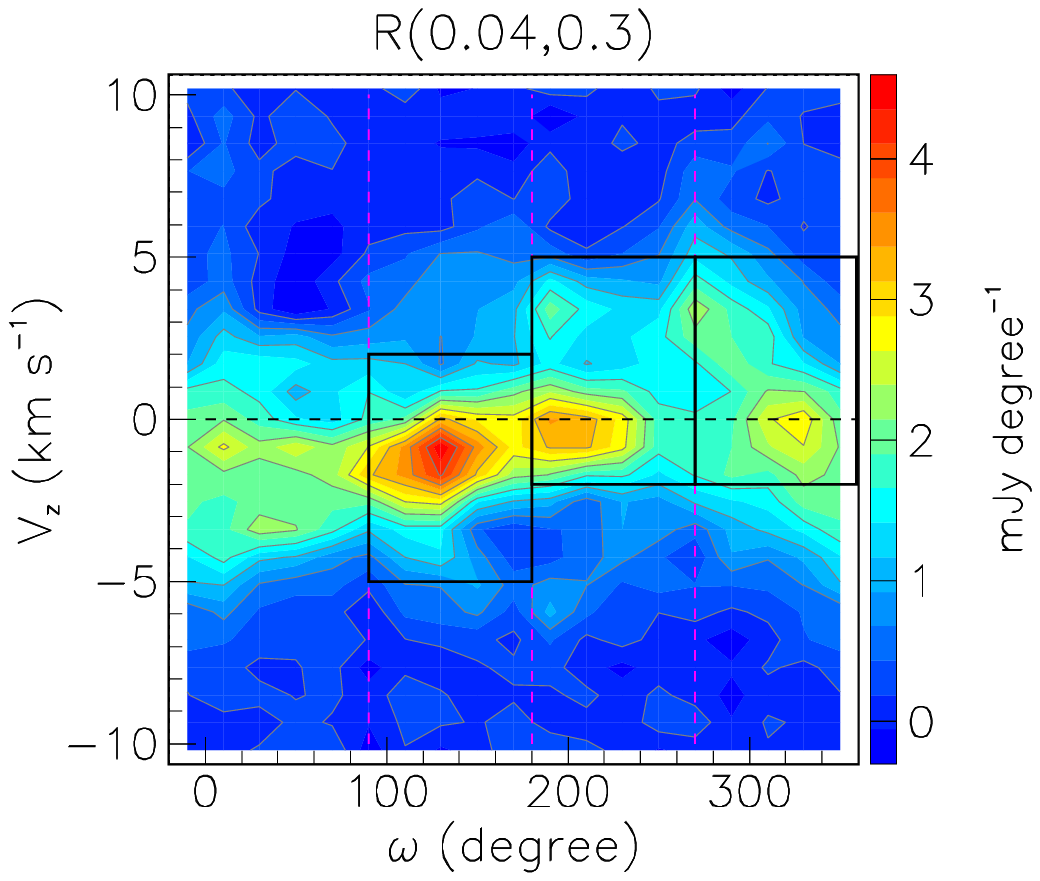}
  \includegraphics[height=3.5cm,trim=.0cm 1cm 0cm 1.3cm,clip]{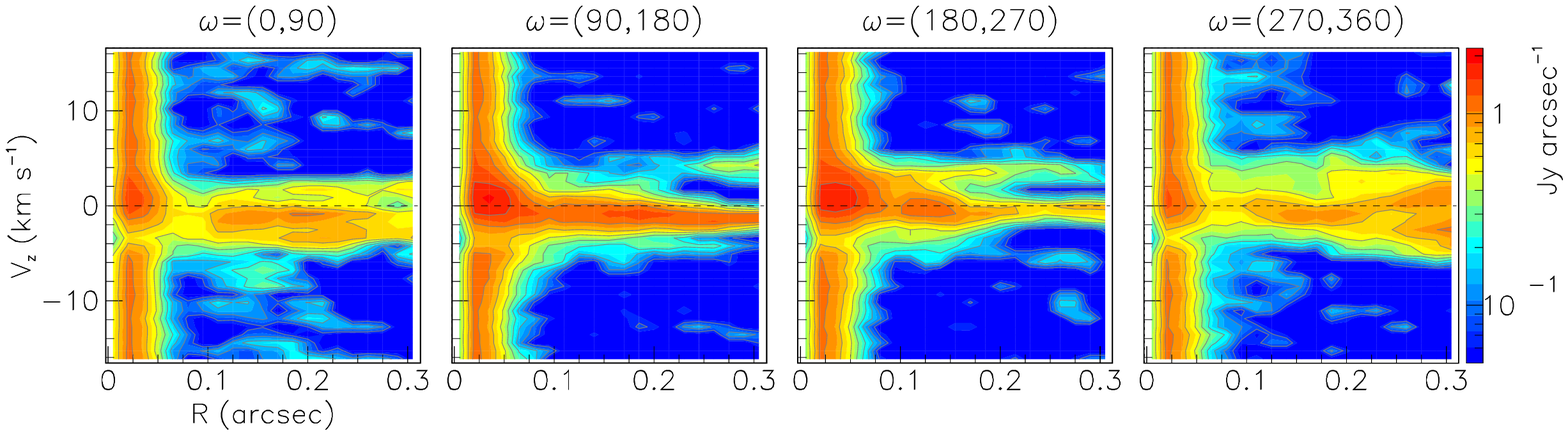}
  
\caption{SO$_2$(13$_{2,12}$-12$_{1,11}$)+H$^{13}$CN(4-3) emission. Same as Figure \ref{fig8}.}
\label{fig12}
\end{figure*}

The $^{29}$SiO(5-4) emission (Figure \ref{fig8}) is seen from the channel maps to be dominantly west and slightly red-shifted. The $V_\text{z}$ vs $\omega$ map locates the bulk of the emission between 0 and 4 \kms\ in $V_\text{z}$ and between 200\dego\ and 340\dego\ in $\omega$. Two blobs of emission, separated by $\sim$50-60\dego\ in $\omega$ and sharing a common mean Doppler velocity of $\sim$2 \kms, are clearly identified from the channel maps and the $V_\text{z}$ vs $\omega$ map. The south-western blob reaches maximum at $R$$\sim$0.1 arcsec and the north-western blob at $R$$\sim$0.2 arcsec. The CO(2-1) emission (Figure \ref{fig9}) displays a qualitatively similar pattern as the $^{29}$SiO(5-4) emission, with red-shifted enhancements in the south-western and north-western quadrants. The SO(5$_5$-4$_4$) (Figure \ref{fig10}), SO$_2$(22$_{2,20}$-22$_{1,21}$) (Figure \ref{fig11}) and SO$_2$-HCN (Figure \ref{fig12}) emissions are relatively stronger in the blue-shifted south-eastern quadrant. In summary, and qualitatively, the five lines display clear enhancements of emission in three angular ranges, well separated from each other, one slightly blue-shifted in the south-eastern quadrant and two slightly red-shifted in the western hemisphere, one south-west and the other north-west. We have checked that such strong anisotropy is not an artefact of the imaging process by simulating visibilities, for the same observational coverage, for a disc source having a radius of 0.5 arcsec and an isotropic brightness decreasing radially as 1/$R$. The brightness of the image, obtained with natural weighting, was found to deviate from the source brightness by less than a few percent.

\begin{figure*}
  \includegraphics[height=4.cm,trim=.5cm 1cm 1.7cm .5cm,clip]{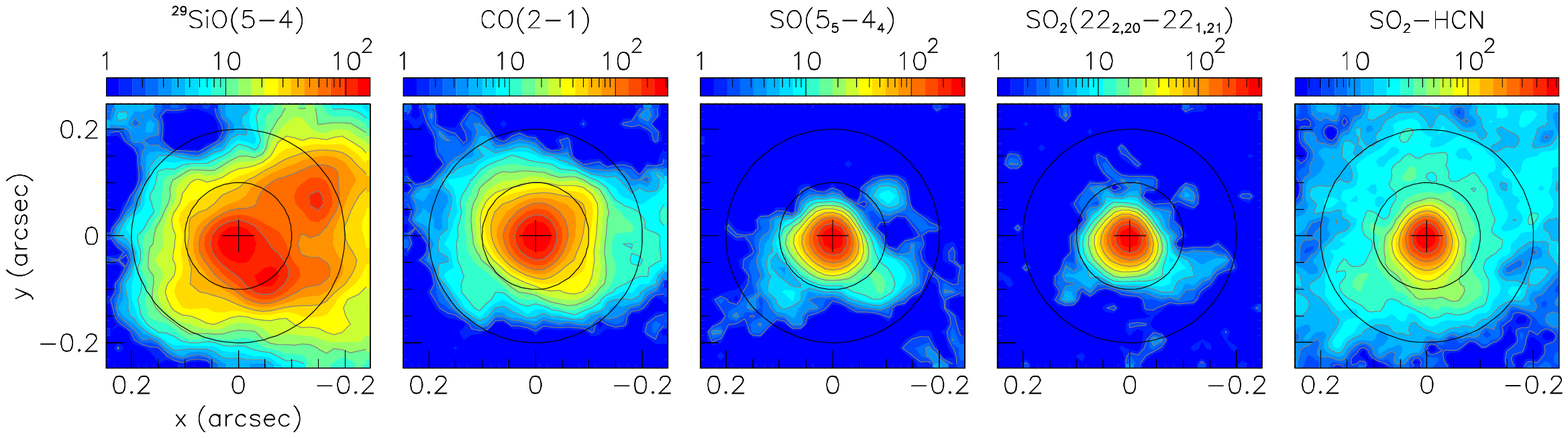}
  \includegraphics[height=3.8cm,trim=.0cm 1cm 1.7cm 1cm,clip]{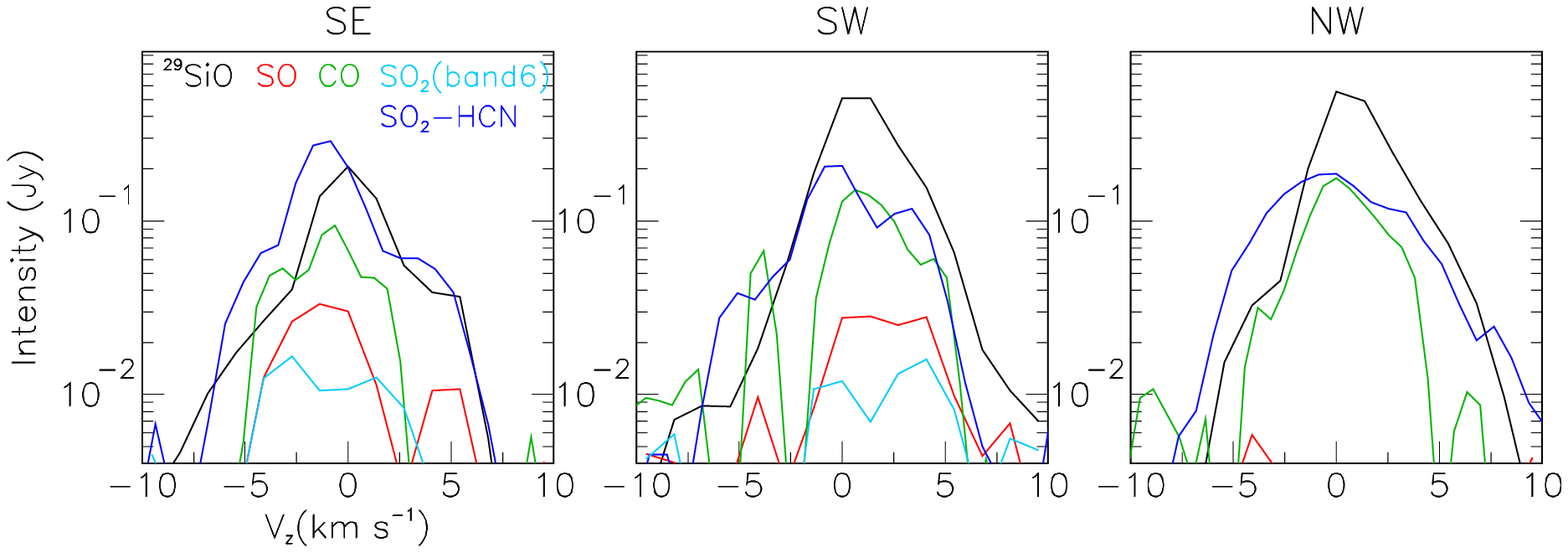}
\caption{Intensity maps for |$V_\text{z}$|<6 \kms\ (up) and line spectra in the ($R$,$\omega$) ranges of the regions SE, SW and NW (down). The position of the star and the radii of 0.1 and 0.2 arcsec are indicated by crosses and black circles respectively. The colour scales are in units of Jy arcsec$^{-2}$ \kms.}
\label{fig13}
\end{figure*}

In order to compare quantitatively the emissions of the five lines, we define three regions, called SE, SW and NW, with angular separations from the centre of the star between 60 and 300 mas, covering $V_z$ and $\omega$ ranges listed in Table \ref{tab4}. They have a same data cube volume of 0.48 arcsec$^2$ \kms\ and are indicated by rectangles on the $V_\text{z}$ vs $\omega$ maps of Figures \ref{fig8} to \ref{fig12}. They contain well the features observed in these figures and take due account of the sizes of the beams and of the continuum emission. Figure \ref{fig13} displays, for each line emission, its intensity map integrated over the interval |$V_\text{z}$|<6 \kms\ and the Doppler velocity spectra measured in each of the three regions defined above. Table \ref{tab4} lists the mean and rms values of the Doppler velocity and position angle distributions measured in each region for each line, together with the associated intensity. 

\begin{table*}
  \caption{Main features of the distributions displayed in Figure \ref{fig13}. The mean and rms values of $V_\text{z}$ are in \kms\ and the intensities in Jy \kms. In all cases a 3$\sigma$ cut has been applied to the data. The lower rows are a reminder of the beam dimensions (in mas) and of the noise (in mJy beam$^{-1}$). For each of the three regions and of the five lines we list the mean values of the Doppler velocity and position angle, <$V_{\text{z}}$> and <$\omega$> calculated over the region using the flux density as weight. Similarly, we list the root-mean-square deviations, Rms($V_{\text{z}}$) and Rms($\omega$) calculated the same way.}
  \label{tab4}
  \begin{tabular}{cccccccc}
    \hline
    Region&&CO(2-1)&$^{29}$SiO(5-4)&SO(5$_5$-4$_4$)&SO$_2$(22$_{2,20}$-22$_{1,21}$)&SO$_2$(13$_{2,12}$-12$_{1,11}$)&Mean \\
    \hline
&<$V_\text{z}$>&
$-$1.2&
$-$0.3&
$-$0.4&
$-$1.4&
$-$1.1&
$-$0.8\\

SE&Rms($V_\text{z}$)&
1.8&
1.5&
1.6&
1.8&
1.3&
1.5\\

$-$5<$V_\text{z}$<2 \kms\ &<$\omega$>&
129\dego&
137\dego&
130\dego&
132\dego&
133\dego&
133\dego\\

90\dego<$\omega$<180\dego&Rms($\omega$)&
25\dego&
25\dego&
24\dego&
23\dego&
24\dego&
24\dego\\

&Intensity&
0.35&
0.71&
0.13&
0.08&
0.81&
$-$\\
\hline
& <$V_\text{z}$>&
1.2&
1.1&
1.7&
1.4&
0.6&
1.1\\

SW&Rms($V_\text{z}$)&
1.7&
1.5&
1.7&
1.8&
1.7&
1.6\\

 $-$2<$V_\text{z}$<5 \kms\ &<$\omega$>&
232\dego&
230\dego&
220\dego&
214\dego&
216\dego&
227\dego\\

180\dego<$\omega$<270\dego&Rms($\omega$)&
24\dego&
23\dego&
19\dego&
24\dego&
24\dego&
22\dego\\

&Intensity&
0.59&
2.19&
0.12&
0.04&
0.61&
$-$\\
\hline

&<$V_\text{z}$>&
0.6&
1.0&
1.0&
$-$&
0.5&
0.8\\

NW&Rms($V_\text{z}$)&
1.6&
1.5&
1.0&
$-$&
1.7&
1.5\\

$-$2<$V_\text{z}$<5 \kms\ &<$\omega$>&
305\dego&
305\dego&
306\dego&
$-$&
312\dego&
305\dego\\

270\dego<$\omega$<360\dego&Rms($\omega$)&
25\dego&
21\dego&
12\dego&
$-$&
25\dego&
22\dego\\

&Intensity&
0.62&
2.19&
0.02&
$-$&
0.53&
$-$\\
\hline
\multicolumn{2}{c}{Beam}&
43$\times$36/33\dego&
47$\times$44/40\dego&
46$\times$38/34\dego&
46$\times$37/34\dego&
49$\times$44/$-$11\dego&
$-$\\
\multicolumn{2}{c}{Noise}&
0.8&
0.7&
0.5&
0.8&
2.1&
$-$\\
\multicolumn{2}{c}{Total intensity}&
1.56&
5.09&
0.27&
0.12&
1.95&
$-$\\
\multicolumn{2}{c}{Share SE/SW/NW (\%)}&
22/38/40&
14/43/43&
48/44/7&
67/33/0&
42/31/27&
$-$\\
\hline
  \end{tabular}
\end{table*}

In order to comment on these results, and to suggest plausible interpretations, we shall use the predictions of a local thermal equilibrium (LTE) regime as a reference with which to compare our results. We recall the relevant arithmetic in the following sub-section.

\subsection{LTE approximation and temperature dependence}

In the LTE approximation, for a given velocity $V_\text{z}$ and a brightness $I_0$ [Jy arcsec$^{-2}$] on the entrance side of a slab of thickness d$z$[arcsec], density $n$[molecules cm$^{-3}$] and temperature $T$[K], the brightness $I$ [Jy arcsec$^{-2}$] at the exit side, reads
\begin{equation}
  I=I_0e^{-\tau}+\varepsilon(1-e^{-\tau})/\tau =\varepsilon/\tau+(I_0-\varepsilon/\tau)e^{-\tau}
\end{equation}

where $\varepsilon$ describes the unabsorbed emission of the slab, and $\tau$ its opacity.
\begin{align}
  \varepsilon&={(hc)/(4\pi\Delta{V})}A_\text{ji}f_\text{pop}n \text{d}z \\
          &=\frac{0.558\,10^6}{\Delta{V}\text{[\kms]}} A_\text{ji}\text{[Hz]} f_\text{pop} n\text{[mol. cm$^{-3}$]} \text{d}z\text{[arcsec]} d\text{[pc]}
\end{align}

Here, $h$ is the Planck constant, $c$ the light velocity, $\Delta{V}$ the FWHM of the line profile, $A_\text{ji}$ the Einstein coefficient, $n$ the density and $d$ the distance of the star from Earth;
\begin{equation}
  f_\text{pop}=(2J+1)e^{-E_\text{u}/T}/Q
\end{equation}
where $J$ and $E_\text{u}$[K] are the angular momentum quantum number and energy of the upper level and $Q=T/k$ is the partition function. Parameters $A_\text{ji}$, $E_\text{u}$ and $k$ are given in Table \ref{tab2} together with the frequency $f$. The opacity is given by the relation ($\Delta{E}$ is the energy of the transition):
\begin{align}
  \tau&=\frac{c^2}{2h} \varepsilon \frac{e^{\Delta E/T}-1}{f^3}\\
  &=2.88\,10^7 \varepsilon \mbox{[Jy arcsec$^{-2}$]}\frac{e^{\Delta E/T}-1}{f\mbox{[GHz]}^3}.
\end{align}

Neglecting opacity and the temperature dependence of $k$, the temperature dependence of $\varepsilon$ is governed by a factor $f_\text{T}=e^{-E_\text{u}/T}/T$: for $T>>E_\text{u}$, $f_\text{T}\sim1/T$ independently from $E_\text{u}$. When comparing two lines, the ratio of the two factors is $f_\text{T1}/f_\text{T2}$=e$^{(E_\text{u2}-E_\text{u1})/T}$. 
To the extent that we deal with temperatures in excess of $\sim$200 K, this ratio is between $\sim$0.7 and 1 for most lines: we cannot expect temperature to cause important differences between the relative line emissions. The only exception is the SO$_2$(22$_{2,20}$-22$_{1,21}$) line, for which $f_\text{T}$(SO$_2$)/$f_\text{T}$(CO)=e$^{-231/T}$, namely  $\sim$0.3 at 200 K and $\sim$0.1 at 100 K.

We also remark that the ratio of opacity to unabsorbed emissivity, $\tau/\epsilon$, is proportional to (e$^{\Delta{E}/T}-1$)/$f^3$. For $\Delta{E}<<T$, $\tau/\epsilon$ is therefore inversely proportional to temperature and to the square of the frequency, namely $\sim$2.5 times smaller for Band 7 than for Band 6 data.

\subsection{Discussion of the results}

We are now in a position to discuss the results summarised in Table \ref{tab4}. Averaging over the five lines, the mean velocity <$V_z$> of the emission of the SE region is blue-shifted by 0.8 \kms, that of the SW region red-shifted by 1.1 \kms\ and that of the NW region red-shifted by 0.8 \kms. Even if the outflows are inclined by as little as 10\dego\ with respect to the plane of the sky, their mean space velocity does not exceed $\sim$6 \kms: we are dealing with slow streams, which barely overcome the star gravity (the escape velocity from a 0.7 solar mass star at an angular distance of 300 mas from its centre is $\sim$6 \kms). In each region, the rms deviations of $V_z$ and $\omega$ from their mean is $\sim$80\% of the value corresponding to a uniform distribution, showing that the selected regions cover well the observed outflows.

A first important remark is that emission is shared very differently between the different regions. While the CO(2-1) and $^{29}$SiO(5-4) lines display similar shares between the three regions (on average, 18\%/41\%/41\% in SE/SW/NW, respectively), the shares of the SO and SO$_2$ lines are very different, with strong dominance in the SE region and, in the case of the Band 6 lines (SO(5$_5$-4$_4$) and SO$_2$(22$_{2,20}$-22$_{1,21}$)), near absence in the NW region. Such important differences challenge interpretation.

\begin{figure*}
  \includegraphics[height=3.8cm,trim=.0cm 1cm 1.7cm 1.cm,clip]{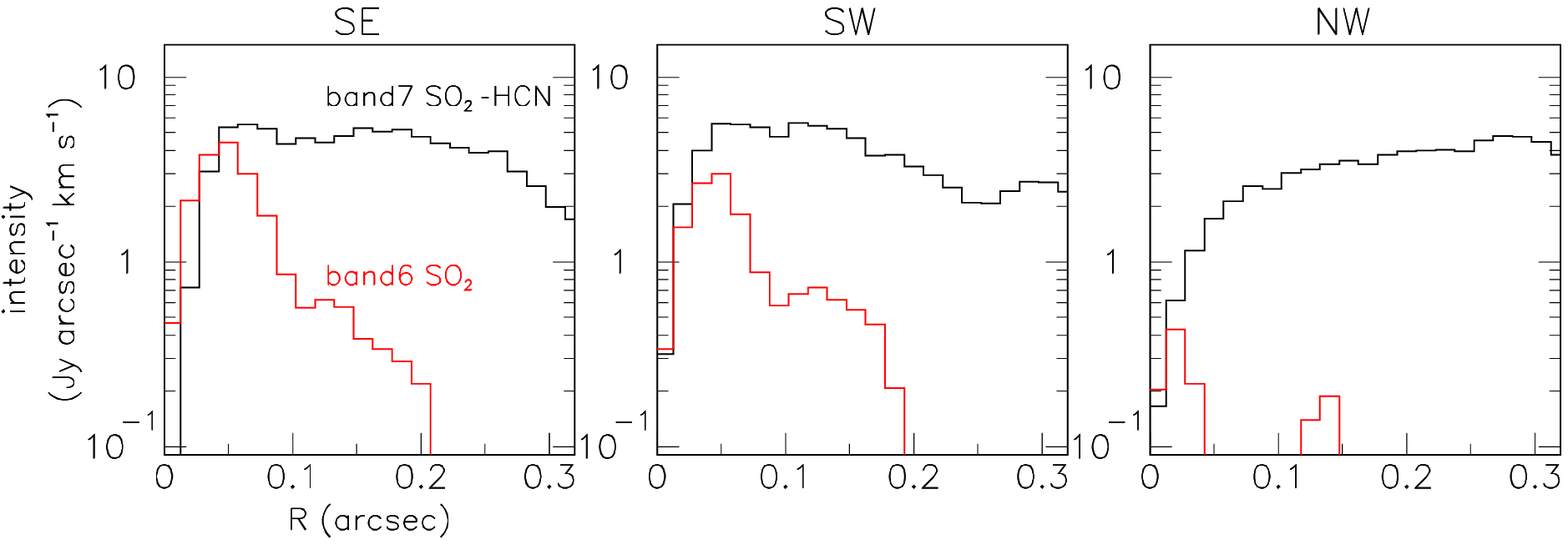}
  \includegraphics[height=3.8cm,trim=.0cm 1cm 1.7cm 1.cm,clip]{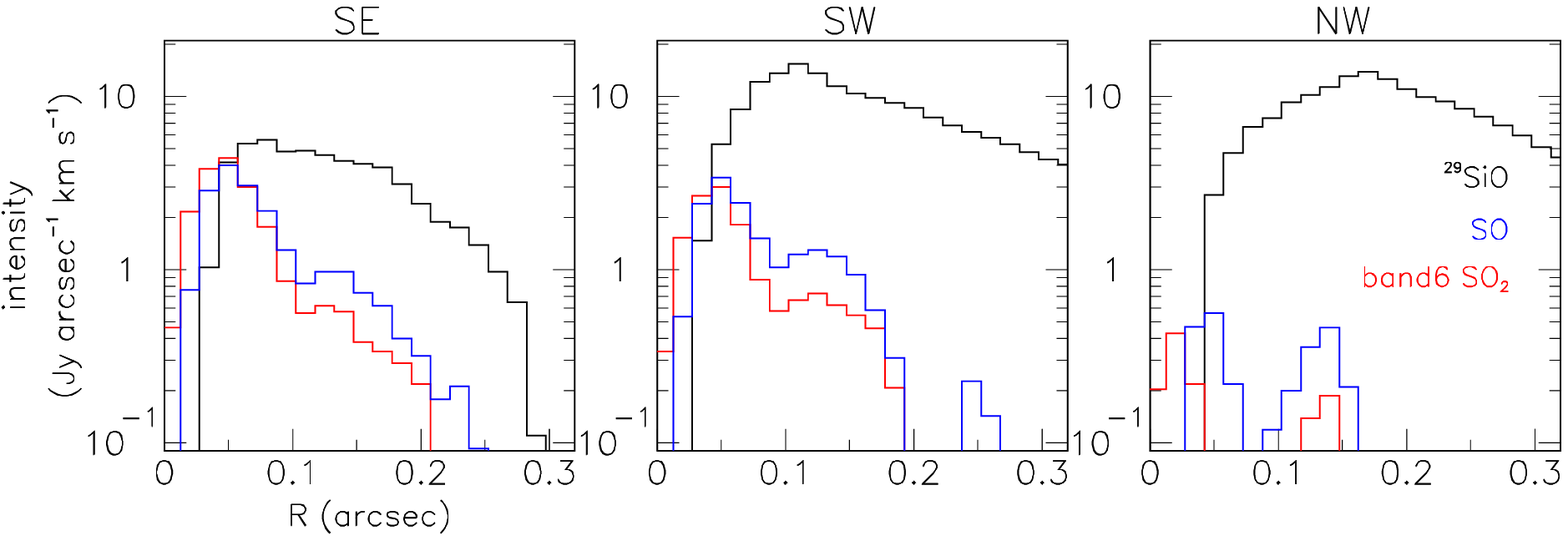}
  \caption{Radial dependence of the intensities (continuum substracted) of the SO$_2$(22$_{2,20}$-22$_{1,21}$) (red) and SO$_2$-HCN (black) lines  (upper panels) and of the SO$_2$(22$_{2,20}$-22$_{1,21}$) (red), SO(5$_5$-4$_4$) (blue) and $^{29}$SiO(5-4) (black) lines (lower panels) in the SE, SW and NW regions defined in Table \ref{tab4}.}
\label{fig14}
\end{figure*}

The Band 7 SO$_2$-HCN line is a blend of the SO$_2$(13$_{2,12}$-12$_{1,11}$) line with the H$^{13}$CN(4-3) line, which is blue-shifted by $\sim$1.1 \kms\ with respect to the SO$_2$ line. Figure \ref{fig14} displays the observed radial dependence of the intensities of the SO$_2$(22$_{2,20}$-22$_{1,21}$) and SO$_2$-HCN emissions. While the latter remains at an approximate level of 2-5 Jy arcsec$^{-2}$ \kms\ over the whole $R$ range (60 to 300 mas) in each of the three regions, the former drops rapidly when $R$ increases and nearly vanishes in the NW region. This suggests that HCN emission dominates over SO$_2$ emission over most of the $R$ range and that SO and SO$_2$ emissions are both strongly depressed in the western hemisphere, nearly absent in the NW region. In the remaining of the article, we take it as granted that such is the case. Accordingly, we interpret the emission of the SO$_2$-HCN line as dominated by emission of the H$^{13}$CN(4-3) line for $R$>$\sim$0.15 arcsec and the emissions of the SO(5$_5$-4$_4$), SO$_2$(22$_{2,20}$-22$_{1,21}$) and SO$_2$(13$_{2,12}$-12$_{1,11}$) lines as being confined within $\sim$0.15 arcsec from the centre of the star, as illustrated in the lower panels of Figure \ref{fig14}. Such confinement is at strong variance with the much broader radial distributions assumed by \citet{Danilovich2016}, \citet{VandeSande2020} and \citet{Massalkhi2020} and its vicinity from continuum emission prevents a reliable study. The similarity between the SO and SO$_2$ emissions supports the assumption that SO$_2$ is essentially produced \citep{Cherchneff2006} via the reaction:
\begin{equation}
   \text{SO+OH} \longrightarrow \text{SO}_2+\text{H}
\end{equation}
and suggests that SO and SO$_2$ molecules are destroyed by the same mechanism. However, the observed confinement is not expected. As remarked by \citet{Massalkhi2020} photodissociation by UV radiation is equally probable for SO and SiO molecules: Figure \ref{fig14} excludes an interpretation of the confinement in terms of photodissociation of the SO and SO$_2$ molecules. Confinement within 1-2 stellar radii (60-150 mas) is predicted by chemical models for carbon-rich stars, the absence of OH molecules at larger distances preventing formation of SO molecules via the reaction \citep{Cherchneff2006}:
\begin{equation}
  \text{S+OH} \longrightarrow \text{SO+H}
\end{equation}
   
However, as soon as C/O falls below unity, SO formation extends well beyond 5 stellar radii (60-150 mas). \citet{Agundez2020} use a chemical equilibrium model that predicts confinement within 5-10 stellar radii for Mira stars, however with uncertainties at the level of an order of magnitude.

To get some insight into the effect of opacity, we display in Figure \ref{fig15} Doppler velocity spectra observed over the stellar disc ($R$<30 mas) and around it (30<$R$<50 mas). In the ring around the stellar disc, where the continuum level is $\sim$17\% of its value over the stellar disc, absorption is weak and indicates that opacity corrections in the radial range covered by Table \ref{tab4} are small. As expected, absorption is much stronger over the stellar disc. In percentage of the continuum level, it reaches $\sim$70\% at $V_z$$\sim$$-$3 \kms\ for $^{29}$SiO(5-4) and $\sim$24\% at $V_z$$\sim$6 \kms\ for CO(2-1). These differences reveal different contributions from out-flowing and in-falling gas, depending on the radial range being probed and are briefly discussed in the next subsection.

In order to perform a more quantitative analysis of the observed intensities listed in Table \ref{tab4}, we use the LTE description as reference. In the radial range covered by Table \ref{tab4}, 60 to 300 mas, we expect the temperature to decrease from $\sim$1300-1500 K to $\sim$300-500 K \citep{Fonfria2020} and the factor e$^{-Eu/T}$ is close to unity. We calculate accordingly, for each of the three regions and for each line, the ratio between the observed intensity (Table \ref{tab4}) and the factor $F=A_{ji}(2J+1)k\text{e}^{-Eu/T}$ calculated at $T$=1000 K using the line parameters listed in Table \ref{tab2}. In the LTE approximation, neglecting absorption, this ratio is proportional to the abundance of the excited molecule, namely to the global density in each region and to the relative molecular abundances. The latter are particularly informative as they reveal important differences between the different lines. We list in Table \ref{tab5}, for each region and for each line, the calculated ratio of the relevant molecular abundance to the $^{29}$SiO abundance chosen as reference.  We also list, in the same table, abundances obtained from single dish observations using broad and isotropic radial distributions. The main results can be summarized as follows:

(i) The observed CO/SiO abundance ratios are within $\pm$30\% equal to the abundance ratios obtained from single dish observations, 30\% larger in the SE region, 30\% smaller in the SW and NW regions. This implies that within the large uncertainties attached to these evaluations, absorption and deviations from an LTE regime do not much influence the evolution of the outflows. Namely the presence of shocks suggested by studies of continuum emission and molecular abundances is confined to the very close neighbourhood of the star, well within 60 mas. Yet, small differences between different regions of the CO/SiO abundance ratio would not be surprising as the SiO emission is more sensitive to temperature than the CO emission, making SiO a good shock tracer.
  
(ii) The observed SO/SiO and SO$_2$/SiO abundance ratios are much smaller in the western hemisphere, and particularly in the NW region, than in the SE region. However, the confinement of the SO and SO$_2$ molecules within $\sim$0.15 arcsec from the centre of the star, for which strong evidence has been obtained, makes a reliable quantitative analysis difficult. As clearly illustrated in Figures \ref{fig10} and \ref{fig11}, the emission of these molecules does not extend radially much beyond continuum emission: better angular resolution and sensitivity would be required for a reliable analysis.

(iii) The observed H$^{13}$CN/$^{29}$SiO abundance ratio, evaluated by assuming that H$^{13}$CN emission dominates over SO$_2$ emission in the NW region, is $\sim$7\%. It corresponds to a relative abundance of H$^{12}$CN of $\sim$10$^{-6}$, $\sim$4 times larger than obtained from single dish analyses. This is consistent with observations in the other regions, but a reliable evaluation of the HCN abundance in the SE region would require observing the emission of other, unblended, lines.
  
\begin{table*}
  \caption{Comparison between observations and LTE predictions for the $\nu$=0 line emissions in the SE, SW and NE regions. Single dish abundances relative to H$_2$ are from (a) \citet{GonzalezDelgado2003}, (b) \citet{Schoier2013} and (c) \citet{Danilovich2016} (value obtained for R Dor) and (g) \citet{Massalkhi2020}. Isotopic ratios are from (e) \citet{DeBeck2018} and (f) \citet{Ramstedt2014}. The factors $F=A_{\text{ji}}(2J+1)k\text{e}^{-Eu/T}$ are calculated at $T$=1000 K. Relative abundances normalised to the $^{29}$SiO value, $A_{\text{SiO}}$, are listed for single dish observations and for the present work in regions SE, SW and NW separately. For the HCN line we quote the values obtained using the observed blended line intensities as upper limits.}
  \label{tab5}
 \begin{tabular}{cccccccc}
   
\multicolumn{2}{c}{Line}&CO(2-1)&
$^{29}$SiO(5-4)&
H$^{13}$CN(4-3)&
SO(5$_5$-4$_4$)&
SO$_2$(22$_{2,20}$-22$_{1,21}$)&
SO$_2$(13$_{2,12}$-12$_{1,11}$)\\
\hline
\multicolumn{2}{c}{Abundance}&
\multirow{2}{*}{2$\times$10$^{-4}$ (a)}&
13$\times$10$^{-6}$ (a)&
2.5$\times$10$^{-7}$ (b)&
6.7$\times$10$^{-6}$ (c)&
5$\times$10$^{-6}$ (c)&
5$\times$10$^{-6}$ (c)\\

\multicolumn{2}{c}{($f_0=n/n_{H2}$)}&&$\times$0.08 (e)&$\times$0.08 (e)&6.6$\times$10$^{-7}$ (g)&<1.1$\times$10$^{-7}$ (g)&<1.1$\times$10$^{-7}$ (g)\\
\multicolumn{2}{c}{$A_{\text{SiO}}$ Single dish}&
192 &
1&
0.019 &
6.4 (c)/0.6 (g)&
4.8 (c)/<0.11 (g)&
4.8 (c)/<0.11 (g)\\

\multicolumn{2}{c}{$F$}&
8.92$\times$10$^{-3}$&
4.53&
14.3&
0.44&
0.069&
0.124\\
\multirow{3}{*}{A$_{\text{SiO}}$ this work}&
SE&
250&
1&
< 0.36&
1.9&
7.2&\\

&SW&
140&
1&
<0.089&
0.6&
1.2&\\

&NW&
140&
1&
<0.076&
0.1&
0&\\

\hline
 \end{tabular}
 \end{table*}

Finally, we remark that the complexity of the observed morpho-kinematics prevents confirmation of two predictions made earlier by \citet{Fonfria2019b} and \citet{Wiesemeyer2009}. The former authors speculated that the high resolution CO(2-1) observations might suggest the presence of rotation in the close neighbourhood of the star. It is difficult, however (see Figure \ref{fig9}), to separate a possible effect of rotation from the contribution of the south-western outflow and evidence for a possible rotation cannot be stated with reasonable confidence. The latter authors have given arguments in favour of the presence of an evaporating Jovian planet orbiting R Leo at some 30 mas from its centre. According to their definition of the orbit parameters, such a planet would have been south-east of the star at the time of the present observations and blue-shifted by some 4-5 \kms, in a region where we observe no feature that might be revealing its presence: again, speculation about the possible presence of a planet cannot be made with reasonable confidence from the present observations.

\subsection{Summary}

\begin{figure*}
  \includegraphics[height=3.9cm,trim=0.cm 0.5cm 1.7cm 1.cm,clip]{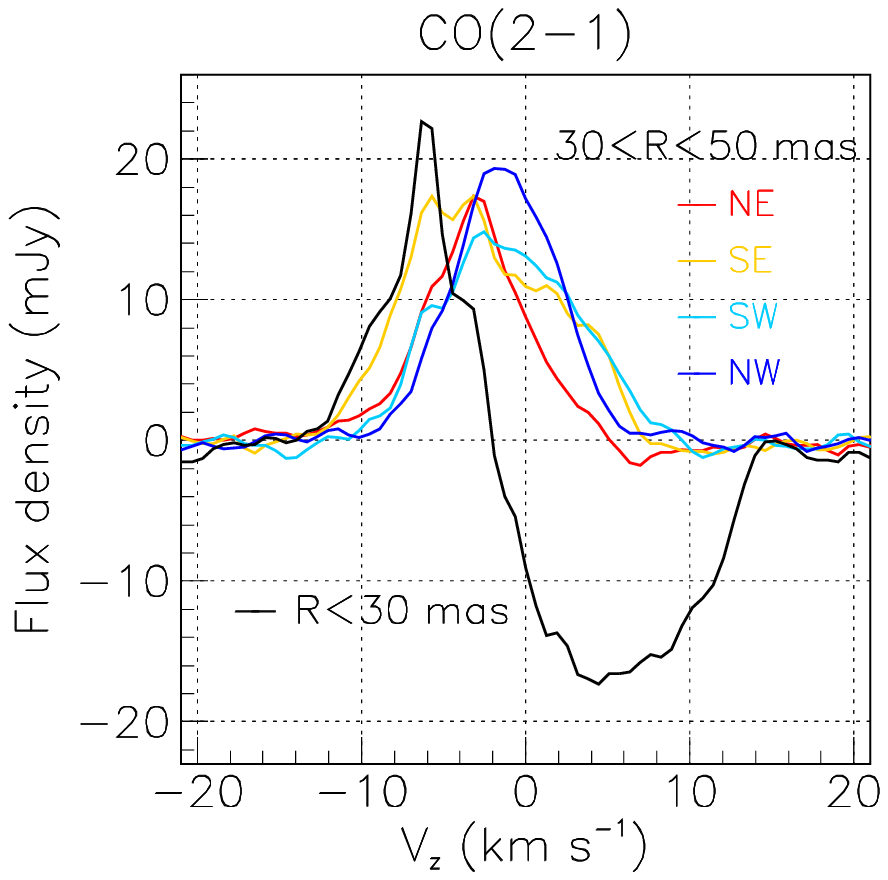}
  \includegraphics[height=3.9cm,trim=1cm 0.5cm 1.7cm 1.cm,clip]{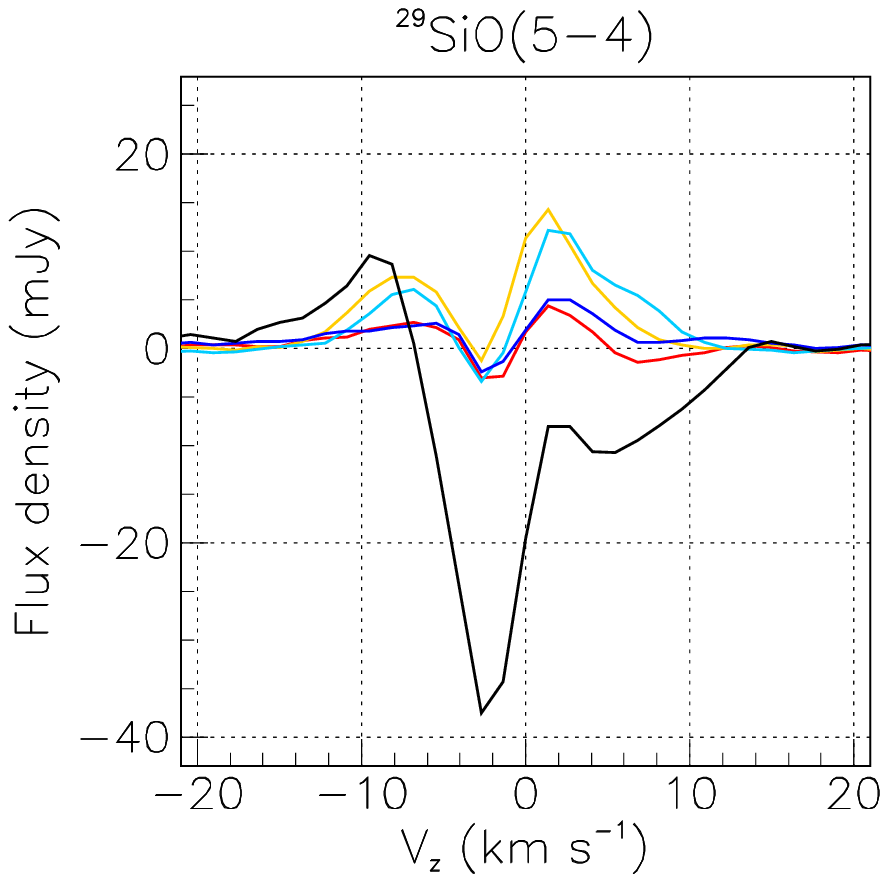}
  \includegraphics[height=3.9cm,trim=1cm 0.5cm 1.7cm 1.cm,clip]{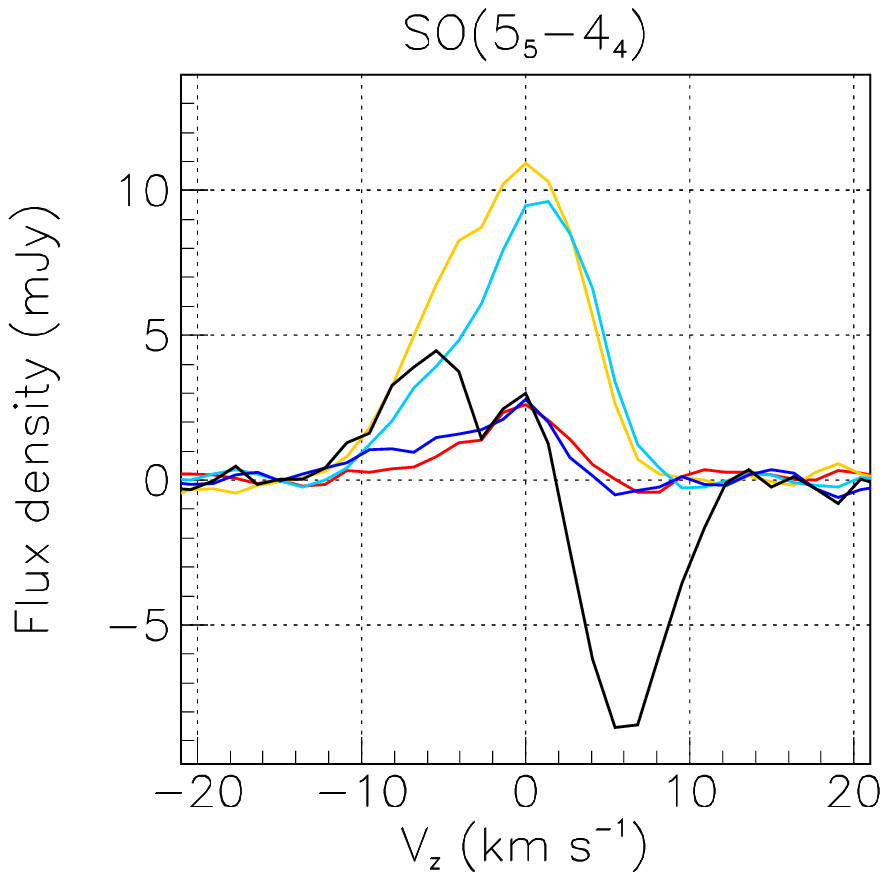}
  \includegraphics[height=3.9cm,trim=1cm 0.5cm 1.7cm 1.cm,clip]{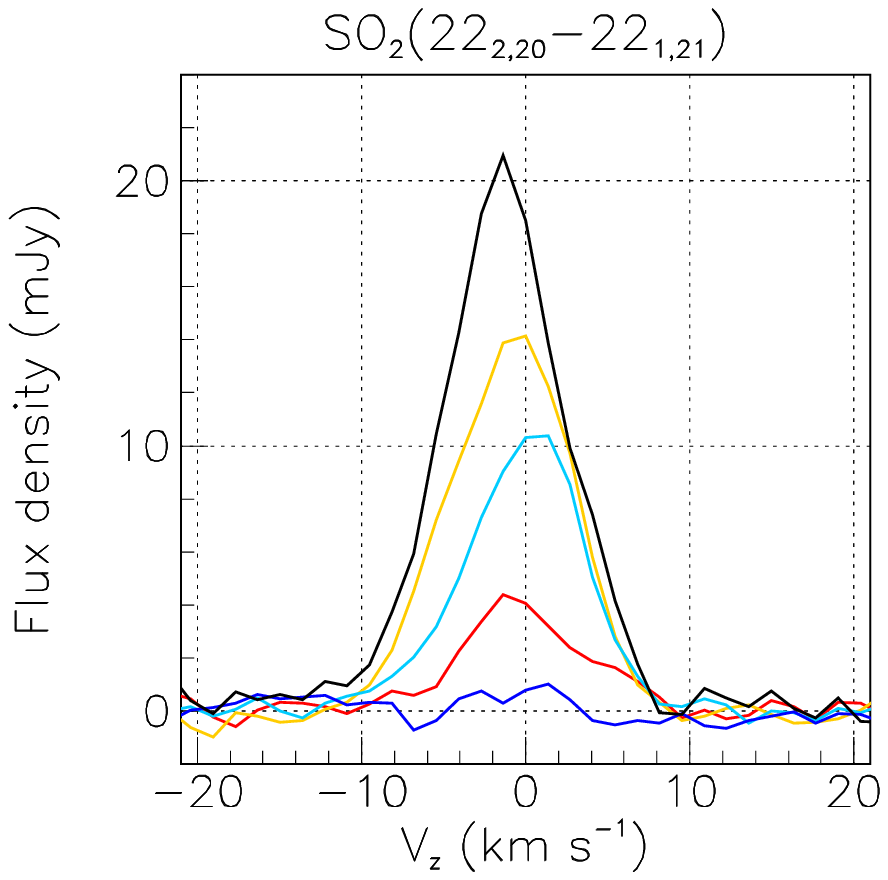}
  \includegraphics[height=3.9cm,trim=1cm 0.5cm 1.7cm 1.cm,clip]{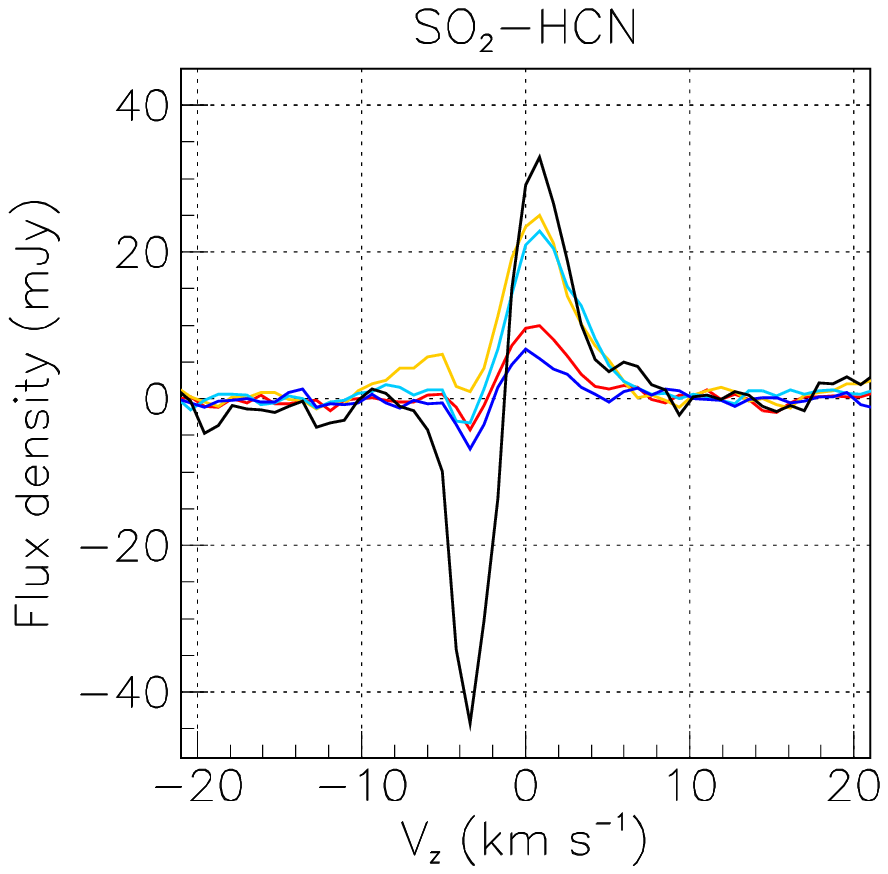}
\caption{Doppler velocity spectra observed above the stellar disc (black, $R$<30 mas) and around it in four different quadrants (30<$R$<50 mas, colour code in the insert of the left panel). Lines are specified on top of each panel. Each spectrum, evaluated from cubes without continuum subtraction, has been shifted downward to have the flux density cancel outside the line.}
\label{fig15}
\end{figure*}

\begin{figure*}
  \includegraphics[height=4.8cm,trim=.0cm 0.5cm 1cm 1.cm,clip]{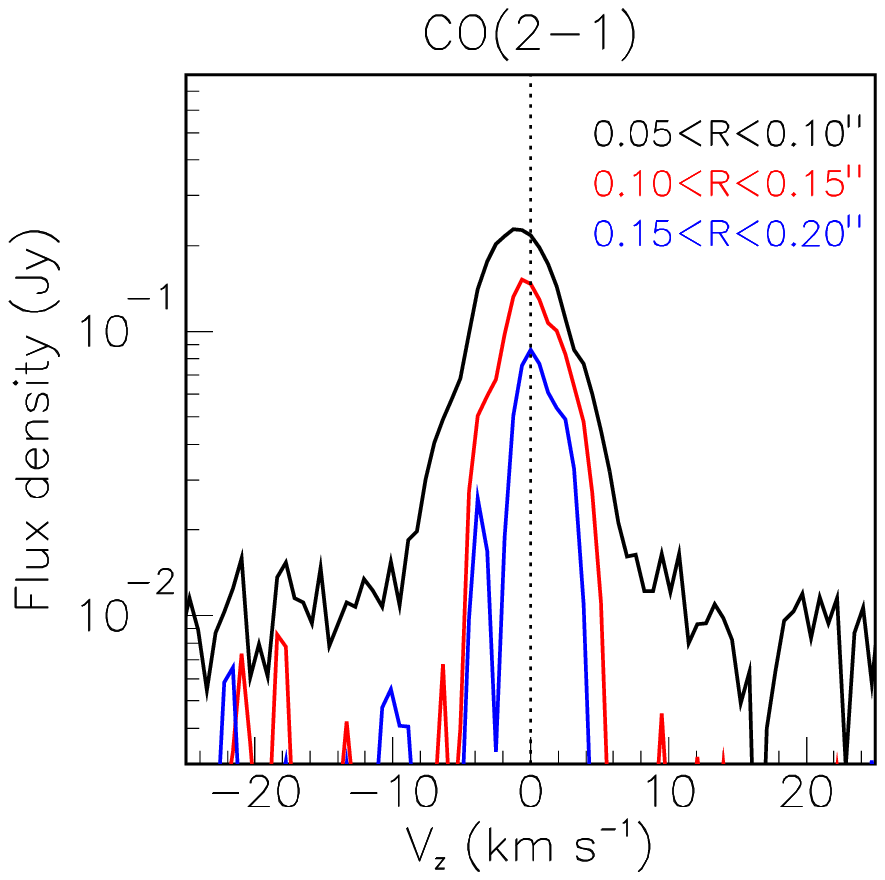}
  \includegraphics[height=4.8cm,trim=.0cm 0.5cm 1cm 1.cm,clip]{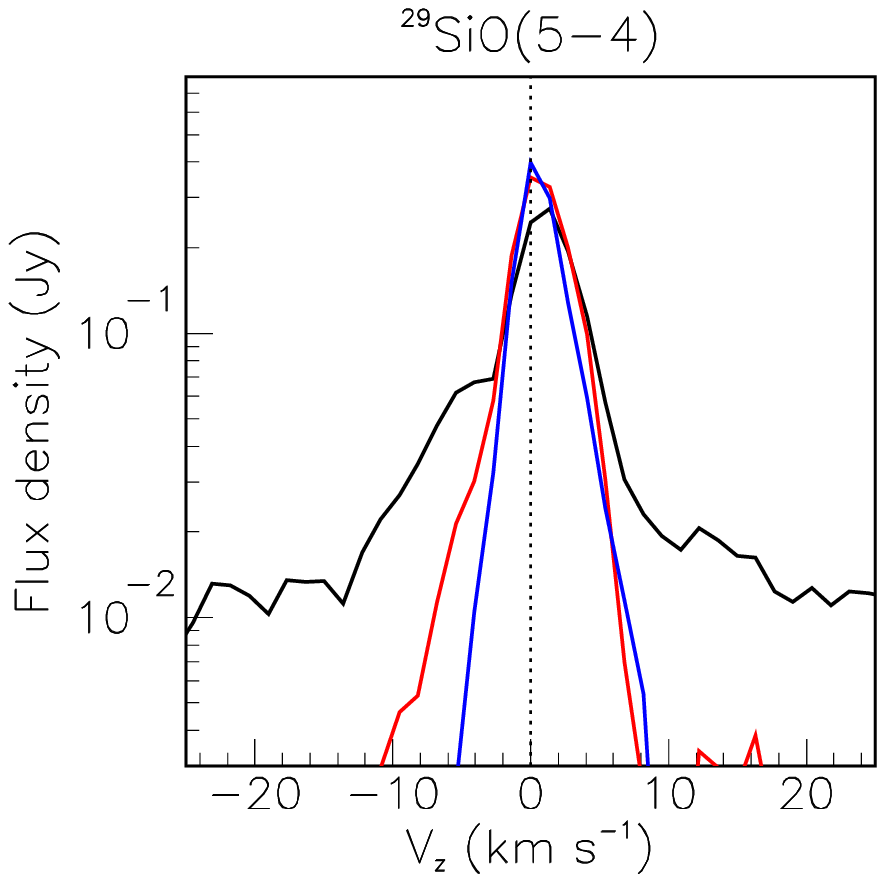}
\caption{Doppler velocity spectra of the CO(2-1) and $^{29}$SiO(5-4) line emissions in 50 mas wide rings as indicated in the insert.}
\label{fig16}
\end{figure*}

In summary, within 300 mas from the centre of the star, we observe a very complex morpho-kinematics giving evidence for anisotropic emission and, within 40 mas from the centre of the star, we find evidence for simultaneous presence of outflowing and in-falling gas. Similar features have been observed previously in several other oxygen-rich AGB stars, such as R Dor \citep{Hoai2020, Nhung2021}, W Hya \citep{Hoai2021,Vlemmings2017} and $o$ Ceti \citep{Wong2016, Khouri2018, Nhung2022a}. The angular and radial distributions of the pattern of in-falling and out-flowing gas, its evolution as a function of stellar phase, and how it is influenced by anisotropy, cannot be accurately predicted in the present state of our understanding. In particular the radial range over which the simultaneous presence of in-falling and out-flowing gas contributes to line broadening cannot be precisely defined. In the present case of R Leo, Figure \ref{fig15} has shown that Doppler velocities reaching beyond 10 \kms\ are observed over the stellar disc for the $^{29}$SiO(5-4), CO(2-1) and SO(5$_5$-4$_4$) lines. They decrease to $\sim$10 \kms\ for $R$ between 30 and 50 mas and reach terminal values of $\sim$5 \kms\ for $R$ beyond $\sim$100 mas (Figure \ref{fig16}). In the case of $o$ Ceti \citep{Nhung2022a}, at a distance of 100 pc instead of 114 pc for R Leo, a similar decrease is observed, from $\sim$20 \kms\ to $\sim$6 \kms\ between $\sim$ 50 mas and $\sim$ 150 mas. However, while $o$ Ceti displays approximate isotropy in this radial range, R Leo shows important anisotropy, with much stronger emission in the southern than northern hemisphere.

The diversity of patterns displayed in Figure \ref{fig15} suggests that the SO$_2$(22$_{2,20}$-22$_{1,21}$) line probes too high temperatures and/or too small abundances to reveal significant absorption, while the CO(2-1) line probes too large distances to reveal important anisotropy. The different patterns displayed by in-falling gas are more difficult to interpret. \citet{Fonfria2019b} have commented on the cases of the CO(2-1) and $^{29}$SiO(5-4) lines, both in the $\nu$=0 and $\nu$=1 lines, and underscored the complexity of the implied morpho-kinematics.

Three slow gas streams have been identified, covering projected angular distances from the centre of the star of up to $\sim$400 mas with rms spread in $V_z$ and $\omega$ of typically 1.5 \kms\ and 22\dego, respectively. Beyond 40 mas angular separation from the centre of the star, the emissions of the CO(2-1), $^{29}$SiO(5-4) and H$^{13}$CN(4-3) molecular lines do not reveal major anomaly when compared with LTE predictions, showing neither strong absorption contributions nor important deviations from an LTE regime. This is at variance with the regime at stake within 40 mas, known to host shocks and turbulences from the study of continuum emission \citep{Vlemmings2019} and of the non-equilibrium chemistry required to explain the molecular abundances of species formed in this radial range \citep{Bieging2000, Schoier2013, Massalkhi2020}. In contrast with the CO(2-1), $^{29}$SiO(5-4) and H$^{13}$CN(4-3) line emissions, the SO(5$_5$-4$_4$) and SO$_2$(22$_{2,20}$-22$_{1,21}$) line emissions display features, which current chemical models fail to explain. While the evidence for similar confinement of both SO and SO$_2$ molecules within $\sim$0.15 arcsec from the centre of the star in each of the three outflows can be stated with confidence, a reliable quantitative study would require new observations and a better understanding of the chemistry at stake.

\section{Vibrationally excited SiO lines}

\begin{figure*}
  \includegraphics[height=5.05cm,trim=0.2cm 0cm 0.3cm .5cm,clip]{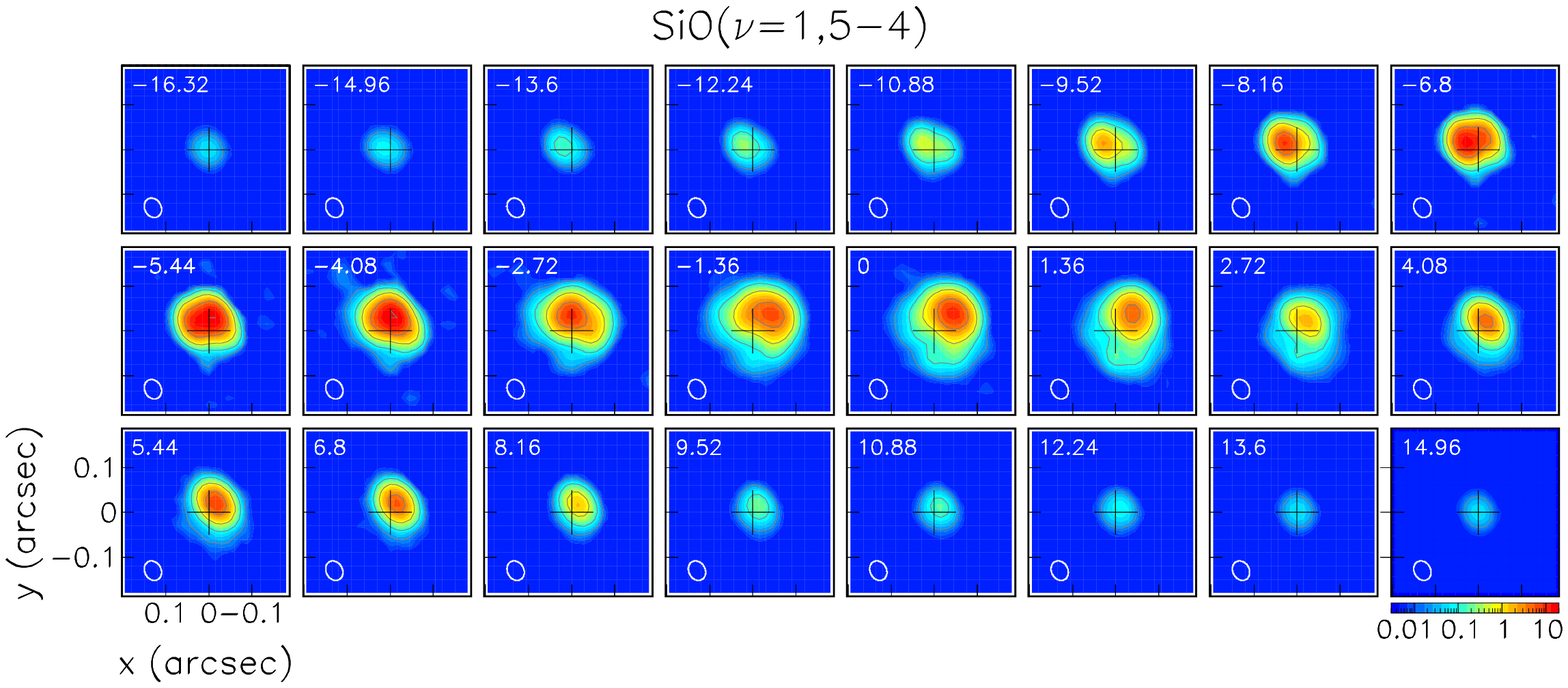}
  \includegraphics[height=5.cm,trim=.0cm 0.7cm 0cm 1.2cm,clip]{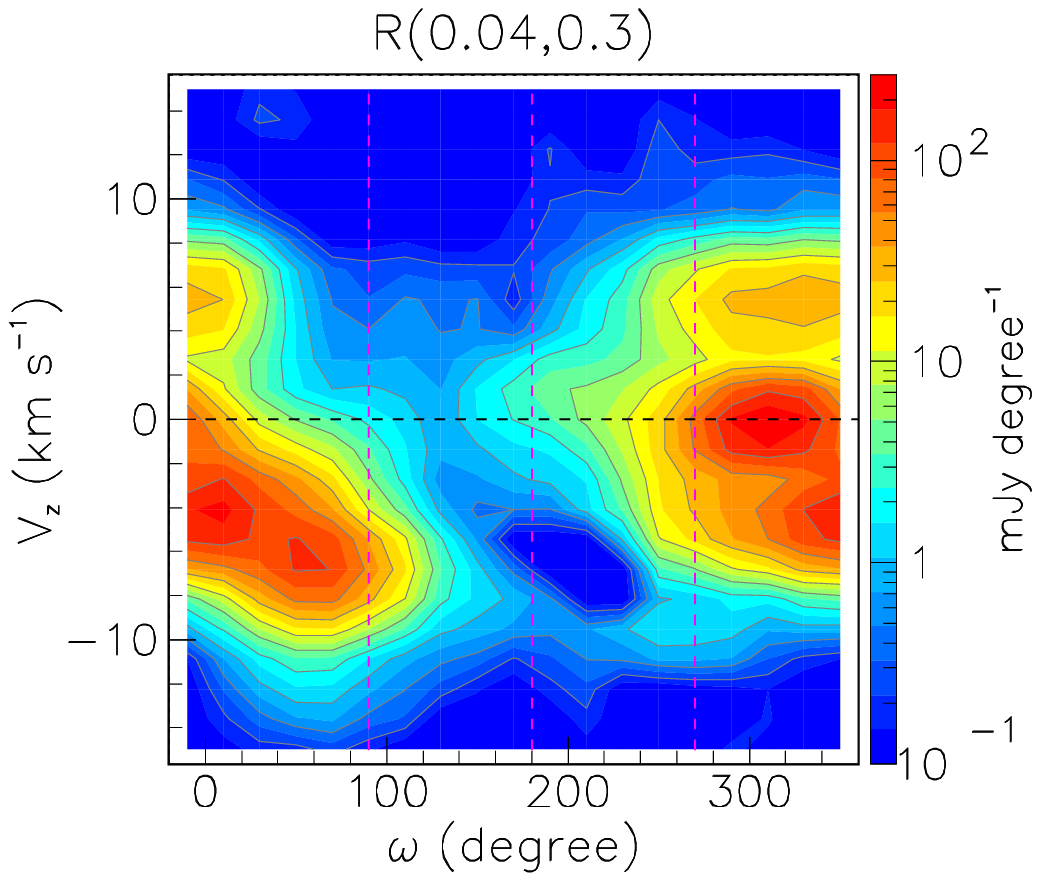}
  \includegraphics[height=3.5cm,trim=0cm 1cm 0cm 1cm,clip]{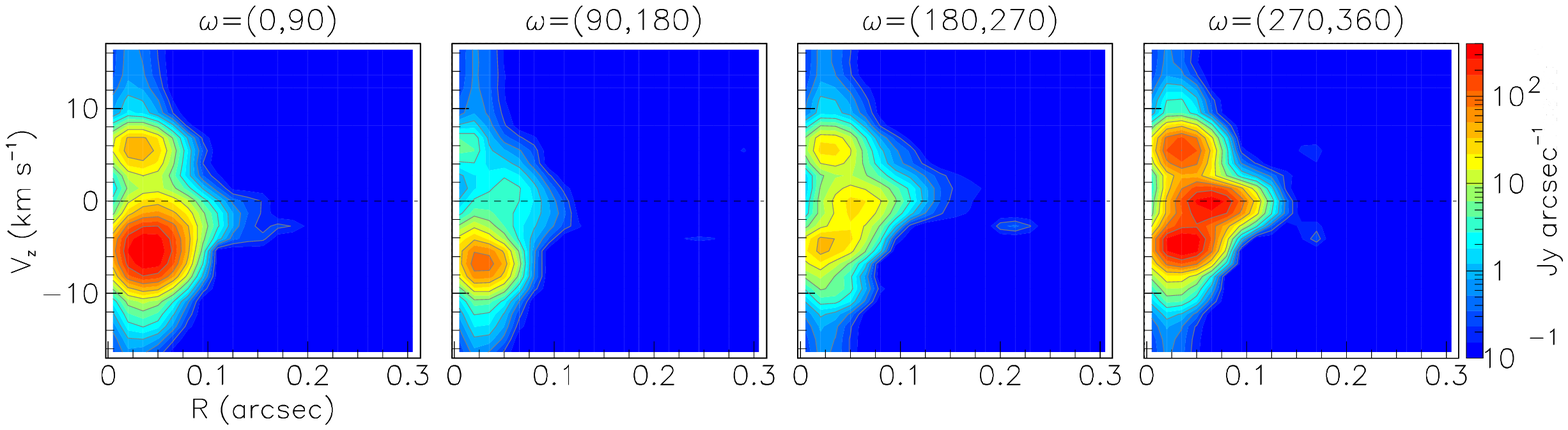}
  \caption{$^{28}$SiO($\nu$=1,5-4) emission.  Upper row: Left, channel maps. The crosses show the position of the star. The colour scale is in units of Jy beam$^{-1}$.  Right, PV map, $V_\text{z}$ vs $\omega$, for 0.04<$R$<0.3 arcsec. Lower row: PV maps, $V_\text{z}$ vs $R$, for 0\dego<$\omega$<90\dego, 90\dego<$\omega$<180\dego, 180\dego<$\omega$<270\dego\ and 270\dego<$\omega$<360\dego, from left to right.}
\label{fig17}
\end{figure*}

\begin{figure*}
  \includegraphics[height=4.8cm,trim=.5cm 0cm 0cm .5cm,clip]{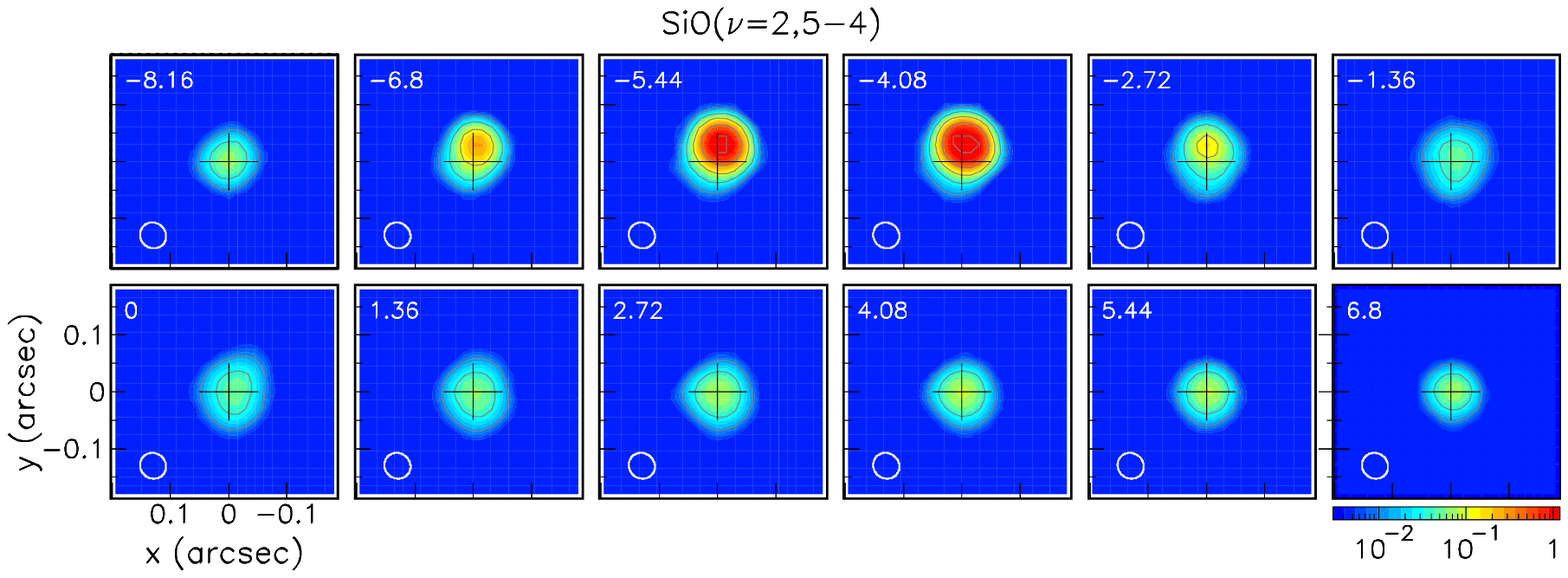}
  \includegraphics[height=4.55cm,trim=.0cm 0.cm 0cm 1.2cm,clip]{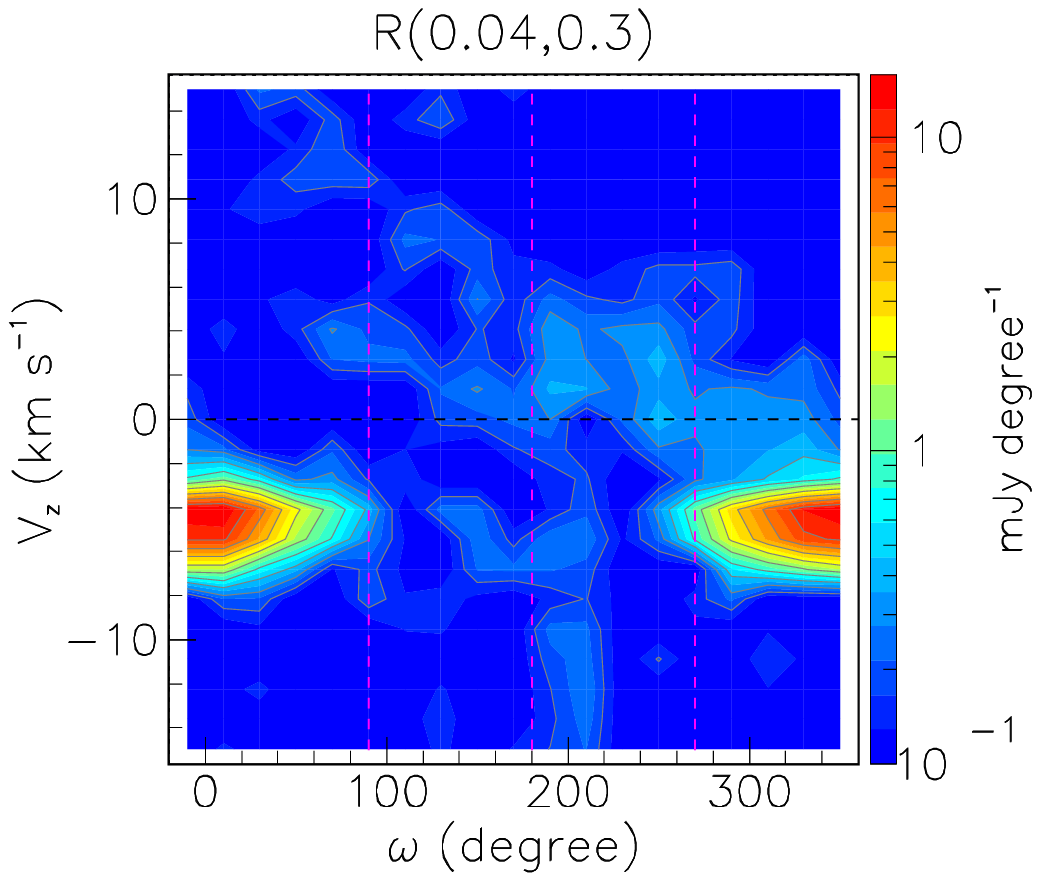}
  \includegraphics[height=3.5cm,trim=.0cm 1cm 0cm 1.3cm,clip]{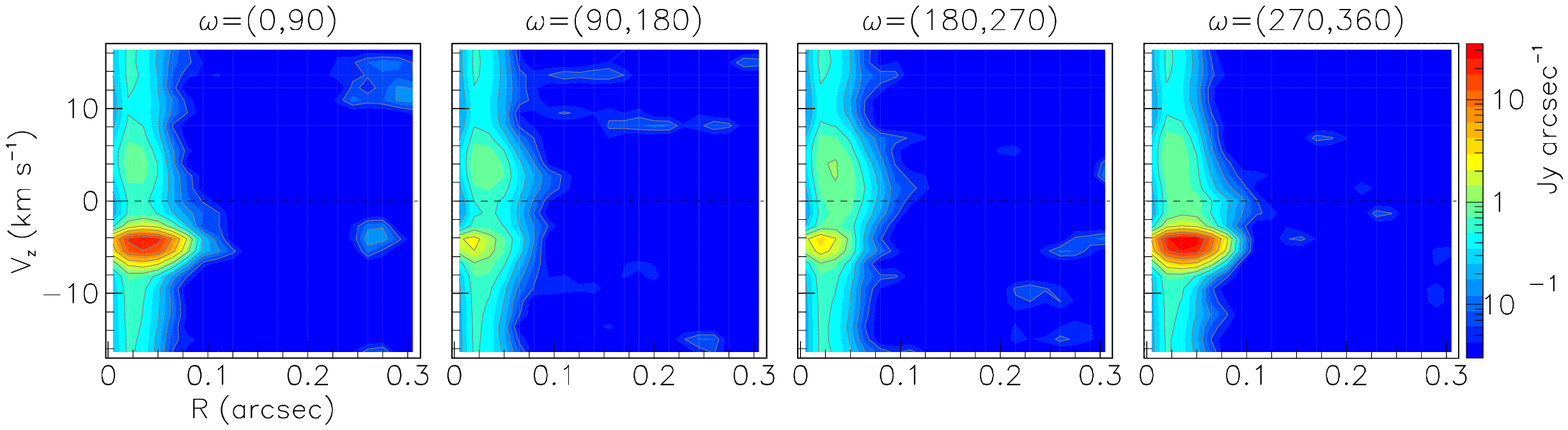}
\caption{$^{28}$SiO($\nu$=2,5-4) emission. Same as Figure \ref{fig17}.}
\label{fig18}
\end{figure*}

\begin{figure*}
  \includegraphics[height=4cm,trim=0cm 0cm 0.5cm 0cm,clip]{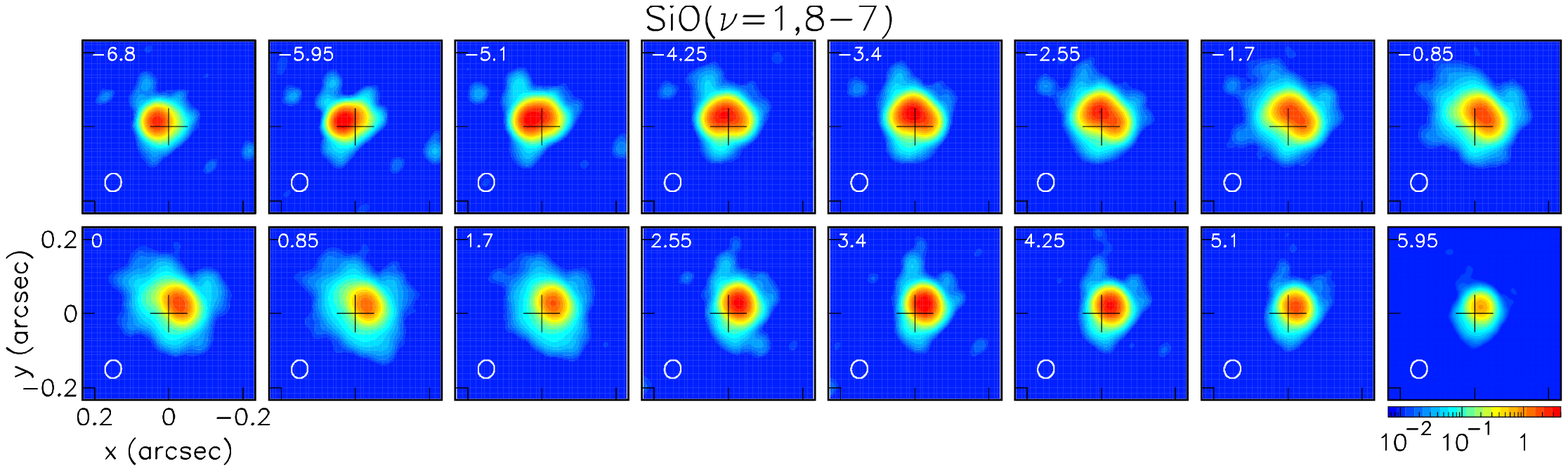}
  \includegraphics[height=4cm,trim=.0cm 0.cm 0cm 1.3cm,clip]{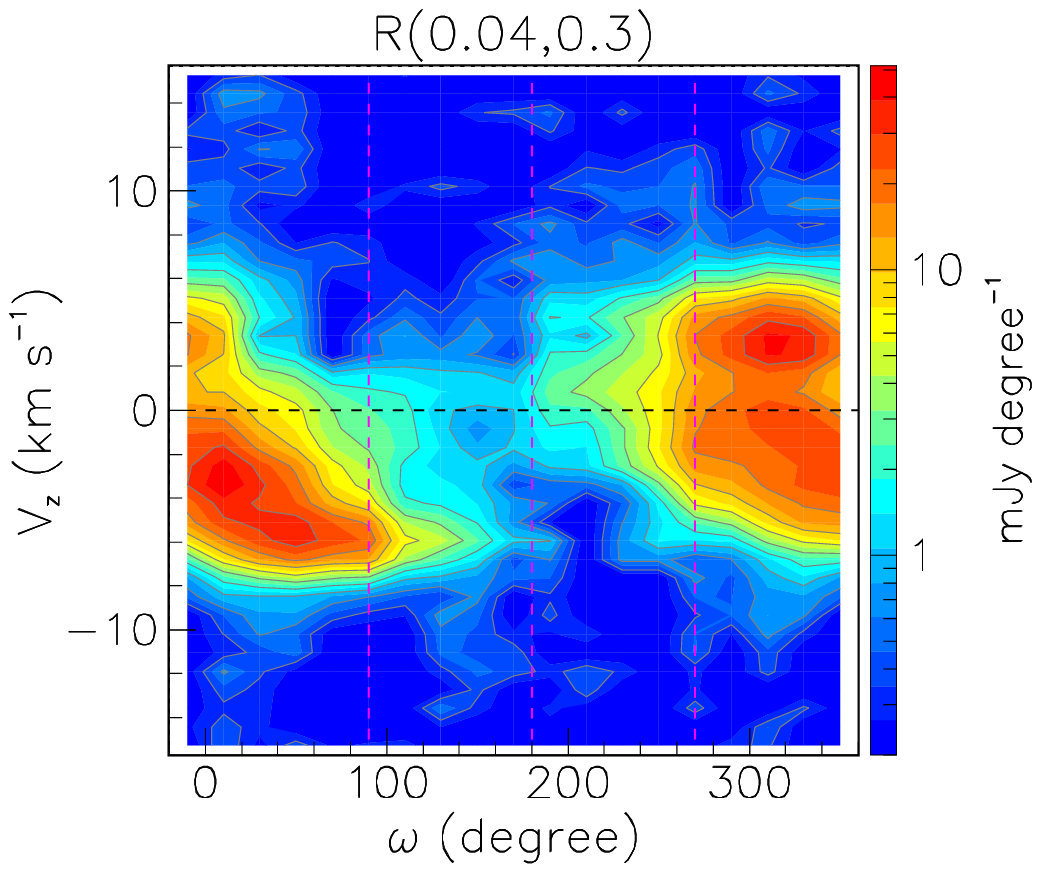}
  \includegraphics[height=3.5cm,trim=.0cm 1cm 0cm 1.3cm,clip]{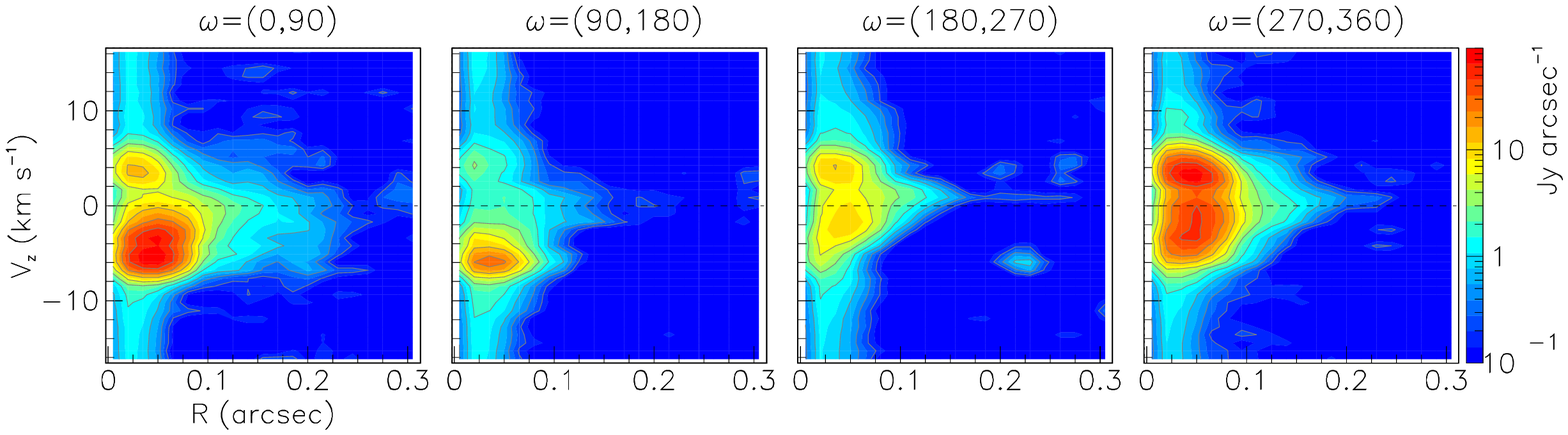}
  \caption{$^{28}$SiO($\nu$=1,8-7) emission. Same as Figure \ref{fig17}.}
  \label{fig19}
\end{figure*}

The emissions of the vibrationally excited SiO lines, $^{28}$SiO($\nu$=1,5-4) and $^{28}$SiO($\nu$=2,5-4) in Band 6 and $^{28}$SiO($\nu$=1,8-7) in Band 7, when compared to that of the $^{29}$SiO($\nu$=0,5-4) line, probe higher temperature regions closer to the star (Table \ref{tab2}). Accounting for the $^{29}$SiO/$^{28}$SiO isotopic ratio, the LTE values of the $J$=5-4 line intensities for $\nu$=0 and $\nu$=1 are equal for a temperature $T$$\sim$700 K and in a ratio $\sim$1/5 for $T$=2000K.
Figures \ref{fig17} to \ref{fig19} display the same information for these lines as Figure \ref{fig8} does for the ground state. The emission of the $\nu$=1 lines is dominantly north, evolving from north-east in the blue-shifted hemisphere to north-west in the red-shifted hemisphere, and is largely confined within 100 mas from the centre of the star, reaching up to 200 mas for the $J$=8-7 line. The emission of the $\nu$=2 line is limited to a small region of the emission of the $\nu$=1 lines, north of the star, with a mean Doppler velocity of $\sim$$-$5 \kms\ and confined within $\sim$80 mas. When inspected along $V_\text{z}$, the emission is seen to consist of three different components at mean Doppler velocities of $\sim$$-$8 to $-$2 \kms, $\sim$$-$2 to 2 \kms\ and $\sim$4 to 8 \kms, respectively.

\begin{figure*}
  \includegraphics[height=5cm,trim=.0cm 0.5cm 1.7cm 1.cm,clip]{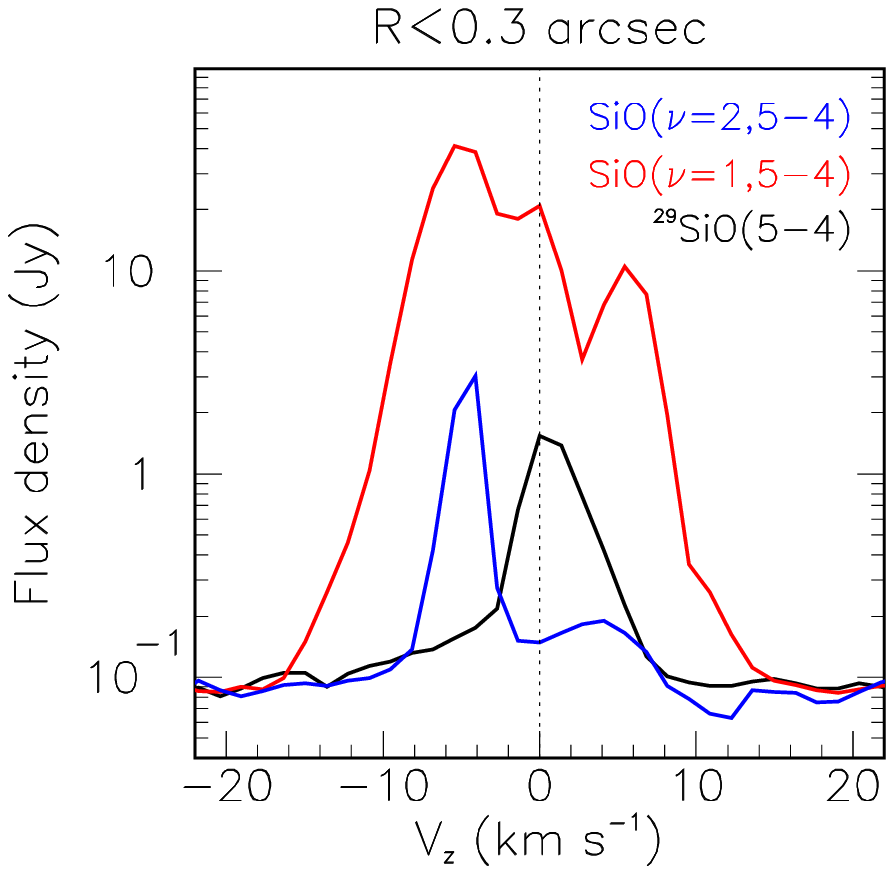}
  \includegraphics[height=5cm,trim=.0cm 0.5cm 1.7cm 1.cm,clip]{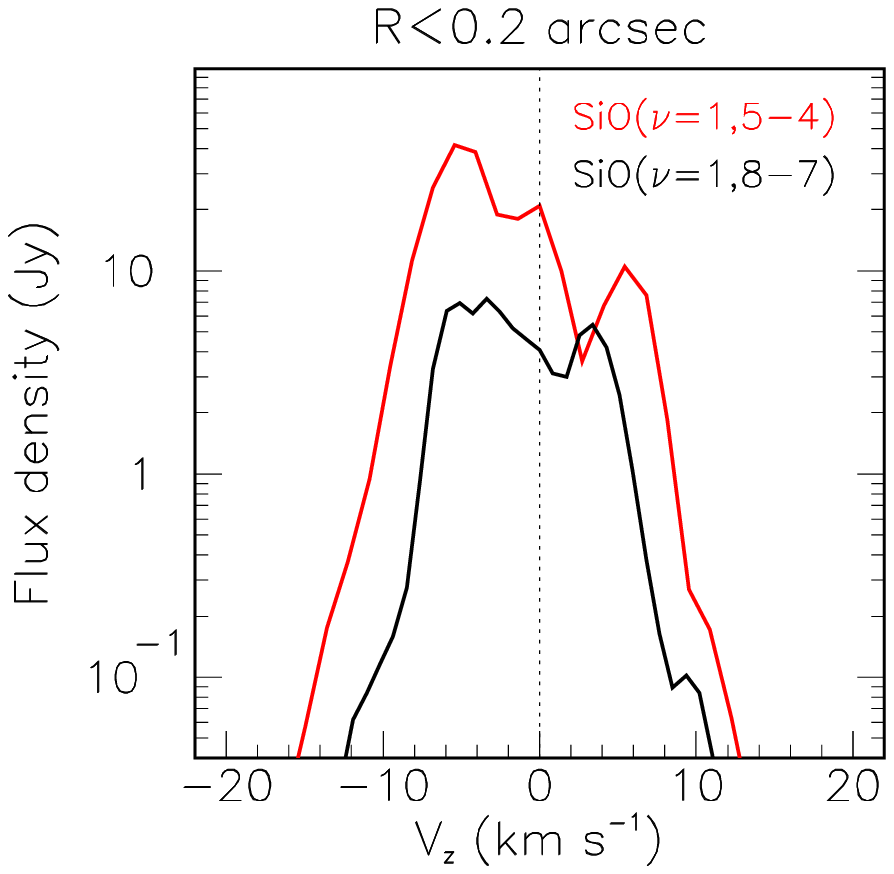}
  \includegraphics[height=5.5cm,trim=-1cm -1.cm 0cm 0.cm,clip]{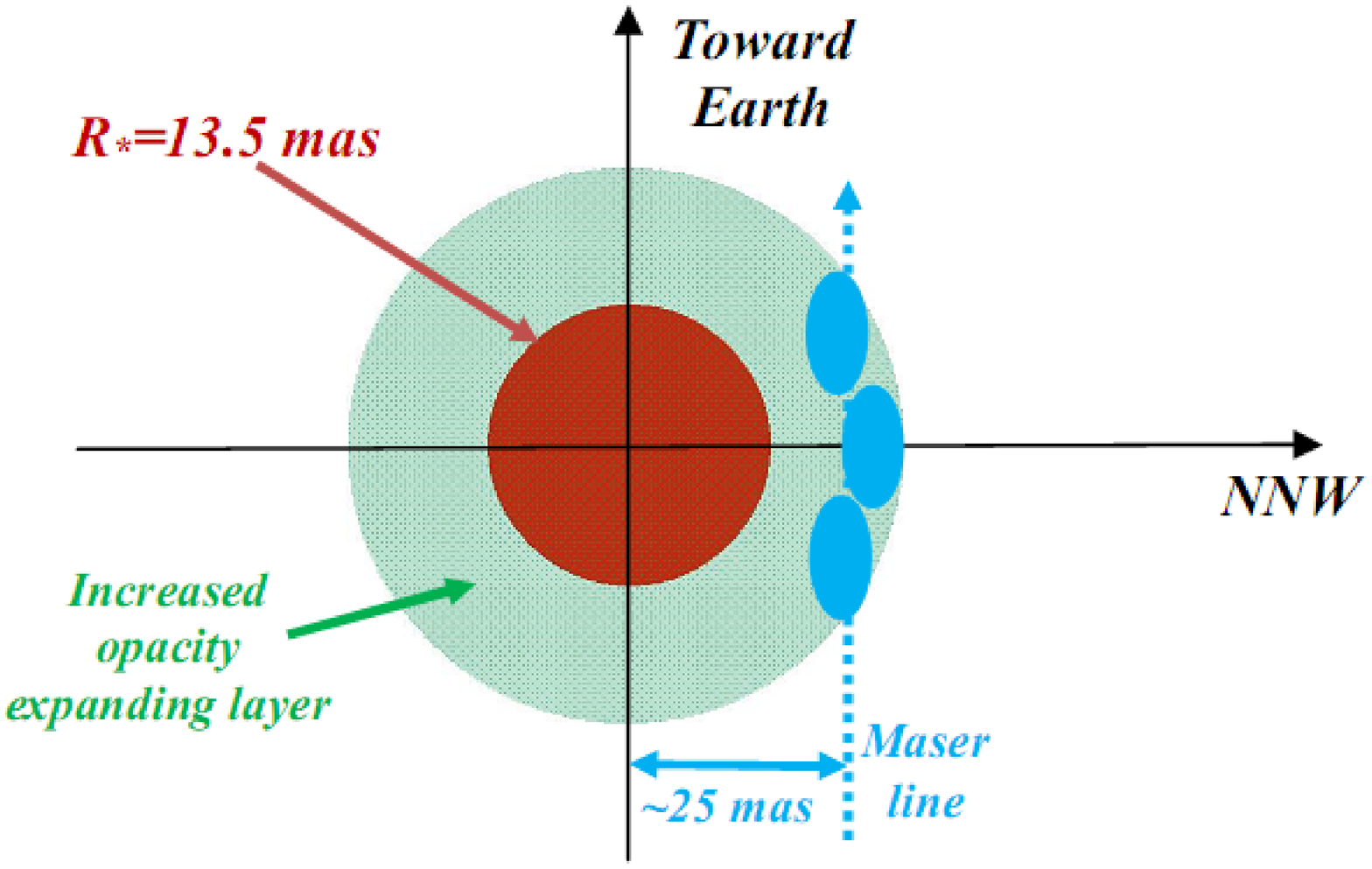}
  \caption{Left: Doppler velocity spectra of the SiO(5-4) line emissions integrated in the circle $R$<0.3 arcsec. Centre: Doppler velocity spectra of the SiO($\nu$=1,5-4) and SiO($\nu$=1,8-7) line emissions integrated in the circle $R$<0.2 arcsec (continuum subtracted). Right: schematic of the suggested geometry.}
  \label{fig20}
\end{figure*}

Figure \ref{fig20} (left) illustrates the peculiarity of the $^{28}$SiO($\nu$=1,5-4) emission by comparing its Doppler velocity spectrum, integrated within $R$<0.3 arcsec, with those of the $\nu$=0 and $\nu$=2 lines: its emission is an order of magnitude more intense and it covers a much broader interval of Doppler velocities, reaching beyond $\sim$$\pm$16 \kms. From Table \ref{tab2}, the LTE approximation predicts the intensity of the $\nu$=1, $J$=8-7 line to be a factor $\sim$7 larger than that of the $\nu$=1, $J$=5-4 line in the optically thin limit and $\sim$2.5 times larger in the optically thick limit, while in fact it is over 3 times smaller (Figure \ref{fig20}, centre). Such features are evidence for masing \citep{Gray2009}, the latter probably being an effect of the wavelength dependence of the stimulated emission probability. 

Indeed, the emission of the $\nu$>0 lines is known to host strong masers, the amplitude of which is highly variable. As was mentioned in the introduction, they have been the subject of numerous outstanding VLBA observations. As remarked by \citet{Cotton2009}, they appear within a few stellar radii of the stellar surface, typically $\sim$30 mas from the centre of the star, between the hot inner envelope and the cooler region where the dust forms. They tend to occur in clumpy, partial rings centred on the star.


The presence of masers does not simply trace local SiO abundance but is a function of the conditions for population inversion and the direction of maser beaming. Their amplification depends on the length of the path over which they can develop, and therefore takes values that are difficult to predict. The blue-and red-shifted emissions observed in the present case may reveal the presence of an expanding layer of gas reminiscent of the contemporaneous observation (October 3 and 11 instead of September 21 and October 27) of the extended atmosphere made by \citet{Vlemmings2019} within 1-2 stellar radii. These authors model the continuum emission with a uniform elliptical stellar disc and find a temperature of $\sim$1900$\pm$100 K at an angular separation of $\sim$22 mas from the centre of the star. The continuum emission probes the same radial range as do the $\nu$=1 and $\nu$=2 SiO lines and these authors model their observations as emission from an increased opacity layer expanding at a radial velocity of $\sim$10.6 \kms. If the actual expansion velocity of the SiO masers seen on the $\nu$=1 line is the same as the 10.6 \kms\ inferred for the continuum, then a $|V_z|\sim$6 \kms\ would place the masing clumps at an inclination angle of 30-40\dego\ with respect to the plane of the sky. The right panel of Figure \ref{fig20} sketches the corresponding geometry. The maximum SiO velocity (Figure \ref{fig16} and left panels of Figure \ref{fig20}) is $\sim$10 \kms, which is consistent with this picture. The clumpy nature of the emission (north in the blue-shifted hemisphere and north-west in the red-shifted hemisphere) cannot be simply interpreted in terms of gas density but simply reveals regions where masers are more intense.

\begin{figure*}
  \includegraphics[height=5cm,trim=.0cm 0.5cm 1.7cm 1.cm,clip]{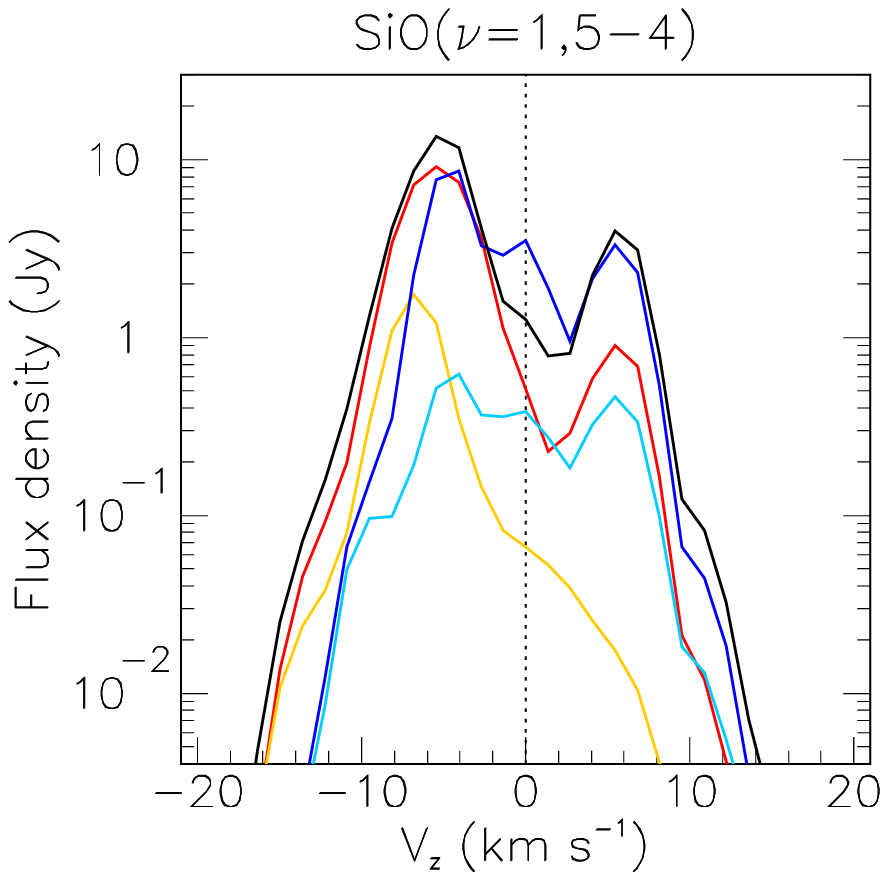}
  \includegraphics[height=5cm,trim=.0cm 0.5cm 1.7cm 1.cm,clip]{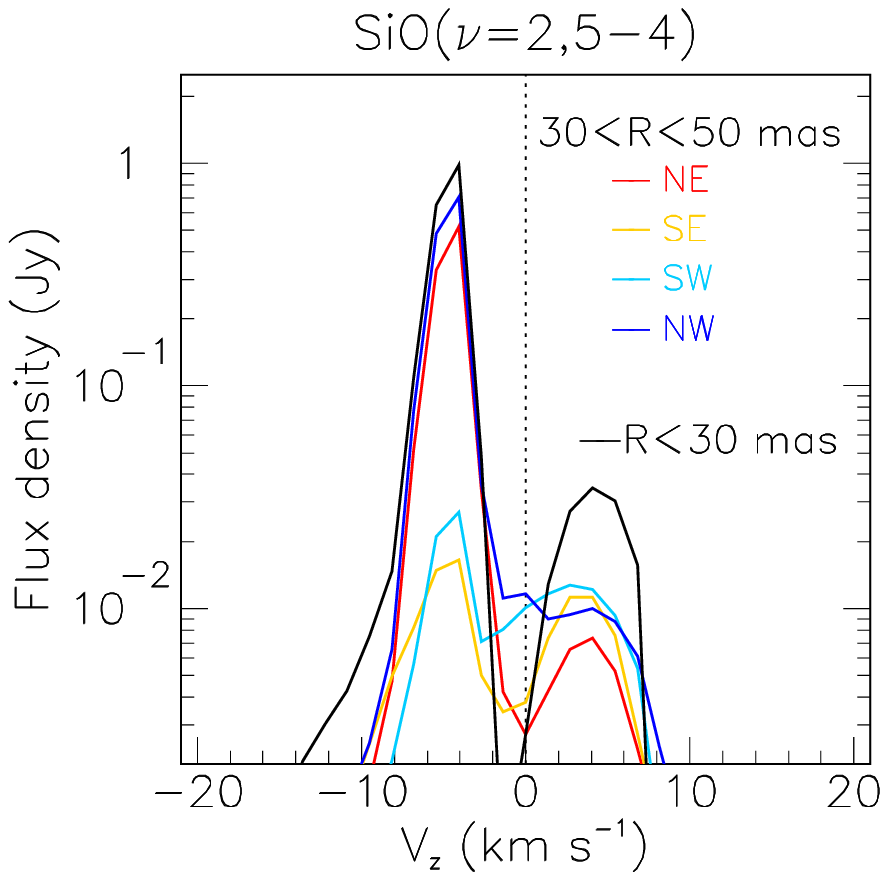}
  \includegraphics[height=5cm,trim=.0cm 0.5cm 1.7cm 1.cm,clip]{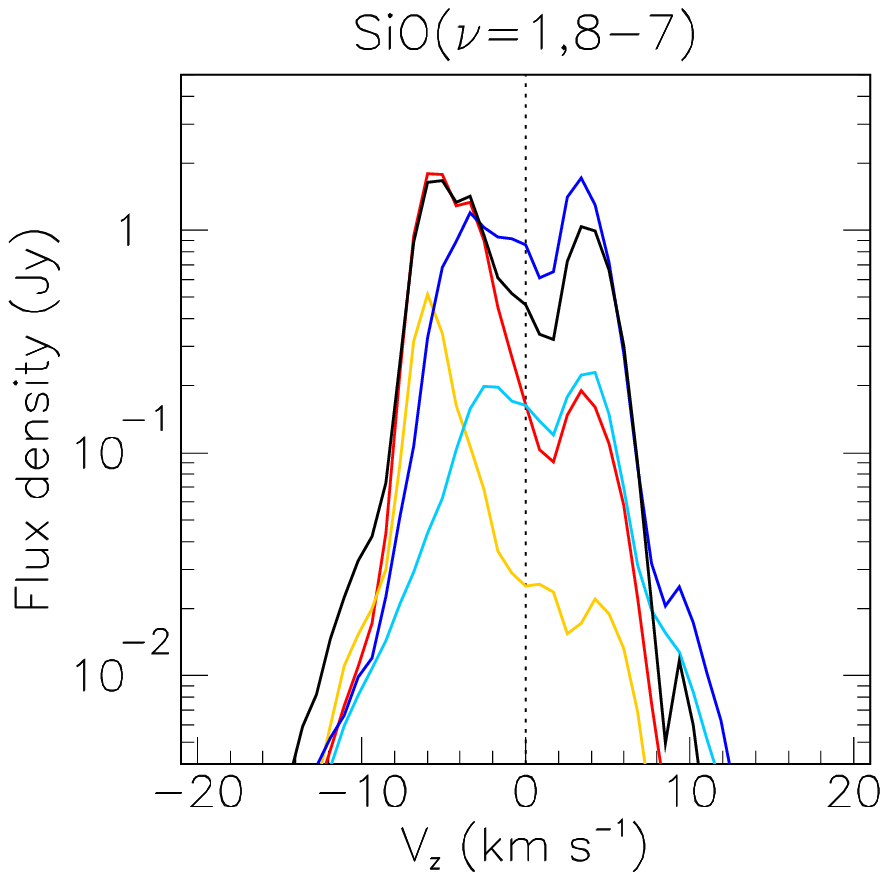}
  \caption{Same as Figure \ref{fig15} for transitions from vibrationally excited states. }
  \label{fig21}
\end{figure*}

No significant absorption can be revealed above and around the stellar disc (Figure \ref{fig21}), in strong contrast with the $\nu$=0 lines (Figure \ref{fig15}), the $\nu$>0 lines probing a region where masing is important and where the gas is too hot to absorb strongly, as was already the case for the SO$_2$(22$_{2,20}$-22$_{1,21}$) line (Section 5.2). 

\section{Conclusions}
A global view of the morpho-kinematics of the circumstellar envelope of R Leo has been presented, which complements earlier studies. In short, it distinguishes between four regions covering different radial ranges: within $\sim$2 stellar radii, using continuum emission and $\nu$>0 maser lines as probes, evidence is found for an expanding shell with a radial velocity of $\sim$10 \kms; between $\sim$2 and $\sim$10 stellar radii, in a region probed in particular by the emission of SO and SO$_2$ $\nu$=0 molecular lines and hosting in its inner part both in-falling and outflowing gas, evidence is found for a non-LTE regime suggesting that shocks play an important role; between $\sim$10 and $\sim$30 stellar radii, in a region probed in particular by the $\nu$=0 emission of SiO, CO and HCN molecules and hosting three distinct outflows, LTE becomes progressively dominant; beyond $\sim$30 stellar radii, in a region probed by the emission of CO molecules, evidence is found for patchy and episodic mass ejection, with an enhanced layer of emission having a mean radial velocity of $\sim$5.5 \kms\ and a mean angular separation from the centre of the star of $\sim$6 arcsec.

At variance with preliminary analyses presented by \citet{Fonfria2019a,Fonfria2019b}, we have shown that continuum emission is nearly circular and gives insufficient evidence to claim, or refute, the possible presence of an evaporating Jovian planet and/or of rotation in the close neighbourhood of the star. Instead, we give evidence for three distinct outflows, each covering a solid angle at steradian scale,which have been described in detail in Section 5. As they cover nearly 400 mas in projected angular distance from the centre of the star, and have a space velocity expected to not significantly exceed $\sim$6 \kms, they have been present for several decades. While it is tempting to associate such outflows with possible convective cells on the surface of the star, this seems an order of magnitude longer time than the expected lifetime of a given granulation pattern, underscoring the lack of detailed understanding of the mechanism at stake. A study of the relative abundances of the molecular species present in these outflows has revealed an unexpected confinement of SO and SO$_2$ molecules within $\sim$0.15 arcsec from the centre of the star, suggesting a complex competition between chemical reactions contributing to the formation and destruction of the parent SO molecules. The similar morpho-kinematics at stake for the emission of the SO and SO$_2$ molecules has given support that the latter are dominantly produced by oxygenation of the former. Evidence for different SO/SiO ratios in each of the three outflows is strong (lower panels of Figure \ref{fig14}) suggesting that at the short distances probed by the SO and SO$_2$ molecules, LTE does not apply and shocks are likely to play a significant role.

In contrast with SO and SO$_2$ molecules, the relative abundances of the CO, SiO and HCN molecules are consistent, within large uncertainties, with expectation.

Very close to the star, evidence for the simultaneous presence of expanding and in-falling gas, reminiscent of observations made in several other oxygen-rich AGB stars, has been presented and commented. A study of vibrationally excited SiO lines, expected to host important masers, has been shown in Section 6 to probe an expanding layer of hot gas, just above the photosphere, consistent with earlier studies by \citet{Vlemmings2019}.

At larger distances from the star, where depletion of the SiO molecules by condensation on dust grains and photo-dissociation of the SiO, SO and SO$_2$ molecules are expected to be significant, the CO line emission traces the mass loss history over the past few centuries, giving evidence for a very patchy morpho-kinematics with inhomogeneity at the level of $\sim$40\%, and revealing an episode of enhanced mass-loss seen as a layer of increased emission at a mean distance of 6 arcsec from the centre of the star and an expansion velocity reaching up to 9 \kms\ and having an average value of $\sim$5.5 \kms. An escape velocity of 6 \kms\ from a 0.7 solar mass star is reached at $\sim$300 mas: some acceleration is therefore necessary after the initial boost. Understanding its nature requires independent information about dust formation. The dust emission coefficient in the mid-infrared is known to be low \citep{Sloan1998} and \citet{Paladini2017}, using VLTI/MIDI observations, remark that R Leo displays no significant 8 $\mu$m silicate feature in its spectrum and speculate that the important brightness asymmetry supports a large grain scenario of dust formation.

We noted the presence of patches of emission inside the shell, in particular of some arcs that may suggest the presence of spirals associated with the wake emission of a companion; however, we have shown that such an interpretation cannot be reliably supported or refuted from the present data alone.

The complexity of the observed morpho-kinematics, which had been underscored by all earlier studies, together with the strong anisotropy of line emission ratios, provide evidence for non-equilibrium chemistry, supporting the role played by shocks and/or density and temperature inhomogeneity in the generation of the nascent wind. But it also makes it hard to develop accurate, comprehensive models with confidence, restricting reliable interpretations to considerations based on very crude and approximate modelling. When observed in sufficient detail, the inner layers of the circumstellar envelopes of oxygen-rich AGB stars are all different. Their observation shows the importance of collecting a sufficient sample to be able to interpret reliably their diversity in terms of state-of the art simulations such as proposed by \citet{Hofner2019}, in particular for being able to describe more precisely the role played by convective cells in defining the pattern of mass ejections. While such simulations provide a useful framework in which to describe the mechanisms at stake in the genesis of the nascent wind, many questions remain unanswered, requiring new observations: How does the pattern of outflows depend on the stellar phase? What distinguishes outflows lasting for a few decades from mass ejections covering only a few years? What causes the CSE to retain, in some cases, a same axial symmetry over centuries? An accurate model of the morpho-kinematics of the observed line emissions in the neighbourhood of the star should be able to define precisely the angular size and time duration of the patches of emission associated with the convective cell granulation, but we are still far from such achievement.

\section*{Acknowledgements}

We are deeply indebted and grateful to the referee, Dr Anita Richards, for a very careful reading of the manuscript and pertinent comments that helped greatly with improving both the form and the substance of the present article. We thank Dr St\'{e}phane Guilloteau for guidance in producing Table 2. This paper makes use of the following ALMA data: ADS/JAO.ALMA\#2016.1.01202.S, ADS/JAO.ALMA\#2017.1.00862.S. ALMA is a partnership of ESO (representing its member states), NSF (USA) and NINS (Japan), together with NRC (Canada), MOST and ASIAA (Taiwan), and KASI (Republic of Korea), in cooperation with the Republic of Chile. The Joint ALMA Observatory is operated by ESO, AUI/NRAO and NAOJ. We are deeply indebted to the ALMA partnership, whose open access policy means invaluable support and encouragement for Vietnamese astrophysics. Financial support from the World Laboratory, the Odon Vallet Foundation and the Vietnam National Space Center is gratefully acknowledged. This research is funded by the Vietnam National Foundation for Science and Technology Development (NAFOSTED) under grant number 103.99-2019.368.

\section*{Data Availability}
The raw data are available on the ALMA archive: ADS/JAO.ALMA\#2016.1.01202.S and ADS/JAO.ALMA\#2017.1.00862.S. The calibrated and imaged data underlying this article will be shared on reasonable request to the corresponding author.



\bibliographystyle{mnras}
\bibliography{rleo_clean}




\bsp	
\label{lastpage}
\end{document}